\documentclass[12 pt]{report}

\usepackage{showlabels}
\usepackage[english]{babel}
\usepackage{graphicx}
\usepackage{amsmath}
\usepackage{amssymb}
\usepackage{multicol}
\usepackage{braket}
\usepackage{latexsym} 
\usepackage{verbatim}
\usepackage{cancel} 
\usepackage{slashed}
\usepackage{hyperref}
\usepackage{color}
\usepackage{tensor}
\usepackage{framed}
\usepackage[usenames,dvipsnames,svgnames,table]{xcolor}
\usepackage{tikz}
\usepackage{cite}
\usetikzlibrary{calc}
\usetikzlibrary{shapes}

\graphicspath{}
\DeclareMathSizes{11.5}{30}{16}{12}

\begin{document}

\title{A short review on Noether's theorems, gauge symmetries and boundary terms}

\author{Max Ba\~{n}ados and Ignacio Reyes \\
 Facultad de F\'{i}sica, Pontificia Universidad Cat\'olica de Chile,  \\ Casilla 306, Santiago, Chile}

        \maketitle

\abstract{ This review is dedicated to some modern applications of the remarkable paper written in 1918 by E. Noether.  On a single paper, Noether discovered the crucial relation between symmetries and conserved charges as well as the impact of gauge symmetries on the equations of  motion. Almost a century has gone since the publication of this work and its applications have permeated modern physics. Our focus will be on some examples that have appeared recently in the literature. This review is aim at students, not researchers. The main three topics discussed are (i) global symmetries and conserved charges (ii) local symmetries and gauge structure of a theory (iii) boundary conditions and algebra of asymptotic symmetries. All three topics are discussed through examples.}

\newpage

\tableofcontents

\newpage
\chapter{Preface}

Emmy Noether's famous paper, \textit{Invariante Variationsprobleme}, was published in \textit{Nachr. d. K¬onig. Gesellsch. d. Wiss. zu G¬ottingen, Math-phys. Klasse} in 1918 \cite{Noether},\cite{Tavel}. In this paper, Noether proves \textit{two} different theorems.  The First Theorem deals with ``global" symmetries (generated by finite Lie groups) and states that these symmetries lead to conserved charges. The Second Theorem applies to local gauge symmetries (infinite dimensional Lie groups), containing arbitrary functions of spacetime (like Einstein's theory of gravity) and shows that these gauge symmetries inevitably lead to relations among the equations of motion (e.o.m. onwards). 

This review is dedicated to the applications of Noether's paper and the fundamental results uncovered by it right at the birth of the `modern physics era'. As a basic outline, we discuss the following aspects of classical field theory:

\begin{enumerate}
\item Noether's theorem for non-gauge symmetries; energy-momentum tensor and other conserved currents
\item Gauge symmetries, hamiltonian formulation and associated constraints
\item Asymptotics conditions, boundary terms and the asymptotic symmetry group
\end{enumerate}

Our focus will be on examples, some of them developed in great detail. We shall leave historical and advanced considerations aside and be as concrete as possible. As our title explains, this work is dedicated mostly to graduate students who, in our experience, often find it difficult to feel familiar with Noether results mainly because in most texts the only example displayed is the Poincare group and associated conserved charges. We shall discuss many examples both in particle mechanics and field theory.

We start with global (``rigid") symmetries in modern language and then proceed with gauge symmetries, trying to be as systematic as possible. In the final chapter we address the role and importance of asymptotic boundary conditions and their associated boundary terms, a subtle point often neglected. 

We assume the reader has a basic knowledge of classical field theory, its Euler-Lagrange equations and the basics of  Hamiltonian mechanics. 

Noether theorem is almost hundred years old and has been discussed in many textbooks. It is impossible to give a full account of the literature available. The treatment of gauge theories in Hamiltonian form was initiated by Dirac long ago.   For the purposes of this review, the book by Henneaux and Teitelboim \cite{HT} is of particular importance. The issue of boundary terms and conserved charges in gauge theories has a more recent development. There are several approaches to this subject. For our purposes the classic work of Regge and Teitelboim \cite{Regge-Teitelboim} is the starting point. See \cite{Jamsin} for a recent discussion. A powerful tool to compute conserved currents was devised in \cite{Barnich}. Other important references include \cite{AbbotDeser,AshDas}; and of course the work that descends from Maldacena's AdS/CFT duality, for example\cite{dHSS} (see also \cite{ABR} for a discussion along the ideas presented here).

~

The authors will welcome all comments about this document that may help to improve its presentation (maxbanados@gmail.com; iareyes@uc.cl). We apologise for not giving a full account of the excellent literature available. This review does not contain original material. It only represents a pedagogical presentation of the subjects covered, aim at students.


\newpage
\chapter{Noether's theorem for global symmetries in classical mechanics and field theory}

Classical mechanics, classical field theory and to some extent quantum theory all descend from the study of an action principle  of the form
\begin{align}\label{action0}
I[q^i(t)]=\int dt\ L(q^i,\dot{q}^i,t)
\end{align}
and its associated Lagrange equations, derived from an extremum principle with fixed endpoints,
\begin{align}\label{eom}
\frac{d}{dt} \left( \frac{\partial L}{\partial \dot{q}^i} \right)-\frac{\partial L}{\partial q^i}=0\ .
\end{align}
The variables $q^i(t)$ can represent the location of particles on real space, fluctuations of fields, in some cases Lagrange multipliers, or auxiliary fields.   
Actions of the form (\ref{action0}) encode a huge number of situations. The importance of action  principles in modern mechanics cannot be overstated. 

In this review we assume some basic knowledge regarding the principles of classical mechanics in both Lagrangian and Hamiltonian form, and jump directly to some of its notable applications, and particularly in the beautiful relation between symmetries and conservation laws, discovered by Noether \cite{Noether}.

\section{Noether's theorem in particle mechanics}\label{particles}

\subsection{The theorem}\label{The theorem}

Noether's theorem is often associated to field theory, but it is a property of any system that can be derived from an action and possesses some continuous (non-gauge) symmetry. In words, to any given symmetry, Neother's algorithm associates a conserved charge to it.

The key idea follows from a relation between \textit{on-shell variations} and \textit{symmetry variations} of an action. In this section we review its formulation, illustrating its features by considering some simple examples.

\subsubsection{Noether symmetries}

The crucial concept exploited by Noether is that of an action symmetry. This concept is subtle and often a source of confusion. Consider the simplest example provided by the action, 
\begin{align}
I[x(t),y(t),z(y)] = \int dt  \left( \frac{m}{2}(\dot {x}^2 +\dot {y}^2 + \dot {z}^2) - H_0 z \right). 
\end{align} 
where $H_0$ is a constant. This action is clearly invariant under constant translations in $x(t)$ and $y(t)$, 
\begin{align}\label{sss}
I[x(t) + x_0, y(t) + y_0, z(t)] = I[x(t),y(t),z(y)].
\end{align} 
This property is the basic example of a Noether symmetry.  
The following important aspect of (\ref{sss}) should not be overlooked: equation (\ref{sss}) holds for {\it all} $x(t),y(t),z(t)$. This seems like a trivial statement in this example but it is a crucial property of action symmetries.  

Symmetries can take far more complicated forms. Let $q^i$ be a set of generalised coordinates. Then for an action $I[q^i(t)]$, a (small) function $f^i(t)$ is a symmetry if $I[q^i(t) + f^i(t)] = I[q^i(t)]$, for all $q(t)$. Symmetries are directions in the space spanned by the $q^i$'s on which the action does not change.  The function $f^{i}(t) $ can be very complicated!  

For Noether's theorem one is interested in infinitesimal symmetries and it is customary to denote them as variations using the Greek letter $\delta$. The functions $f^{i}(t)$ in the example above will be denoted as $f^i(t) = \delta_s q^i(t)$ (the subscript $s$ denotes symmetry), and $I[q^i + \delta_s q^i(t)]$ is expanded to first order. This notation may cause some confusion because, apparently, implies a relation between $q^{i}(t) $ and $\delta_s q^{i}(t)$. 
The reader must keep in mind that $q^i(t)$ and $\delta_s q^i(t)$ are totally independent functions. $\delta_s q^i(t)$ defines directions on which the action does not change. 

So far we have mentioned the strong version of a symmetry where the action is strictly invariant. Noether's theorem accepts a weaker version. From now on we define a symmetry as a function $\delta_s q^{i}(t) $ such that, for {\it any} $q^i(t)$, the action is invariant,
\begin{eqnarray}\label{sym}
\delta I[q^{i}(t),\delta_s q^i(t)] &\equiv & I[q^i(t) + \delta_s q^i(t)] - I[q^i(t)] = \int dt \frac{dK}{dt},
\end{eqnarray}  
up to a boundary term that we denote by $K$. We shall see that the boundary term $K$ does not interfere with the existence of a conserved charge, and that for many important examples it is non-zero and indeed contributes to the charge.  We have also introduced the notation   $\delta I$ (the variation of the action under the symmetry) which is a function of both, the configuration $q^i(t)$ and the symmetry $\delta_s q^i(t)$\footnote{Let us get this point clear. Given a function $f(x)$ one can consider the variation $f(x+dx)-f(x)= f'(x)dx \equiv df(x,dx)$: the variation depends on both, the point $x$ and the magnitude of the variation $dx$.}.

Any function $\delta_s q^i(t)$ that satisfies (\ref{sym}) represents a symmetry. Eqn. (\ref{sym}) must be understood as an equation for $\delta_s q^i(t)$. If, for a given action $I[q^i(t)]$, we find all functions $\delta_s q^{ i}(t) $ satisfying (\ref{sym}), then we have solved the equations of motion of the problem. 
The central force problem in mechanics and the conformal particle (see below) are examples where the equations are solved by looking at symmetries. We stress, for the last time, that $\delta_s q^i(t)$ is a symmetry if and only if it satisfies (\ref{sym}) for arbitrary $q^i(t)$. 

\subsubsection{Some examples and comments}

Invariance under rotations is an important example. The central force problem
\begin{align}\label{pp}
I[\vec{r}(t)] = \int dt\, \left( {m \over 2}\, \dot{\vec{r}}\,^2 - V(r) \right)
\end{align} 
is invariant under  $\vec{r} \rightarrow \vec{r}\,'$ where  
\begin{equation}
\vec{r}\ '(t) = R\, \vec{r}(t)
\end{equation}  
and $R$ is a constant orthogonal matrix $R^{T} = R^{-1}$. To put this symmetry in Noether's language we first consider its infinitesimal version.  For small angles $\alpha$ one has $R\, \vec{r} = \vec{r} + \vec{\alpha}\times\vec{r}$ and therefore
\begin{equation}\label{rot}
\vec{r}\,'(t) = \vec{r}(t) + \vec{\alpha} \times \vec{r}(t)   \ \ \ \Rightarrow \ \ \ 
\delta_s \vec{r}(t) = \vec{\alpha} \times \vec{r}(t), 
\end{equation} 
The reader is encouraged to check explicitly that  the action (\ref{pp}) satisfies $I[\vec{r}+ \vec{\alpha} \times \vec{r}] = I[\vec{r}]$, for any $\vec{r}(t)$. In this example $K=0$.

A simple example of a symmetry with $K \neq 0$ is the invariance of the same action (\ref{pp}) under a different symmetry 
\begin{equation}\label{dr}
\delta_s \vec{r}(t) = - \epsilon\, \dot{\vec{r}}(t)
\end{equation} 
where $\epsilon$ is constant. (See below for an interpretation.) By direct calculation, 
\begin{align}\label{}
\delta I[\vec{r},\delta_s \vec{r}] &= \int dt \, \left( m\dot {\vec{r}}\cdot \delta_s \dot{\vec{r}} - \nabla V  \delta_s \vec{r} \right) \nonumber \\
 &= \int dt\, \epsilon ( - m\dot{\vec{r}}\cdot \ddot{\vec{r}} + \nabla V \cdot \dot{\vec{r}}  ) \nonumber\\
 &= \int dt {d \over dt}  \left(-\epsilon {m \over 2} \dot {\vec{r}}\,^2  + \epsilon V(\vec{r}) \right). \label{ttt}
\end{align}
Hence, the action varies up to a boundary term, which is this case is equal to $K = -\epsilon  \left({m \over 2} \dot {\vec{r}}\,^2  -  V(\vec{r}) \right)$.  

The symmetry (\ref{dr}) is related to spacetime translations in the following way. To simplify the figure, consider the 1-dimensional case where the coordinate is called $q(t)$.  
\begin{figure}\label{figu1}
\centerline{\includegraphics[scale=0.4]{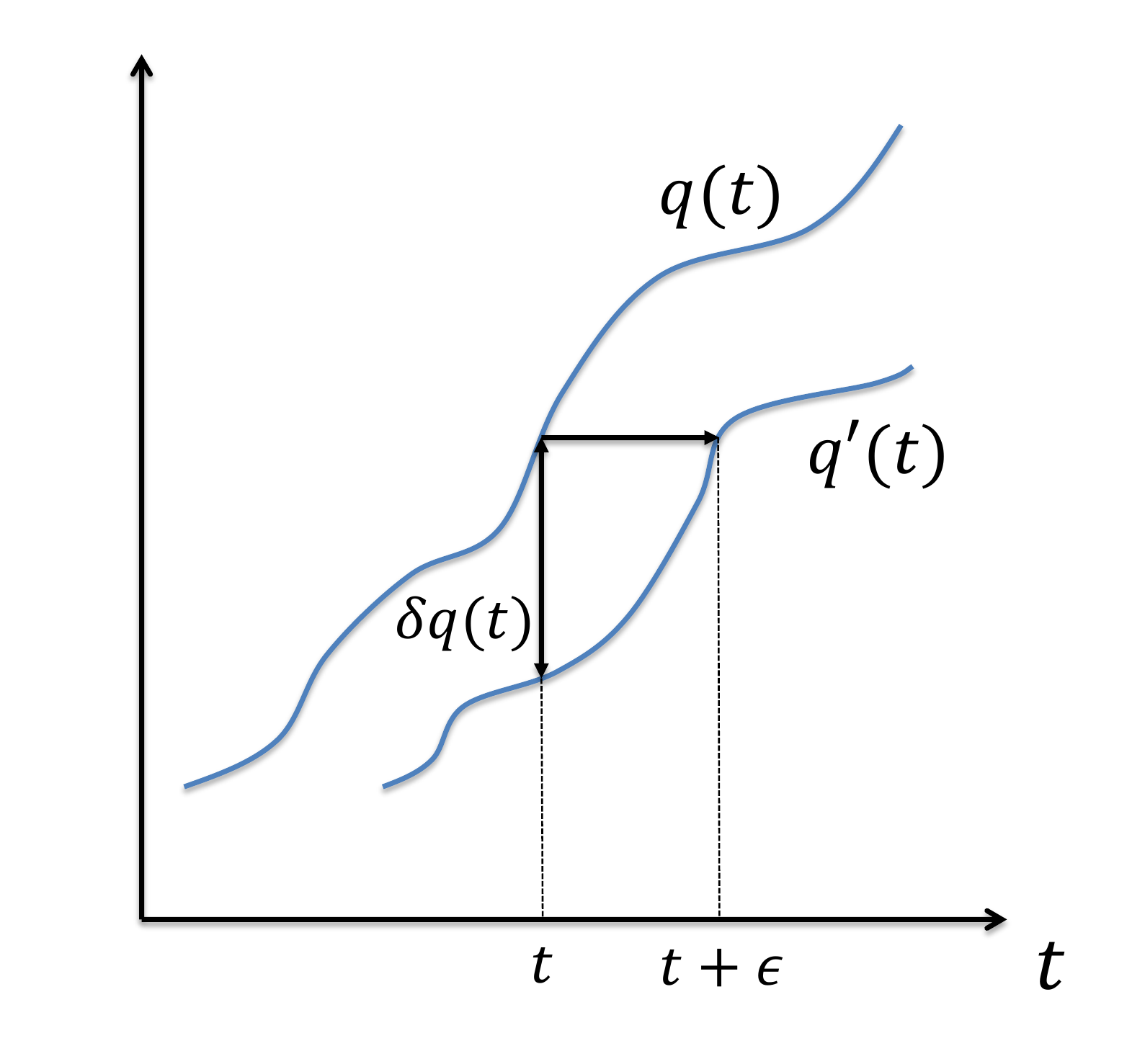}}
\caption{ Two functions $q(t)$, $q'(t)$ related by a time translation.  At any given time $t$, $\delta q(t)$ represents the difference between both functions.}
\end{figure}
 In figure (\ref{figu1}) we have drawn two functions $q(t)$ and $q'(t)$ which are related by a time translation of magnitude $\epsilon$. Directly from the picture we find that the values of $q'(t)$ are related to those of $q(t)$ via:
\begin{equation}
q'(t+\epsilon) = q(t)\ .
\end{equation} 
If $\epsilon$ is treated as an infinitesimal quantity, this equation can be written as
\begin{equation}\label{trans}
q'(t) + \epsilon \dot q(t) = q(t) \ \ \  \Rightarrow \ \ \  \delta_s q(t) = -\epsilon \dot q(t)
\end{equation} 
where $\delta_s q(t) = q'(t) - q(t)$ is the difference of the two functions evaluated at the same argument $t$ (see the figure). Two comments are in order:
\begin{enumerate}
\item The symmetry is represented by the function $\delta_s q(t) = -\epsilon \dot q(t)$ which involves only one time.  We have transmuted the time translation into a deformation of the function $q(t)$. 
\item Since  $\delta_s q(t) = q'(t) - q(t)$ is simply the difference of two functions evaluated at the same time $t$ (see figure), it follows directly that $\delta_s \Big( {d \over dt}q(t)\Big) = {d \over dt}\delta_s q(t)$.  
\end{enumerate}

This will be our \textit{modus operandis} throughout the text. Symmetries will always be deformations of the fields, not the coordinates. Note that from the point of view of the action $I = \int dt L$, the time $t$ is a ``dummy variable" that can be changed at will, without affecting the value of $I$; just like when doing the integral $\int_{t_1}^{t_2}  e^{t^2} t\, dt$ one is allowed to replace $u = t^2$ simplifying greatly the value of the integral (adjusting the limits). This kind of transformations have nothing to do with Noether theorem or conserved charges. Of course there is nothing wrong with changing variables inside an integral, but mixing these kind of transformations with true symmetries --deformations of the dynamical variables-- makes the theorem far more complicated than it needs to be\footnote{It is useful to elaborate this point in field theory. It is often said that gauge transformations $A_\mu'(x)= U^{-1} A_\mu U + U^{-1} \partial_\mu U$ are local because they involve only the point $x$, while changes of coordinates $ A_\mu'(x') = {\partial x^\nu \over \partial x^{'\mu} } A_\nu(x)$ 
are non local because they involve two points $x$ and $x'$. The interpretation where $x$ and $x'$ are the same place in different coordinate systems (passive) or different places (active) has been discussed many times. In our discussion this distinction is irrelevant. The infinitesimal versions are, 
\begin{eqnarray}
\delta_{\mbox{\scriptsize{gauge}}}A_\mu(x) &:=& A'_{\mu}(x) - A_\mu(x) = D_{\mu}\lambda(x) \nonumber\\
\delta_{\mbox{\scriptsize{diff}}} A_{\mu}(x) &:=& A'_{\mu}(x) - A_\mu(x) = \xi^\nu(x) A_{\mu,\nu}(x) - \xi^\nu_{,\mu}(x) A_\nu(x)\label{ddiff}
\end{eqnarray} 
where all fields and parameters are evaluated at the same point $x$.}.

\subsubsection{On-shell variations}

We now discuss a different type of action variation, the on-shell variation. This variation is somehow the opposite to a symmetry. For symmetries, the variations $\delta_s q^i(t)$ are constrained to satisfy an equation, while the ``fields" $q^i(t)$ are totally arbitrary. For on shell variations, the fields $q^i(t)$ are constrained to satisfy their Euler-Lagrange equations while the variations $\delta q^i(t)$  are arbitrary. 

Let $\delta q^i(t)$ be an arbitrary infinitesimal deformation of the variable $q^i(t)$. Then, for an action of the form $I[q] = \int dt L(q,\dot q)$ the variation $\delta I[q] \equiv I[q+\delta q]-I[q]$ can be expressed as,
\begin{align}
\delta I[q^i,\delta q^i] &= \int dt  \left( \frac{\partial L}{\partial q^i}\delta q^i+\frac{\delta L}{\delta \dot{q}^i}\delta \dot{q}^i \right) \nonumber\\ 
&=\int dt\left( \frac{\partial L}{\partial q^i} - \frac{d}{dt} \left( \frac{\partial L}{\partial \dot{q}^i} \right) \right) \delta q^i + \int dt {d \over dt} \left(  {\partial L \over \partial \dot q^i}\delta q^i \right) .
\end{align} 
where in the second line we have made an integral by parts. This equation contain a powerful piece of information. 
If $q^i(t)$ satisfies its Euler-Lagrange equations \eqref{eom}, the bulk contribution vanishes and the variation is a total derivative,
\begin{eqnarray}\label{dosI2}
\delta I[\bar q^{i}(t),\delta q^i(t)] &\equiv & I [\bar q^i + \delta q^i]  - I[\bar q^i] =  \int dt {d \over dt} \left(  {\partial L \over \partial \dot q^i}\delta q^i \right). 
\end{eqnarray} 
The bar over $\bar q^{i}(t)$ reminds that this variation is evaluated on a solution to the Euler-Lagrange equations. On the other hand, (\ref{dosI2}) is valid for \textit{any} $\delta q^i(t)$.  This variation depends on the particular solution chosen $\bar q^i$, and the arbitrary variation $\delta q^i$. 

\subsubsection{Noether's theorem} The combination of a symmetry with an on-shell variation gives rise to Noether theorem. Recapitulating, a symmetry is defined by the equation
\begin{equation}\label{sym3}
\delta I[q^i(t),\delta_s q^i(t)] = \int dt {dK \over dt},  
\end{equation} 
and, an on-shell variation is defined by
\begin{equation}\label{os2}
\delta I[\bar q^i(t),\delta q^i(t)]   =\int dt {d \over dt} \left(  {\partial L \over \partial \dot q^i}\delta q^i \right). 
\end{equation} 
Both variations are boundary terms but for very different reasons. (\ref{sym3}) is a boundary term because   $\delta_s q^i$ satisfies a particular equation, while (\ref{os2}) is a boundary term because $\bar q^i(t)$ satisfies a particular equation. On the other hand, $q^{ i}(t) $ in (\ref{sym3}) is totally arbitrary, while $\delta q^i(t)$ in (\ref{os2}) is totally arbitrary. 

Inserting $q^i(t)=\bar q^i(t)$ into (\ref{sym3}) and $\delta q^{i}(t)= \delta_s q^i(t)$ into (\ref{os2}) the left hand sides of these two equations are equal. Subtracting, the left hand sides cancel, and from the right hand sides we obtain the conservation law, 
\begin{equation}\label{qn}
\boxed{ {d \over dt} Q =0 \ \ \ \mbox{   with   } \ \ \   Q =  K -  \frac{\partial L}{\partial \dot{q}^i}\delta_s q^i .  }
\end{equation} 
This is Noether's First theorem: Given a symmetry $\delta_s q^i(t)$, the combination $Q$ showed in (\ref{qn}) is conserved. 

It is impossible to master Noether theorem without making several examples. The derivation is very simple but very few students, if any, understands it at first sight. We shall discuss many applications of this extraordinary result from 2-dimensional conformal field theory to Maxwell's electrodynamics. 

Our first two examples are the symmetries of the central force potential (\ref{pp}) discussed above. The action (\ref{pp}) is invariant under rotations with $K=0$. The conserved quantity is therefore  
\begin{eqnarray}
Q_\alpha  &=& -m\,\dot{\vec{r}} \cdot \delta_s \vec{r} \nonumber \\
&=&- m\, \dot{\vec{r}} \cdot (\vec{\alpha}\times \vec{r}) \nonumber \\
&=& -\vec{\alpha} \cdot \Big( m\, \vec{r}\times \dot{\vec{r}} \Big)\ . \label{angular}
\end{eqnarray}  
Since $\vec{\alpha}$ is constant and arbitrary we conclude the conservation of angular momentum $\vec{L} = m \vec{r}\times \dot{\vec{r}} $. For the time translation invariance (\ref{dr}), we showed in \eqref{ttt} that $K = -\epsilon L$. We then conclude that the conserved quantity is, 
\begin{eqnarray}
Q_\epsilon = -\epsilon\left( {m \over 2} \dot{\vec{r}}\,^{2} - V \right) + \epsilon\, m\, \dot{\vec{r}}\,^2 = \epsilon E
\end{eqnarray} 
where $E$ is the total energy. 

The 4 conserved charges $E = {m \over 2} \dot r^2 + V(r)$ and $\vec{L} = m \vec{r}\times \dot{\vec{r}}$ of the central force problem allows a simple solution to the equations of motion $m\ddot{\vec{r}} = -\nabla V$. These are first integrals and the full solution is found by doing one integral. We now discuss an example where the full solution is found purely by looking at symmetries.

\subsection{The `conformal' particle} \label{confpart}

A remarkable application of Noether's theorem  is a `conformal' particle. In this example one can solve the e.o.m. without even having to write them down.  Just by looking at the symmetries and making use of Noether's theorem, one can completely integrate the dynamics. 

Consider a particle of mass $m$ under the influence of an inverse quadratic potential
\begin{align}\label{Iconformal}
I[x]=\int dt \left(\frac{1}{2}m\dot{x}^2-\frac{\alpha}{x^2}  \right).
\end{align}
The equation of motion is 
\begin{equation}\label{con10}
m \ddot x = {2\alpha \over x^3}\ .
\end{equation} 
We shall find its full solution by looking at symmetries of the action. We first note that the Lagrangian is not explicitly time dependent hence the total energy  is conserved:
\begin{align}\label{EWeyl}
E=\frac{1}{2}m\dot{x}^2+\frac{\alpha}{x^2}\ .
\end{align}
This equation provides an algebraic relation between $x(t)$ and $\dot x(t)$. We shall now find a second equation of this type, which will fix $x$ and $\dot x$ completely.    

The second equation follows from the \textit{Weyl} symmetry of this system:
\begin{align}\label{dilationsym}
t\rightarrow t'=\lambda t\hspace{1cm} x\rightarrow x'(t')=\sqrt{\lambda}\ x(t).
\end{align}
for constant $\lambda$. Indeed, under this transformation, $dx/dt\rightarrow d(\sqrt{\lambda} x)/d(\lambda t)=\frac{1}{\sqrt{\lambda}}dx/dt$, and the action remains invariant,
\begin{align}\label{}
I\rightarrow \int \lambda dt \left( \frac{1}{2}m \frac{\dot{x}^2}{\lambda}-\frac{\alpha}{\lambda x^2} \right)=I \ .
\end{align}

Now, the transformation \eqref{dilationsym} is not useful for Noether's theorem as it stands.  We need to convert it into an infinitesimal variation acting on $x(t)$ at some time $t$. Let $\lambda=1+\epsilon$ with $\epsilon\ll 1$. We expand the transformation of $x(t)$ in (\ref{dilationsym}) to first order in $\epsilon$:
\begin{align}\label{}
x'\left( (1+\epsilon)t \right)&\approx \left( 1+\frac{\epsilon}{2} \right)x(t)\hspace{.5cm}\Rightarrow\hspace{.5cm} x'(t)+\dot{x}(t)\epsilon t\approx x(t)+\frac{\epsilon}{2}x(t)
\end{align}
from where we extract
\begin{align}\label{}
\delta_s x(t)=x'(t)-x(t)  = -\epsilon t\dot{x} + \frac{\epsilon}{2} x\ .
\end{align}
This transformation acts only on $x(t)$, and is a Noether symmetry of the action:
\begin{align}\label{}
\delta I[x] &= \int dt \left( \frac{1}{2}m\delta (\dot{x}^2)-\alpha \delta \left( \frac{1}{x^2} \right) \right)\nonumber \\
&=\epsilon \int dt\left[ -m \left( \frac{1}{2}\dot{x}^2+t\dot{x}\ddot{x} \right)+\alpha \frac{x-2t\dot{x}}{x^3} \right]\nonumber\\
&=\epsilon\int dt \frac{d}{dt}\left[ -m \left( \frac{t\dot{x}^2}{2} \right) +  \frac{\alpha t}{x^2} \right]\nonumber\\
&=\epsilon\int dt \frac{d}{dt} \left[ -t L \right]\ ,
\end{align}
so the boundary term is $K=-tL$. Thus, the conserved quantity associated to Weyl symmetry (up to a sign and an $\epsilon$ factor, which are both irrelevant constants)
\begin{align}\label{QWeyl}
Q = \frac{1}{2}mx\dot x-\left( \frac{1}{2}mt\dot{x}^2+\frac{\alpha t}{x^2} \right)\ .
\end{align}

Equations (\ref{EWeyl}) and (\ref{QWeyl}) are two \textit{algebraic} equations for $x(t)$ and $\dot x(t)$. From them we obtain $x(t)$ as a function of time, in terms of two integration constants $E$ and $Q$, as it should for a second order e.o.m. with one degree of freedom. This completely solve the problem. It is left as an exercise to solve these equations, find $x(t)$ and prove that it satisfies (\ref{con10}). 
 
Note that, in general, at least one of the conserved charges must be an explicit function of time, as (\ref{QWeyl}), otherwise there would be no dynamics.   

It is a good exercise to check explicitly,
\begin{align}\label{}
\frac{d Q}{dt}&=\frac{1}{2}m\left( \dot{x}^2+x\ddot{x} \right)-\frac{1}{2}m\left( \dot{x^2}+2t\dot x \ddot x \right)+\alpha \left( \frac{x^2-2t x\dot x}{x^4} \right)\nonumber\\
&=\left( x-2t\dot{x} \right) \left[ \frac{1}{2}m\ddot x-\frac{\alpha}{x^3} \right]=0\ ,
\end{align}
due to the e.o.m. \eqref{con10}.

\subsection{Background fields and non-conservation equations}

In this paragraph we discuss the  role of background fields, and how to deal with them within the framework of Noether's theorem. Noether's theorem applied to point particles sheds light into many aspects of symmetries which sometimes are subtle in field theory. Background fields is an example. 

Consider the following action,
\begin{equation}\label{IB0}
I_{B}[x(t),y(t),z(t)] = \int \left[ {m \over 2} (\dot{x}^{\,2}+\dot{y}^{\,2}+\dot{z}^{\,2}) -  B\,z(t)  \right]dt \ , 
\end{equation} 
where $B$ is a constant. Let us not worry about the origin/relevance of this theory but only about its symmetries. 

The interaction term clearly breaks spherically symmetry. Rotations in the $x/y$ plane remain a symmetry, but the full $O(3)$ symmetry is broken. Now, let $\vec{B} \equiv B \hat z$, we also collect $x,y,z$ into the vector $\vec{r}$ are rewrite the action as
\begin{equation}\label{IB}
I_{\vec{B}}[\vec{r}(t)] = \int \left[ {m \over 2}\, \dot{\vec{r}}\,^{2} -  \vec{B}\cdot \vec{r} \right]dt . 
\end{equation} 
This is exactly the same action, simply written in a more elegant way. Both $\vec{r}$ and $\vec{B}$ are vectors whose components are referred to some set of axes. Applying a rotation of the axes, the scalar products $\vec{r}\cdot \vec{B}$ and $\dot{\vec{r}}^{\, 2} $ remain invariant and so the full action is invariant. If $R$ denotes the rotation matrix we have 
\begin{equation}\label{Rr}
I_{R\vec{B}}[R\vec{r}]=I_{\vec{B}}[\vec{r}] 
\end{equation} 
Does this ``symmetry" imply conservation of angular momentum? No, of course it does  not ($\vec{L}$ is not conserved for this system, as can be easily checked from the equations of motion). What is wrong? There is nothing wrong. We just need to be careful with the role of different variables and transformations. 

Equation (\ref{Rr}) is a mathematical identity and could be called a ``symmetry" on its own right. However, it does not imply a conservation equation because it involves the variation of a background quantity. Let us apply Noether algorithm to (\ref{Rr}) to understand what is going on. We first consider an infinitesimal transformation with angle $\vec{\alpha}$. The components of $\vec{r}$  and  $\vec{B}$ vary as 
\begin{equation}\label{bf}
\delta_s \vec{r} =\vec{\alpha}\times \vec{r}, \ \ \ \ \ \ \ \  \delta_s \vec{B} = \vec{\alpha}\times \vec{B}\ .
\end{equation} 
and one can quickly check that,
\begin{equation}\label{dibf}
\delta I[\vec{r},\delta_s \vec{r}]= 0,
\end{equation} 
(that is, $K=0$). Of course, this is simply the infinitesimal version of (\ref{Rr}). 

We now compute the on-shell variation, i.e. a generic variation of all quantities followed by use of the e.o.m., 
\begin{eqnarray}\label{dl}
\delta I[\bar{\vec{r}},\delta_s \vec{r}] &=& \int (-m \ddot{\vec{r}} - \vec{B})\cdot \delta \vec{r} + {d \over dt} (m \dot{\vec{r}}\cdot \delta \vec{r}) - \delta \vec{B} \cdot \vec{r}\; dt \nonumber\\
&=&  \int \vec{\alpha} \cdot\left( {d \over dt} \vec{L} +   \vec{r} \times \vec{B}\right) dt. \label{dL1}
\end{eqnarray}  
and we observe that it is not a total derivative!  This is the crucial point. Since the ``symmetry" involves the variation of a background quantity, the on-shell variation is not a total derivative. We can proceed with the same logic  anyway. Since the symmetry variation is zero, we obtain, 
\begin{equation}
{d \over dt} \vec{L} = -\vec{r} \times \vec{B},
\end{equation} 
which is the correct torque equation! Noether's algorithm is very intelligent indeed. It has detected that $\vec{L}$ is not conserved, and has given  us the rate of change of the would-be-conserved charge.  

Dynamical variables (functions of time varied in an action
principle) and background quantities (masses, charges or even vectors, tensors but not varied in the action principle) play very different roles. The vector $\vec{B}$ in (\ref{IB}) is an example of a background quantity. Whenever a symmetry involves the variation of a background quantity, Noether theorem will not deliver a conserved quantity. 

From Noether's point of view, we say that (\ref{IB}) is not invariant under rotations because $I_{\vec{B}}[R\vec{v}] \neq I_{\vec{B}}[\vec{r}]$. The equality (\ref{Rr}) represents a {\it passive} transformation, a transformation where all vectors remain fixed and only the axes are rotated. But `passive transformations' is not what Noether symmetries are about. Noether symmetries are those transformations such that, for given values of the background quantities, the action is invariant up to a total derivative. These symmetries give rise to conserved quantities.   

~

Some other examples and comments. Consider particle of charge $q$ is in the presence of the field of a much heavier charge $Q$. The action is:
\begin{equation}\label{Ip}
I[\vec{r}_1] = \int {m_1 \over 2} \dot{\vec{r}}_1^{\, 2} -  {q Q \over |\vec{r}_1 - \vec{r}_2 |} \; dt  \ .
\end{equation} 
Here $\vec{r}_1$ is a dynamical variable while $\vec{r}_2$, the coordinate of $Q$, is fixed. This action is not invariant under translations $\vec{r}_1 \rightarrow \vec{r}_1 + \vec{a}$ with constant $\vec{a}$. As a consequence linear momentum is not conserved. On the other hand $\vec{r}_1 \rightarrow \vec{r}_1 + \vec{a} $ together with $\vec{r}_2 \rightarrow \vec{r}_2 + \vec{a} $ is a ``symmetry". The time variation of the momentum can be computed as we did in the example below  (exercise!). 

One can restore the symmetry by allowing the second charge to move. The action is now,
\begin{equation}\label{Ip2}
I[\vec{r}_1,\vec{r}_2] = \int {m_1 \over 2} \dot{\vec{r}}_1^{\,2} -  {q Q \over |\vec{r}_1 - \vec{r}_2 |}   +{m_2 \over 2} \dot{\vec{r}}_2^{\, 2} \; dt.
\end{equation} 
Although the kinetic pieces are invariant under independent translations $\vec{r}_1 \rightarrow \vec{r}_1 + \vec{a}_1 $, $\vec{r}_2 \rightarrow \vec{r}_2 + \vec{a}_2 $, the full action is only invariant under the diagonal symmetry with $\vec{a}_2 = \vec{a}_1$. As a consequence, only the {\it total} linear momentum (center of mass momentum)
\begin{equation}
\vec{p}_T = m_1 \dot{\vec{r}}_1 +  m_2 \dot{\vec{r}}_2
\end{equation} 
is conserved (this is easily derivable from the equations of motion). 

The same applies to angular momentum. The action (\ref{Ip2}) is only invariant under {\it simultaneous} rotations of both coordinates by the same angle. As a consequence only the total angular momentum $\vec{L} = \vec{L}_1 + \vec{L}_2$ is conserved.  If the interaction term was, for example, ${\beta \over  |\vec{r}_1| | \vec{r}_2| }$, then the Lagrangian would remain invariant under independent rotations and therefore $\vec{L}_1$, $\vec{L}_2$ would be conserved independently. But such potential, depending on the relative distances to a third point, does not seem to be physically sound. 

Needless to say, all these results are easily derivable from the equations of motion. Our goal here has been to analyse them in the light of Noether's theorem. In field theory the equations of motion are more complicated and it is often quicker and simpler to apply Noether's ideas directly. But special care must be observed to get the right results.

\section{ Noether's theorem in Hamiltonian mechanics: Symmetry generators and Lie algebras} \label{inverse}

Noether's theorem is applicable to any action, Lagrangian, Hamiltonian, or any other.  But the Hamiltonian action has extra structure. Consider a general Hamiltonian action, 
\begin{equation}\label{ha}
I[p_i,q^j] = \int dt ( p_i \dot q^i - H(p,q) )\ ,
\end{equation} 
and its Poisson bracket 
\begin{align}\label{Poissonpoint}
[ F,G ]&=\frac{\partial F}{\partial q^i} \frac{\partial G}{\partial p_i}- \frac{\partial F}{\partial p_i} \frac{\partial G }{\partial q^i}\ .
\end{align}
The equations of motion are
\begin{eqnarray}
\dot q^i = & {\partial H \over \partial p_i} & = [q^i,H], \\
\dot p_i =  & -{\partial H \over \partial q^i} & = [p_i,H].
\end{eqnarray} 
The time derivative of a function $G(p,q,t)$ (which can be explicitly time dependent) of the canonical variables is expressed in terms of Poisson brackets as, 
\begin{eqnarray}
{dG(p,q,t) \over dt} &=& {\partial G \over \partial q^i} \dot q^i + {\partial G \over \partial p_i} \dot p_i  +\frac{\partial G}{\partial t} \nonumber \\
&=&  {\partial G \over \partial q^i} {\partial H \over \partial p_i}  - {\partial G \over \partial p_i} {\partial H \over \partial q^i} +\frac{\partial G}{\partial t} \nonumber \\
&=& [G,H] +\frac{\partial G}{\partial t} \ . \label{GH}
\end{eqnarray} 

We are interested in conserved charges. Eqn. (\ref{GH}) means that a conserved charge $Q(p,q,t)$ satisfies
\begin{align}\label{qh1}
{dQ(p,q,t) \over dt}=0 \ \ \ \ \  \Rightarrow \ \ \ \ \  [Q,H]+\frac{\partial Q}{\partial t}=0.
\end{align}
In many examples the quantity $Q$ has no explicit time dependence, and then \eqref{qh1} reduces to having zero Poisson bracket with the Hamiltonian\footnote{ As an example, take a rotational-symmetric system $H=\frac{1}{2}\left( p_x^2+p_y^2 \right)$ and consider the quantity $Q=xp_y-yp_x$. Then, the Poisson bracket is simply $[ H,Q ]=-p_xp_y+p_yp_x=0$ where we never used the e.o.m: it vanishes automatically. }.

We now prove the following important results. 
\begin{enumerate}
\item \underline{Noether's inverse theorem:} If $Q$ is a conserved charge, then the following transformation
\begin{align}\label{invNoe}
\delta_s q^i=[q^i,\epsilon Q] = \epsilon{\partial Q \over \partial p_i}\hspace{1cm},\hspace{1cm} \delta_s p_i=[p_i,\epsilon Q] = - \epsilon{\partial Q \over \partial q^i}\ ,
\end{align}
is a symmetry of the action. This is the inverse theorem because we first assume the existence of the charge and then build the symmetry. 

\item \underline{The Lie algebra of symmetries:} The set of all conserved charges $Q_a$ ($a=1,2,...N$) satisfies a closed Lie algebra,   
\begin{equation}
[Q_a,Q_b] = f^{\ c}_{ ab}\, Q_c.
\end{equation} 
\end{enumerate} 
The proof of these two statements is extremely simple. We first show that if $Q$ satisfies (\ref{qh1}) then (\ref{invNoe}) is a symmetry. Varying the action (without using the e.o.m anywhere!)
\begin{align}\label{}
\delta I &=\int dt \left( \delta_s p\ \dot{q}+p\frac{d}{dt} \delta_s  q - \frac{\partial H}{\partial p}\delta_s  p-\frac{\partial H}{\partial q}\delta_s  q \right) \nonumber \\
&= \int dt \left( -\epsilon\frac{\partial Q}{\partial q}\dot{q}+\frac{d}{dt}\left( p\ \delta_s  q \right)-\epsilon\dot{p}\frac{\partial Q}{\partial p}+\epsilon\frac{\partial H}{\partial p}\frac{\partial Q}{\partial q}-\epsilon\frac{\partial H}{\partial q}\frac{\partial Q}{\partial p} \right) \nonumber \\
&= \int dt \left(  \epsilon\left(- \frac{dQ}{dt} + \frac{\partial Q}{\partial t} +[Q,H] \right)+\frac{d}{dt}(p\delta_s  q) \right) \nonumber \\
&= \int dt \frac{d}{dt}\Big(-\epsilon Q+p\delta_s  q \Big)\ ,
\end{align}
which is a total derivative, as required for a symmetry. In the last line we have used the assumption (\ref{qh1}). It is now direct to calculate the on-shell variation and using Noether theorem discover that the conserved charge associated to this symmetry is, not surprisingly,  exactly $Q$ (exercise!). For a more general form of the theorem, see \cite{deriglazov2010}.

~

Suppose now that we have two conserved charges $Q_1$ and $Q_2$ (both satisfy \eqref{qh1}). Then it is a very simple exercise to show explicitly that the commutator $[Q_1,Q_2]$ is also conserved:
\begin{align}\label{}
\frac{d}{dt}[Q_1,Q_2]&=\left[ [Q_1,Q_2] , H \right] + \frac{\partial }{\partial t} [Q_1,Q_2]=0
\end{align}
Thus $[Q_1,Q_2]$ is also a conserved charge, and as a consequence generates another symmetry. Now, $[Q_1,Q_2]$ may be zero, may be a new charge, or it may be proportional to $Q_1$ or $Q_2$. In any case, the conclusion is that a complete set of conserved charges $Q_a=Q_1,Q_2,Q_3,...$ must satisfies a Lie algebra
\begin{equation}
[Q_a,Q_b] = f^{\ c}_{ ab}\, Q_c.
\end{equation} 
for some structure constants $f^{\ c}_{ ab}$. This algebra is of most importance in the quantum theory. 

It is left as an exercise to extend this proof to the case where the charge depends explicitly on time. 
The conservation equation then  reads, 
\begin{equation}\label{tQ}
{dQ \over dt} = [Q, H ] + {\partial Q \over \partial t} =0.
\end{equation} 
Given two charges that satisfy (\ref{tQ}), their Poisson bracket satisfies the same equation.  We now discuss the conformal particle, as an explicit example.

\subsection{The conformal particle in Hamiltonian form}

The `conformal particle' of Section \ref{confpart} has two conserved charges, with one of them depending explicitly on time. Since the system carries only one degree of freedo (two integration constants) these two charges completely solve the equations of motion. 

This system in Hamiltonian form has extra structure. See \cite{dealfaro} for a detailed treatment, and \cite{britto2001,michelson,chamon} for some applications to the AdS/CFT correspondence. The Hamiltonian action is, 
\begin{equation}
I = \int  \left(  p\dot q - \left( {p^2 \over 2 m} + {\alpha \over q^2} \right) \right) dt
\end{equation} 
and three conserved charges (see \cite{dealfaro}), 
\begin{eqnarray}
H   &=& \frac{p^2}{2m}+\frac{\alpha}{q^2} \\
Q   &=& -t H + {1 \over 2} pq \\
K   &=& t^2 H + 2t Q - {m \over 2} q^2
\end{eqnarray} 
can be found: $H,Q,K$ satisfy (\ref{tQ}). Of course, given that this theory has only 2 integration constants, there exists a relation between the three charges, indeed, 
\begin{equation}
2KH+2Q^2+m\alpha =0.
\end{equation} 
The interesting aspect of the Hamiltonian formulation is that $H,Q,K$ satisfy the sl(2,$\Re$) algebra, 
\begin{eqnarray}
~ [Q,H] &=& H, \\
~  [Q,K] &=& -K,  \\
~  [H,K] &=& 2Q, 
\end{eqnarray} 
which is relevant for the AdS$_2$/CFT$_1$ correspondence.

\newpage

\section{ Noether's theorem in  Field theory. Derivation and examples.}

\subsection{The proof}

Just as for particle mechanics, the key to Noether's theorem is the concept of action symmetries. The simplest field theory example is provided by
\begin{equation}\label{Iphi0}
I[\phi(x)] = {1 \over 2} \int d^{4}x\,  \partial_\mu \phi \partial^\mu\phi.  
\end{equation} 
which is clearly invariant under the constant translation $\phi(x)\rightarrow \phi(x) + \phi_0$, indeed, $I[\phi(x) + \phi_0] = I[\phi(x)]$. The symmetry now acts on the field $\phi(x)$. 
  
The set of symmetries of a field theory action is defined as the set of all infinitesimal functions $\delta_s\phi(x)$ such that, for \textit{arbitrary} $\phi(x)$,  
\begin{equation}\label{symdef}
\delta I[\phi,\delta_s\phi] \equiv  I[\phi + \delta_s \phi] - I[\phi] = \int d^dx\, \partial_\mu K^\mu   \ \ \ \ \ \ \ \ \ \ \forall\ \phi
\end{equation} 
for some $K^\mu$. We emphasize:
\begin{itemize}
\item Eq. (\ref{symdef}) is an equation for the function $\delta_s\phi$, not for $\phi$: $\delta_s \phi$ is a symmetry provided (\ref{symdef}) holds for {\it all} $\phi$. 

\item The coordinates play no role. The action is a functional of $\phi(x)$ and the coordinates are dummy variables which are `summed over'. The definition of symmetry, eq. (\ref{symdef}), does not involve changes in the coordinates in any way. Even for symmetries associated to spacetime translations, rotations, etc, they can always be written as some transformation $\delta_s\phi(x)$ acting on the field\footnote{Noether's symmetries and in general actions symmetries have various formulations some more complicated than others. It is possible to formulate symmetries acting directly on the coordinates, and many books choose this path. We argue that this is not necessary and introduces complications.  All symmetries can be understood directly as a transformation of the field satisfying (\ref{symdef}). Even the symmetry of general relativity -general covariance- is best understood as a transformation acting on the metric and not the coordinates (see below).}. We shall discuss this issue in detail in the next paragraph.
\end{itemize}

The derivation of Noether theorem in field theory follows exactly  the same path as in particle mechanics, so we shall be brief. 

Field theories in $d$ dimensions are described by actions of the form $I[\phi(x)]=\int d^dx\ \mathcal{L}(\phi,\partial_\mu\phi)$ and the corresponding Euler-Lagrange equations are
\begin{align}\label{fielde.o.m.}
 {\cal E}(\phi(x)) \equiv \partial_\mu \left( \frac{\partial\mathcal{L}}{\partial \phi,_\mu} \right)-\frac{\partial \mathcal{L}}{\partial \phi}=0\ .
\end{align}
(We will use the notation $\phi,_{\mu}\equiv\partial_\mu\phi$ alternatively.) The on-shell variation is computed as
\begin{align}\label{fov}
\delta I[\bar\phi,\delta\phi] &= \int d^4x\left( \frac{\partial \mathcal{L}}{\partial \phi}\delta \phi+\frac{\partial \mathcal{L}}{\partial \phi,_\mu}\delta \phi,_\mu  \right) \nonumber \\
&= \int d^4x\left( \left[ \frac{\partial \mathcal{L}}{\partial \phi}- \partial_\mu \left( \frac{\partial\mathcal{L}}{\partial \phi,_\mu} \right) \right]\delta \phi \right) + \int d^4x\ \partial_\mu \left( \frac{\partial\mathcal{L}}{\partial \phi,_\mu} \delta \phi \right)\nonumber \\ 
&=  \int d^4x\ \partial_\mu \left( \frac{\partial\mathcal{L}}{\partial \phi,_\mu} \delta \phi \right)\ ,
\end{align}
where in the last line we have used that the field $\bar \phi$  satisfies its Euler-Lagrange equations. 

Now, (\ref{symdef}) is valid for any $\phi$, in particular for $\bar \phi$. Eqn. (\ref{fov}) is valid for any $\delta \phi$, in particular for $\delta_s\phi$. Thus, inserting $\bar\phi$ into (\ref{symdef}) and $\delta_s\phi$ into (\ref{fov}) the left hand sides are equal. Substracting both equations we obtain the conserved current equation
\begin{equation}\label{J}
\boxed{ \partial_\mu J^\mu = 0 \ \ \ \ \ \mbox{where}  \hspace{1cm}  J^\mu\equiv  \frac{\partial \mathcal{L}}{\partial \phi,_\mu  } \delta\phi(x) - K^\mu  }  
\end{equation}  
This is Noether's first theorem in field theory. \\

Before moving to the examples, we show how to build a conserved charge from a conserved current. 
In a space+time splitting, the continuity equation becomes  
\begin{align}\label{}
\partial_t J^0=-\nabla\cdot \vec{J}\ .
\end{align}
Integrating both sides of this equation and using the divergence theorem, 
\begin{equation}
 \int_V d^3x\ \partial_t J^0 = -\int_V d^3x\ \nabla\cdot \vec{J}=-\int_{\partial V} \vec{J}\cdot d\vec{A} \ .
\end{equation} 
If the container $V$ is large enough (and assuming field configurations such that $\vec{J}$ drops to zero faster than the growth of the surface area) the last integral vanishes, yielding the conserved charge
\begin{align}\label{}
Q=\int_V d^3x\ J^0(x) \hspace{.5cm} \mbox{with}  \hspace{.5cm}  \hspace{.5cm}  \frac{d}{dt}Q=0\ .
\end{align}
Actually, this widespread phrase that ``fields fall off sufficiently rapidly at infinity" will turn out to be \textit{false} in gauge theories. We shall come back to this in Section \ref{boundary}.

\subsection{Symmetries act on fields: Lie derivatives}

We have emphasized that the natural interpretation of a symmetry is as a transformation acting on the field, not the coordinates. The coordinates are dummy variables and in fact one can change them at will without affecting the value of the integral. This is a basic theorem of calculus. The action's {\it form} may change if the coordinates are changed, but not its value. This discussion can become a bit subtle and complicated. An easy solution is to realize that for Noether theorem one never need to change the coordinates. Only the fields transform.  

For example, going back to the scalar field action (\ref{Iphi0}) one suspects that, besides the symmetry under $\phi \rightarrow \phi + \phi_0$, this action is also invariant under constant spacetime translations $x^{\mu} \rightarrow x^\mu + \epsilon^\mu $ because there is no explicit dependence on the coordinates. Apparently, then, we face two very different kind of symmetries, some acting on the fields, some acting on the coordinates. This is not correct. Within the action, all  symmetries can be expressed as transformations acting on the fields, even symmetries whose origin are variations of the coordinates. 

Spacetime translations $x^\mu \rightarrow x'^{\mu} = x^\mu + \epsilon^\mu $ are understood as a transformation of the field as follows: Given $\phi(x)$ one builds a new field (`the translated field') $\phi'(x)$ whose values are 
\begin{eqnarray}
\phi'(x) &=& \phi(x - \epsilon) \nonumber\\
     &\simeq &  \phi(x) - \epsilon^\mu \partial_\mu \phi(x).
\end{eqnarray} 
where in the second line we retained first order in $\epsilon^\mu$. The variation of the field which is associated to a translation of coordinates is then 
\begin{equation}\label{dp}
\delta \phi(x) = -  \epsilon^\mu \partial_\mu \phi(x).
\end{equation} 
This is a local relation involving only the point $x$. We say that the action (\ref{Iphi0}) is invariant under constant spacetime translations because its variation with respect to (\ref{dp}) is, 
\begin{eqnarray}
\delta I[\phi] &=&   \int d^4x\  \partial_\mu \phi\, \partial^\mu(-\epsilon^\nu \partial_\nu\phi) \nonumber\\
 &=& -{1 \over 2} \int d^4x\ \partial_\nu(\epsilon^\nu \partial_\mu\phi \, \partial^\mu\phi). 
\end{eqnarray} 
a boundary term. No transformation on the coordinates was done. Only a transformation on the field. Note that a $x^\mu$-independent potential $U(\phi)$ would not spoil this symmetry 
because $\delta U = U'(\phi) \delta \phi = -  U'\epsilon^\mu \partial_\mu\phi = - \partial_\mu (\epsilon^\mu U) $, is also a boundary term. This symmetry would be spoiled if the potential depends explicitly on the coordinates because in that case   $ {\partial U(\phi,x)\over \partial\phi} \partial_\mu\phi \neq \partial_\mu U(\phi,x)$.

~

Constant translations are a very special kind of transformation of $x^\mu$. One may wonder whether other important symmetries, e.g., rotations ($\delta\vec{r} = \vec{\omega}\times \vec{r}$), dilatations ($\delta \vec{r} = \lambda \vec{r}$), etc\footnote{For example,  Maxwell's Lagrangian is (classically) invariant under conformal transformations. We study this case in detail below.}, can also be written as a transformation of the field. At the same time, Nature is not composed merely by scalar fields. The general question is, how can a generic transformation $\delta x^\mu = \epsilon^\mu(x) $ be interpreted as a variation of the field, if the field is a vector, a tensor, etc?   

The solution to this problem is encoded in the transformation laws for tensors
\begin{align}\label{}
\phi'(x')&=\phi(x)\hspace{4cm}\mbox{Scalar}\label{Scalar}\\
V'^\mu(x')&=\frac{\partial x'^\mu}{\partial x^\nu}V^\nu (x)\hspace{2.9cm}\mbox{Vector}\\
A'_\mu(x')&=\frac{\partial x^\nu}{\partial x'^\mu}A_\nu (x)\hspace{2.9cm}\mbox{$(0,1)$ tensor}\label{A'mu=Amu}\\
g'_{\mu\nu}(x')&=\frac{\partial x^\alpha}{\partial x'^\mu} \frac{\partial x^\beta}{\partial x'^\nu}g_{\alpha\beta}(x) \hspace{2cm}\mbox{$(0,2)$ tensor}\label{g'munu=gmunu}\\
&\ \vdots  \nonumber
\end{align}
These equations express the variations of components of various fields when a given transformation of the coordinates is applied. 
For Noether's theorem we need the infinitesimal version of these formulae. We start by calculating the infinitesimal version of the Jacobians appearing in the above transformations laws:
\begin{align}\label{}
x'^\mu=x^\mu+\xi^\mu(x)\ \ \ \ \Rightarrow\ \ \ \ \frac{\partial x'^\mu}{\partial x^\nu}=\delta^\mu_\nu+\partial_\nu \xi^\mu(x)\ \ ,\ \ \frac{\partial x^\mu}{\partial x'^\nu}=\delta^\mu_\nu-\partial_\nu \xi^\mu(x')\ .
\end{align}
With these formulae at hand we go back to the above transformations laws and expand to first order case by case.  

\begin{itemize}

\item {Scalar Field:} Expanding $\phi'(x+\xi) = \phi(x)$ to linear order in $\xi$ we derive 
\begin{align}\label{}
\delta \phi(x)=\phi'(x)-\phi(x)=-\xi^\nu(x) \partial_\nu \phi (x)\ .
\end{align}
This is same formulae we use before, but not that now $\xi^\mu(x)$ is an arbitrary function of $x^{\mu} $. 
\item {Vector field:} Expand $V'^\mu(x')=\frac{\partial x'^\mu}{\partial x^\nu}V^\nu(x)$ to linear order. In this case we need to expand the left and right hand sides:
\begin{align}\label{}
V'^\mu(x)+\xi^\nu \partial_\nu V^\mu(x)  = V^\mu(x)+ (\partial_\nu \xi^\mu)V^\nu(x) \ ,
\end{align}
from where we derive
\begin{align}\label{}
\delta V^\mu(x)=V'^\mu(x)-V^\mu(x)=(\partial_\nu \xi^\mu)V^\nu-\xi^\nu \partial_\nu V^\mu \ .
\end{align}

\item {$(0,1)-$tensor:} in the same way as above, the transformation law \eqref{A'mu=Amu} expanded to linear order at both sides give 
\begin{align}\label{}
A'_\mu(x)+\xi^\alpha \partial_\alpha A_\mu &=A_\mu(x)-A_\alpha \partial_\mu \xi^\alpha\ ,
\end{align}
from where it follows,
\begin{align}\label{}
\delta A_\mu(x)=A'_\mu(x)-A_\mu(x)=-\xi^\alpha \partial_\alpha A_\mu-A_\alpha \partial_\mu \xi^\alpha\ .
\end{align}
 
\item {$(0,2)-$tensor:} following the same rules as above, we have at first order in $\xi$, according to \eqref{g'munu=gmunu}
\begin{align}\label{}
 g'_{\mu\nu}(x)+\xi^\alpha \partial_\alpha g_{\mu\nu}(x)   &=\left( \delta^\alpha_\mu-\partial_\mu \xi^\alpha \right) \left( \delta^\beta_\nu-\partial_\nu \xi^\beta \right)g_{\alpha\beta}(x)\\
&=g_{\mu\nu}(x)-g_{\mu\beta}(x)\partial_\nu \xi^\beta-g_{\alpha\nu}(x) \partial_\mu\xi^\alpha \ ,
\end{align}
from where we deduce
\begin{align}\label{Liegmunu}
\delta g_{\mu\nu}(x)=g'_{\mu\nu}(x)-g_{\mu\nu}(x)=-\xi^\alpha \partial_\alpha g_{\mu\nu}(x)-g_{\mu\beta}(x)\partial_\nu \xi^\beta-g_{\alpha\nu}(x) \partial_\mu\xi^\alpha \ .
\end{align}
\end{itemize}

The different variations obtained by this procedure are called in the mathematical literature \textbf{Lie Derivatives}. The Lie derivative along a vector $\xi^{\mu}(x) $ is an operation acting on tensors having good transformations laws under coordinate transformations (without involving a connection). In this way, the Lie derivative along $\xi^\mu(x)$ of a scalar field, a 1-form and tensor are, respectively,  
\begin{eqnarray}
{\cal L}_{\xi} \phi &=& -\xi^\nu \partial_\nu \phi (x) \label{liephi} \\
{\cal L}_{\xi} A_\mu &=& -\xi^\alpha \partial_\alpha A_\mu-A_\alpha \partial_\mu \xi^\alpha \label{lieA} \\
{\cal L}_{\xi} g_{\mu\nu} &= & -\xi^\alpha \partial_\alpha g_{\mu\nu}(x)-g_{\mu\beta}(x)\partial_\nu \xi^\beta-g_{\alpha\nu}(x) \partial_\mu\xi^\alpha \ .
\end{eqnarray} 
The Lie derivative acting on objects with other index structure can be found in a similar way.

\subsection{Energy-momentum tensor. Scalars and Maxwell theory}

The energy momentum tensor is the Noether current associated to the symmetry under constant spacetime translations $x^\mu \rightarrow x^\mu + \epsilon^\mu$.   The reason that this `current' is really a tensor and not a vector field is simple to understand. There is one current associated to time translations $t \rightarrow t + \epsilon^0$, another current for space translations in the $\hat x$ direction $x \rightarrow x + \epsilon^1$, etc. Altogether, there are four symmetries and therefore  four currents. Of course we don't split the calculation in this way but compute the current associated to $x^\mu \rightarrow x^\mu + \epsilon^\mu$ in one step.  The current that follows is of course linear in $\epsilon^\mu$ so it must have the form $J^\mu = T^{\mu}_{\ \nu}\epsilon^\nu$. The coefficient $T^{\mu}_{\ \nu}$ is the energy-momentum tensor and conservation of $J^{\mu} $ implies $\partial_\mu T^{\mu}_{\ \nu}=0$. Put another way, symmetry under spacetime translations implies the conservation of \textit{four} independent currents $\tensor{T}{^\mu_\nu}$, one for each $\nu$. 

We compute now the energy-momentum  tensor $\tensor{T}{^\mu_\nu}$ for two specific examples: a scalar field, and Maxwell's theory, which is particularly interesting.     

In relativistic field theories the energy-momentum tensor is often defined as the functional derivative of the action, $T_{\mu\nu}=\frac{1}{\sqrt{g}} \frac{\delta S}{\delta g^{\mu\nu}}$, which is very convenient for many purposes. With a pedagogical motivation we shall always stick to its definition as the Noether conserved current associated to invariance under space-time translations.  

\subsubsection{Stress tensor for scalar fields}

Consider a Lagrangian density for a scalar field ${\cal L}(\phi,\partial_\mu\phi)$ which does not depends explicitly on the coordinates. For example, ${\cal L} = {1 \over 2}\partial_\mu \phi \partial^\mu \phi $ with equations of motion $\nabla^2 \phi =0$\footnote{A typical situation where ${\cal L}$ depends on $x^\mu$ is when the field is coupled to external sources $J(x)$, for example   ${\cal L} = {1 \over 2}\partial_\mu \phi \partial^\mu \phi + J(x)\phi $ with equations $\nabla^2 \phi = J(x)$.}. But the explicit form of the Lagrangian is not relevant here.    

If the Lagrangian does not depend on $x^\mu$ explicitly, then it is invariant under 
\begin{equation}
\delta \phi(x) = -\epsilon^\mu \partial_\mu \phi(x)\ ,
\end{equation} 
which of course corresponds to the Lie derivative \eqref{liephi}. Indeed, the variation,
\begin{align}\label{}
\delta \mathcal{L}&= \frac{\partial \mathcal{L}}{ \partial \phi} \delta \phi+\frac{\partial \mathcal{L}}{ \partial \phi,_\rho}\delta \phi,_\rho = - \epsilon^\sigma \left[  \frac{\partial \mathcal{L}}{ \partial \phi} \phi,_\sigma+\frac{\partial \mathcal{L}}{ \partial \phi,_\rho}\phi,_{\rho\sigma}  \right]= -\partial_\sigma(\epsilon^\sigma \mathcal{L} )
\end{align}
is a boundary term. The first equality is the chain rule. In the second equality we simply replace the symmetry remembering $\partial_\mu \delta \phi = \delta \partial_\mu \phi$ since our variations are the difference of two functions. The crucial step is  the third equality which requires that the Lagrangian does not depend explicitly  on $x^\mu$\footnote{It is perhaps worth to give a concrete example. The function $f(x(t))= x(t)^2$ depends on time only through $x(t)$. We say that ${\partial f \over \partial t}$ =0  while ${df \over dt} = f'(x) \dot x = 2 x \dot x $.  On the other hand, the function $g(x(t)) = t \, x(t)^2$ depends on $t$ explicitly and $ {dg \over dt} = x^2 + 2t x \dot x  \neq g'(x) \dot x$.  } (and $\epsilon^\mu$ is constant).  

The Noether current generated with space-time translations becomes
\begin{align}\label{}
J^\mu 
= \epsilon^\rho \left[  \frac{ \partial \mathcal{L}}{\partial \phi,_\mu} \partial_\rho \phi -\delta^\mu_\rho \mathcal{L}  \right] \equiv \epsilon^\rho T^\mu_{\ \rho} \ ,
\end{align}
so the energy-momentum tensor is
\begin{align}\label{Tmunuscalar}
T^\mu_{\ \nu} (x) =    \frac{ \partial \mathcal{L}}{\partial \phi,_\mu} \partial_\rho \phi -\delta^\mu_\rho \mathcal{L} \hspace{1cm}\mbox{with}\ \ \ \partial_\mu T^\mu_\rho=0\ .
\end{align}
For the free scalar action ${\cal L} = {1 \over 2}\partial_\mu \phi \partial^\mu \phi $ this yields, 
\begin{equation}
T_{\mu\nu} = \partial_\mu\phi \partial_\nu\phi - {1 \over 2} \eta_{\mu\nu} \partial_\alpha\phi \partial^ \alpha \phi.
\end{equation} 

\subsubsection{Stress tensor for Maxwell's theory}

Let us now study electrodynamics, described by the action
\begin{equation}\label{Imaxw}
I[A_\mu] =-{1 \over 4} \int d^4x\ F_{\mu\nu} F^{\mu\nu}, \ \ \ \ \ \   F_{\mu\nu} = \partial_\mu A_\nu - \partial_\nu A_\mu  \ .
\end{equation} 
This theory is invariant under spacetime translations $x^\mu \rightarrow x^\mu + \epsilon^\mu$ with constant $\epsilon^\mu$, and we can compute the energy-momentum tensor just as we did for the scalar field. Additionally, Maxwell's theory is also invariant under gauge transformations $\delta A_\mu = \partial_\mu \lambda(x)$. We shall see now how both symmetries can be combined to give a nice energy-momentum tensor which is symmetric and gauge invariant, the Belinfante tensor.  

As a motivation for next section we mention that Maxwell's theory in fact possesses a much larger group of Noether symmetry, namely, the conformal group. We display this symmetry explicitly in the next section and compute the associated conformal currents and charges.

As we see from \eqref{lieA}, the action of a {\it constant} translation on a 1-form $A_\mu$ is 
\begin{equation}\label{dA1}
\delta_0 A_\mu = -\epsilon^\nu \partial_\nu A_\mu\ ,
\end{equation} 
where $\delta_0$ indicates that this is the variation that we would in principle use. It is left as an exercise to show that this transformation changes the Maxwell action by a boundary term. 

Before continuing with this calculation,  
we notice that there is an evident problem with this attempt: the variation (\ref{dA1}) does not have good properties  under gauge transformations (it is not gauge invariant). One can anticipate that it's conserved current will not have good properties under gauge transformations either. This problem has been extensively discussed in the literature.  We shall jump the discussion and go straight into the solution \cite{Jackiw}.

Instead of \eqref{dA1}, consider a transformation that combines a constant spacetime translation together with a particular gauge transformation, 
\begin{align}\label{dAgaugeinvar}
\delta A_\mu = -\epsilon^\alpha \partial_\alpha A_\mu+ \partial_\mu \left( \epsilon^\alpha A_\alpha \right) =  F_{\mu\alpha} \epsilon^\alpha\ ,
\end{align}
with constant $\epsilon$. In the literature this is called an `improved translation'. It is gauge invariant because the derivatives of $A_\mu$ appear through only $F_{\mu\nu}$.   

We now compute the variation of the Lagrangian \eqref{Imaxw} under this transformation (using Bianchi's identity)
\begin{align}\label{}
\delta F^2&=2F^{\mu\nu}\delta F_{\mu\nu} =\ 2F^{\mu\nu} \epsilon^\alpha\left( \partial_\mu F_{\nu\alpha}-\partial_\nu F_{\mu\alpha} \right)=2F^{\mu\nu} \epsilon^\alpha\left( \partial_\mu F_{\nu\alpha}+\partial_\nu F_{\alpha\mu} \right)\nonumber \\
&=-2F^{\mu\nu}\epsilon^\alpha \partial_\alpha F_{\mu\nu}=\partial_\alpha \left(- \epsilon^\alpha F^2 \right)\ ,
\end{align}
which is a boundary term, showing that the improved transformation \eqref{dAgaugeinvar} is indeed a symmetry. The variation of the Lagrangian is $\delta \mathcal{L}=-\frac{1}{4}\partial_\alpha \left(-\epsilon^\alpha F^2 \right)= \partial_\alpha \left(- \epsilon^\alpha \mathcal{L} \right)$ so $K^\alpha=-\epsilon^\alpha \mathcal{L}$. With this, the conserved current is
\begin{align}\label{}
J^\mu&=\frac{\partial \mathcal{L}}{\partial A_{\rho,\mu}} \delta A_{\rho} -K^\mu =-F^{\mu\rho} \epsilon^\sigma F_{\rho\sigma} + \epsilon^\mu \mathcal{L}  \nonumber \\
&= \epsilon^\sigma \left[- F^{\mu\rho}  F_{\rho\sigma} +  \delta^\mu_\sigma \mathcal{L}  \right] =- \epsilon^\sigma \left[  F^{\mu\rho}  F_{\rho\sigma} + \frac{1}{4} \delta^\mu_\sigma F^{\alpha \beta}F_{\alpha \beta} \right] 
\end{align}
and we obtain the electromagnetic energy-momentum tensor,
\begin{align}\label{Tem}
 T^\mu_{\ \sigma}=  - F^{\mu\rho}  F_{\sigma\rho} + \frac{1}{4} \delta^\mu_\sigma F^{\alpha \beta}F_{\alpha \beta}  \ .
\end{align}
This tensor is gauge invariant (since $F$ is) and has zero trace (associated to the scale invariance of the classical theory, see below). It is left as an exercise to prove, by direct computation, that the energy-momentum tensor \eqref{Tem} is indeed a conserved current. 

\subsection{Maxwell's electrodynamics and the conformal group}

We have seen that Maxwell's theory is invariant under constant spacetime translations. They have four associated currents which form the energy-momentum tensor. 

We shall now show that Maxwell's theory is invariant under a much larger group, the conformal group. Constants translations are a small subgroup within the conformal group. 

Consider again the transformations $x^\mu \rightarrow x^ \mu + \xi^ \mu(x) $ where $\xi^{\mu}(x)$ is now not constant. As described in \eqref{lieA}, this transformation acts directly on the field generating a variation, 
\begin{equation}\label{dA0}
\delta_0 A_\mu(x) = -\xi^\nu \partial_\nu A_\mu - \partial_\mu \xi^\nu A_\nu .  
\end{equation} 
Once again this transformation is not gauge invariant but this can be fixed by adding a gauge transformation just as in \eqref{dAgaugeinvar}. So we consider the variation (\ref{dA0}) plus a gauge transformation with parameter $\xi^\nu A_\nu$ 
\begin{eqnarray}\label{da}
\delta A_\mu(x) &=& -\xi^\nu \partial_\nu A_\mu - \partial_\mu \xi^\nu A_\nu  + \partial_\mu (\xi^\nu A_\nu)  \nonumber\\
 &=& F_{\mu\nu}\, \xi^\nu(x)\ .
\end{eqnarray} 
We see that the gauge transformation cancels the term with derivatives of $\xi^\mu$, and at the same time all derivatives of $A_\mu$ appear through $F_{\mu\nu}$. This variation is called an `improved diffeomorphism'. The same  can be done for Yang-Mills fields. (Exercise: prove that the commutator of two transformations (\ref{da}) with parameters $\xi^\mu_{1}$ and $\xi^{\mu}_{2} $ give a gauge transformation with parameter $F_{\mu\nu} \xi^{\mu}_{1}\xi^\nu_{2} $ plus another transformation (\ref{da}) with parameter $\xi^\mu = \xi_1^{\nu} \xi^ \mu_{2,\nu} - \xi_2^{\nu} \xi^ \mu_{1,\nu} $).   

This \textit{improvement} of the energy-momentum tensor may seem arbitrary, and even more, problematic: when we add an extra piece to $\delta A_\mu$ this should in principle modify the conserved current and it's associated charge. This in fact does not occur, because as we shall see in Section \ref{Hamgauge}, gauge symmetries are not generated by ``physical charges" but instead by \textit{constraints} which vanish on-shell. Thus the ``charge" associated to a gauge symmetry is always zero. Therefore adding an extra gauge transformation to $\delta A_\mu$ does not alter the conserver current, which is the energy-momentum tensor. 

Maxwell's theory is not invariant under (\ref{da}) for arbitrary $\xi^\mu(x)$\footnote{Invariance under a general $\xi^{\mu}(x)  $ is the symmetry of general relativity. When Maxwell's theory is coupled to a dynamical metric it becomes invariant under general diffeomorphisms. The particular form (\ref{da}) is very useful even in curved spacetimes.}.     
But it is invariant under all transformations that belong to the \textbf{conformal group}: variations (\ref{da}) such that the vectors field $\xi^\mu(x)$ satisfies the particular equation, 
\begin{align}\label{cg}
\xi_{\mu,\nu}+\xi_{\nu,\mu}=\frac{1}{2}\eta_{\mu\nu}\xi^\alpha_{,\alpha} 
\end{align}
(in $3+1$ dimensions). Any solution $\xi^\mu(x)$ to this equation provides a Noether symmetry of electrodynamics with an associated Noether current and conserved charge. We compute these quantities below. 

Equations (\ref{cg}) are called conformal killing equations and can be solved in general. Before computing the currents, we list the solutions to (\ref{cg}). To understand the names given to the various solutions it is useful to recall that, from the point of view of the coordinates $x^\mu$, the vectors fields  $\xi^{\mu}(x)$ appear in $x^\mu \rightarrow x'^\mu = x^\mu + \xi^\mu(x)$, or, what is the same, $\delta x^\mu = \xi^{\mu}(x) $. 

 The vectors $\xi^{\mu}(x) $ that solve (\ref{cg}) are split into the four categories that constitute the conformal group: 
\begin{itemize}
\item[--] Constant translations ($4$ generators). This is the simplest solution to (\ref{cg}), 
\begin{align}\label{trans1}
\xi^\mu = \xi^\mu_0. 
\end{align}
We already know they represent symmetries of Maxwell's theory. 

\item[--] Lorentz transformations (6 generators). Slightly less evident, the following vectors 
\begin{align}\label{Lor}
\xi^\mu(x)=\epsilon^\mu_{\ \nu} x^\nu \hspace{1cm},\hspace{1cm} \epsilon_{\mu\nu}=-\epsilon_{\nu\mu}
\end{align}
also solve (\ref{cg}). 

\item[--] Dilatations (1 generator). A constant rescaling by $\lambda$
\begin{align}\label{dil}
\xi^\mu(x) &= \lambda x^\mu
\end{align}
is also a solution. 
\item[--] Special conformal transformations (4 generators). The least obvious solution but easily shown to solve (\ref{cg}),  
\begin{align}\label{spec}
\xi^\mu(x) &= 2x^\nu b_\nu x^\mu -b^\mu x^\nu x_\nu\ .
\end{align}

\end{itemize}

Let us now prove that the \textit{improved} transformation (\ref{da}) leaves Maxwell's lagrangian invariant (up to a boundary term), if the vector $\xi^\mu(x)$ satisfies (\ref{cg}).  Our aim is to write the variation of the Lagrangian as $\delta \mathcal{L}=\partial_\mu K^\mu+f(\xi)$, for some function $f$ which depends on $\xi$ but \textit{does not depend on the field} $A_\mu$, because we wish to impose some restriction over the kind of transformations of coordinates (namely, that they be conformal), but not on the dynamical fields! Starting from $\mathcal{L}=\frac{1}{4}F^2$, we have
\begin{align}
\delta \mathcal{L}
&=F^{\mu\nu}\partial_\mu \left( \xi^\rho F_{\nu\rho} \right) \nonumber \\
&=F^{\mu\nu}  F_{\nu\rho} \partial_\mu \xi^\rho +F^{\mu\nu} \xi^\rho \partial_\mu F_{\nu\rho} \nonumber \\
&=F^{\mu\nu}  \tensor{F}{_\nu^\rho} \partial_\mu \xi_\rho +\frac{1}{2}F^{\mu\nu} \xi^\rho \left( \partial_\mu F_{\nu\rho} +\partial_\nu F_{\rho\mu} \right) \nonumber \\
&=-F^{\mu\nu}  \tensor{F}{^\rho_\nu} \partial_\mu \xi_\rho -\frac{1}{2}F^{\mu\nu} \xi^\rho \partial_\rho F_{\mu\nu} \nonumber \\
&=-F^{\mu\nu}  \tensor{F}{^\rho_\nu} \partial_\mu \xi_\rho -\frac{1}{4} \partial_\rho \left( F^{\mu\nu}F_{\mu\nu} \right) \xi^\rho \nonumber \\
&=-\frac{1}{2} F^{\mu\nu}  \tensor{F}{^\rho_\nu} \left( \partial_\mu \xi_\rho +\partial_\rho \xi_\mu \right) + \frac{1}{4} F^{\mu\nu}F_{\mu\nu} \partial\cdot \xi -\frac{1}{4} \partial_\rho \left( \xi^\rho F^2  \right) \nonumber \\
&=-\frac{1}{2} F^{\mu\nu}  \tensor{F}{^\rho_\nu} \left( \partial_\mu \xi_\rho +\partial_\rho \xi_\mu - \frac{1}{2} \eta_{\mu\rho}\partial\cdot \xi  \right) - \partial_\rho \left( \xi^\rho \mathcal{L}  \right) \ , \label{aaa}
\end{align}
where we have used the Bianchi identity, and the rest is straightforward calculation. Thus, the transformation $\delta x^\mu=\xi^\mu$ is a symmetry of Maxwell's action provided that the diffeomorphism satisfies the conformal Killing equation \eqref{cg}.

To derive the conserved current associated to conformal symmetry, we simply repeat Noether's algorithm as explained above: for transformations satisfying the conformal Killing equation \eqref{cg}, the symmetry variation of the Lagrangian is simply \eqref{aaa} 
\begin{align}\label{}
\delta \mathcal{L} &= -\partial_\rho \left( \xi^\rho \mathcal{L}  \right) \ ,
\end{align}
while the on-shell variation \eqref{fov}  reads:
\begin{align}\label{}
\delta_{\mbox{\scriptsize os}} \mathcal{L}&=\partial_\rho \left( \frac{\partial \mathcal{L}}{\partial (\partial_\rho A_\alpha )} \delta A_\alpha \right) \nonumber \\
&=\partial_\rho \left( F ^{\rho\alpha}  \xi^\beta F_{\alpha \beta}  \right) \ .
\end{align}
Equating both derivatives, we find the conserved current associated to conformal symmetry:
\begin{align}\label{}
J^\rho&=\xi^\beta F^{\rho\alpha} F_{\alpha\beta} +\frac{1}{4}\xi^\rho F^2 \nonumber \\
&=\xi^\beta \left( F^{\rho\alpha} F_{\alpha\beta}+\frac{1}{4}\delta^\rho_\beta F^2 \right)\hspace{1cm} \mbox{with}\ \ \ \partial_\rho J^\rho=0\ ,
\end{align}
where again $\xi^\beta$ is not arbitrary but must belong to any of \eqref{trans}-\eqref{spec}. As an aside, it happens that this classically conserved current does not survive quantisation: the $\beta-$functions don't vanish because scaling and special conformal transformations are broken in the quantum theory. 

\subsection{Phase invariance and probability conservation for Schr{\"o}dinger's equation}

Schr{\"o}dinger's equation can be derived from the following field theory Lagrangian
\begin{equation}\label{Sch}
I = \int dt \int d^3r \left( i\hbar \psi^*\, \dot \psi - {\hbar^2 \over 2m} \nabla \psi^* \cdot \nabla\psi - V(\vec{r}) \psi^*\psi \right).
\end{equation}
We find the equations of motion by extremizing the action. Since $\psi$ is complex, $\psi$ and $\psi^*$ can be thought as independent. Varying with respect to $\psi^*$ we find
\begin{eqnarray}
\delta I &=& \int dt\int d^3r \left[ i\hbar \delta\psi^* \dot \psi - {\hbar^2 \over 2m} \nabla\delta\psi^* \cdot \nabla\psi - V \delta\psi^* \psi \right] \nonumber\\
&=& \int dt\int d^3r\, \delta\psi^*\left[ i\hbar \dot \psi + {\hbar^2 \over 2m} \nabla^2\psi - V \psi \right] + B\ , \nonumber
\end{eqnarray}
where $B$ is a boundary term evaluated at spatial infinity where we assume the wave function vanishes (but again, we will come back to this subtle issue in section \ref{boundary}!). Hence, for arbitrary variations $\psi^*$ we find the equations of motion,
\begin{eqnarray}\label{Scheq}
i\hbar \dot \psi = -{\hbar^2 \over 2m} \nabla^2\psi + V \psi = H\psi \ .
\end{eqnarray}
It is left as an exercise to find the equations of motion for $\psi$ and show that they are related to the  (\ref{Scheq}) by complex conjugation. 

\subsubsection{Phase invariance and conservation of probability}

This action (\ref{Sch}) is invariant under phase transformations for constant $\alpha$:
\begin{equation}
\psi \rightarrow \psi e^{i\alpha}.
\end{equation}
Let us compute its associated Noether charge. First we need the infinitesimal symmetry transformation,
\begin{equation}\label{dpsiph}
\delta\psi = i\alpha\psi, \ \ \ \ \ \delta\psi^* = -i\alpha\psi^*\ .
\end{equation}
It is direct to see that the action is strictly invariant under these transformations (i.e. boundary term $K=0$), so Noether's charge follows only from the on-shell variation \eqref{fov}, by replacing the symmetry-variation \eqref{dpsiph}
\begin{align}\label{}
  \delta_{\mbox{\scriptsize os}} I &= \int dt \int d^3r \, i\hbar  {d \over dt} (\psi^* \delta\psi)  \nonumber\\
&= -\alpha \hbar \int dt {d \over dt}\int d^3r \,  (\psi^* \psi) \label{dosISch}\ ,
\end{align}
where we have assumed again that the field vanishes sufficiently rapidly at infinity so that additional boundary terms vanish. The associated Noether charge is thus
\begin{equation}\label{QSch}
Q = \int d^3r\, \psi^*\psi \ ,
\end{equation}
representing the total probability of finding the particle in space.  Note finally that the fact that $Q$ is conserved allows to fix it to be equal to one, $Q=1$, as one normally does in quantum mechanics. Had $\int \psi^*\psi$ not been conserved, the definition of probability would be different. 

As an exercise, one can carry out Noether's procedure in the case were the action is defined on a finite volume $V$ with surface $S$, and one does not assume that the field vanishes there, and show that in that case the correct conservation equations take the form
\begin{equation}
{\partial \rho \over\partial t} + \vec{\nabla} \cdot \vec{J} =0 \ ,
\end{equation}
where one must find $\rho$ and $\vec{J}$ as functions of $\psi$ and $\psi^*$. 

For some further developments on this subject, see \cite{deriglazov2009singular}.


\subsection{Two-dimensional conformal field theories: Scalar, Liouville, Dirac and bc system}

Two dimensional conformal field theories (CFTs) have received huge attention since the seminal work by Belavin, Polyakov and Zamolodchikov \cite{BPZ}. 
There exist many families of CFTs classified by their central charges. This subject has been treated in great detail in many reviews \cite{Ginsparg} and also in textbooks \cite{DiFran},\cite{Stefan}.

Our goal in this section will be to derive the energy momentum tensor for the most basic examples of CFTs, namely, the scalar field, Liouville field, Dirac fields and the bc system, arising in string theory. The energy momentum tensor is a central object in two dimensional CFTs. However, it if often derived by tricks that don't apply to all systems. Here we shall follow Noether's theorem and show that it always gives the right result with no ambiguities. Even for the Liouville field, that initially caused some confusion, Noether's theorem gives the so called \textit{improved tensor} in one step. 

For each example, our strategy is always the same: (i) find the correct variation for the given problem;  (ii) compute the Noether current (Virasoro charges) (iii) check that they generate the correct variations through Poisson brackets; (iii) compute the algebra (Virasoro) and find the value of the central charge. 
In the spirit of providing specific examples, we shall do this for scalars, fermions, the Liouville and the bc system. 

2d conformal field theory has infinitely many generators. We emphasize that this does not mean that the symmetry is a gauge symmetry. It only means that there are infinitely many Noether's symmetries and infinitely many conserved charges.  

The applications to string theory are a special case because in the worldsheet action
\begin{equation}
I[X^\mu,h_{\alpha\beta}] = \int \sqrt{h} h^{\alpha\beta} \partial_\alpha X^\mu \partial_\beta X^\nu \eta_{\mu\nu}  \ ,
\end{equation} 
the field $X^\mu$ and the metric $h^{\mu\nu} $ are varied. This action is not only conformally invariant but fully diffeomorphism invariant on the worldsheet. The variation over $h_{\alpha\beta}$ also makes the energy momentum tensor equal to zero. The conformal generators are zero because they generate a subalgebra of diffeos. 

\subsubsection{Free scalar field} \label{scalarCFT}

In most applications of 2d CFT's, it is most convenient to work in complex coordinates $z=e^{t-ix},\bar z=e^{t+ix}$ (for a more detailed treatment, see \cite{DiFran}). Note that in these coordinates, $x$ is the angular variable in the complex plane, and time $t$ relates to the radius of $z$, such that the infinite past $t=-\infty$ gets mapped into the origin $z=0$, while $t=\infty$ maps to complex infinity. 

We start by describing the simplest CFT, namely a scalar field in 2 Euclidean dimensions with action,
\begin{equation}\label{CFT0}
I[\phi] = {1 \over 4\pi }\int dz d\bar z\ \partial\phi \bar\partial\phi \ ,
\end{equation}
whose the equations of motion are
\begin{equation}\label{eqphi}
\partial\bar\partial \phi=0\ .
\end{equation}
The conformal symmetry is evident. Under
\begin{equation}\label{CT}
z \rightarrow f(z), \ \ \ \ \ \ \bar z \rightarrow \bar f(\bar z) \ ,
\end{equation}
were $f(z)$ is holomorphic ($\bar f(\bar z)$ antiholomorphic) the Jacobian in the volume element gives,
\begin{equation} \label{dwdw}
dzd\bar z \rightarrow  |\partial f|^2 dzd\bar z \ ,
\end{equation}
and the derivatives change in a similar way,
\begin{equation}
\partial\phi \bar\partial\phi \rightarrow {1 \over |\partial f|^2} \partial\phi \bar\partial\phi \ ,
\end{equation}
and therefore the action \eqref{CFT0} is left unchanged\footnote{Note that (\ref{CT})  produces the following transformation in the 2-dimensional metric,
\begin{equation}
ds^2 = dz d\bar z \rightarrow |\partial f|^2  dz d\bar z . \nonumber
\end{equation}}.  In the language of complex analysis, mappings of the kind \eqref{CT} are called \textit{conformal} transformations, since they preserve angles locally, or equivalently from \eqref{dwdw} transform the flat metric into itself multiplied by some overall local factor. This transformation acted on the coordinates of the manifold. As we have emphasized along this review, we now change the logic and express this transformation as a transformation that acts on the fields only leaving the coordinates untouched. As shown before, this point of view is important to find the associated conserved charges.

Consider the infinitesimal conformal transformation
\begin{equation}
z' = z + \epsilon(z)\hspace{.8cm},\hspace{.8cm} \bar z' = \bar z + \bar \epsilon(\bar z) \ ,
\end{equation}
where $\epsilon(z),\bar\epsilon(\bar z)$ are small arbitrary holomorphic and antiholomorphic functions respectively. The scalar field $\phi$ satisfies, by \eqref{Scalar}:
\begin{eqnarray}\label{dphi}
\phi(z,\bar z) &=&\phi'(z',\bar z')  \nonumber\\
         &=&  \phi'(z + \epsilon,\bar z + \bar\epsilon) \nonumber\\
         &=& \phi'(z,\bar z) + \epsilon\partial \phi(z,\bar z) + \bar\epsilon\bar\partial\phi (z,\bar z) \ ,
\end{eqnarray}
and from here we derive the transformation,
\begin{equation}\label{infCT}
\delta \phi(z,\bar z) = - \epsilon\partial \phi(z,\bar z) - \bar\epsilon\bar\partial\phi
(z,\bar z ).
\end{equation}
We check now that the action \eqref{CFT0} is invariant under the conformal transformation \eqref{infCT} {\it without} moving the coordinates,
\begin{align}
   \delta I  &= - \frac{1}{4\pi}\int d^2z\left[ \partial (\epsilon \partial\phi + \bar\epsilon\bar\partial\phi)\bar\partial\phi  + \partial\phi \bar\partial (\epsilon \partial\phi + \bar\epsilon\bar\partial\phi) \right] \nonumber\\
  &=-\frac{1}{4\pi}\int d^2z \left[ \bar \partial \left( \bar \epsilon \partial \phi \bar \partial \phi \right) +  \partial \left( \epsilon \partial\phi \bar \partial \phi \right)  \right]   \label{dIcft} \ ,
\end{align}
and thus a symmetry. (Observe that the result can be expressed as $\partial_\mu K^\mu$.)

Let us compute the Lie algebra of conformal transformations. One can regard $z$ and $\bar z$ as independent variables and consider one half of the
transformation,
\begin{equation}
\delta_1 \phi = -\epsilon_1(z) \partial\phi \ .
\end{equation}
Acting with a second transformation,
\begin{equation}
\delta_2 \delta_1 \phi = \epsilon_2 \partial \epsilon_1 \partial\phi + \epsilon_2
\epsilon_1 \partial^2 \phi
\end{equation}
we derive,
\begin{equation}
[\delta_1,\delta_2]\phi = (\epsilon_1\partial\epsilon_2 - \epsilon_2\partial\epsilon_1)
\partial\phi\ .
\end{equation}
As expected, the algebra \textit{closes}: the commutator of two transformations is equal to a new transformation with parameter
\begin{equation}
\epsilon_2\partial\epsilon_1 - \epsilon_1\partial\epsilon_2\ .
\end{equation}
This equation defines the structure constants of this problem. A more explicit form of the structure constants follows by expanding $\epsilon(z)$ in a
Laurent series,
\begin{equation}\label{epsilonn}
\epsilon(z) = \sum_{n\in Z} \epsilon_n \, z^{n+1}\ .
\end{equation}
The transformation of the field \eqref{infCT} can be written as,
\begin{eqnarray}
\delta \phi &=& -\sum \epsilon_n z^{n+1} \partial \phi \nonumber\\
 &=& \sum \epsilon_n L_n \phi \ ,
\end{eqnarray}
where we have defined the operator ,
\begin{equation}
L_n = - z^{n+1} \partial \ ,
\end{equation}
which acts on functions.  Let us compute the Poisson bracket of two operators\footnote{Note
that this procedure is standard in first quantized theories. For example, the wave
function in quantum mechanics changes under constant translations $x\rightarrow x+a$ in the form $\delta\psi(x) = - a {d \over dx} \psi$.  One then identifies the translation
operator as ${d \over dx} $, which can be expressed in terms of the hermitian momentum
operator $p = -i {d \over dx}$, and have an associated abelian algebra, $[p,p]=0$. },
\begin{eqnarray}
  [L_n,L_m]\phi &=& z^{n+1} \partial (z^{m+1}\partial \phi ) - z^{m+1} \partial (z^{n+1}\partial\phi) \nonumber\\
   &=& (m+1) z^{n+m+1} \partial\phi - (n+1) z^{n+m+1}\partial\phi \nonumber\\
   &=&  (n-m) L_{n+m}\phi  \label{classV} \ 
\end{eqnarray}
giving a nice representation for the structure constants. This algebra is
known as ``classical Virasoro algebra". In the quantum theory, as well as some classical examples considered below (Liouville theory), (\ref{classV}) develops a central term at the right hand side.

We end this paragraph commenting that the other half of the symmetry $\delta \phi = -
\bar\epsilon \bar\partial\phi$ has associated operators $\bar L_n$ which satisfy the
same algebra \eqref{classV}. The two copies of this algebra (one for the holomorphic and another for the antiholomorphic part) are called the ``conformal
algebra".

The action \eqref{CFT0} possesses an infinite dimensional symmetry, the conformal
symmetry and this symmetry gives rise to conserved charges via
Noether's theorem. Furthermore, in the canonical (Hamiltonian) picture, the charges generate the corresponding symmetry transformations via Poisson brackets, and in the quantum theory, via
commutators.

Let us workout the Noether charges associated to the holomorphic conformal
transformations. We consider the variation of the action (\ref{CFT0}) under
\begin{equation}\label{dphi2}
\delta_s\phi = -\epsilon(z) \partial\phi \ ,
\end{equation}
with $\epsilon(z)$ holomorphic. Under this transformation, the symmetry-variation is given in \eqref{dIcft}
\begin{equation}\label{dsym}
\delta I[\phi,\delta_s\phi] = {1 \over 4\pi } \int d^2z\  \partial (-\epsilon
\partial\phi \bar\partial\phi).
\end{equation}

On the other hand, the on-shell variation of the action is
\begin{equation}\label{dgen}
\delta I[\bar\phi,\delta\phi] = {1 \over 4\pi } \int  d^2z \left[ \partial(\delta\phi \bar\partial\phi) +  \bar\partial (\delta\phi \partial\phi)  \right]\ .
\end{equation}
Note here that the bar on $\bar\phi$ here refers not to complex conjugation but to the classical solution of the e.o.m. 

Inserting (\ref{dphi2}) into the on-shell variation and $\bar \phi$ into the symmetry variation
The left hand sides are equal. Subtracting both equations one derives the conservation law, 
\begin{equation}
\bar\partial \big( \epsilon \left( \partial \phi \right)^2 \big)=0\ ,
\end{equation}
where we have dropped the bar over the field. Since $\epsilon(z)$ only depends on $z$, the parameter can be extracted and we find our conservation law,
\begin{equation}\label{T}
\bar\partial T = 0,  \ \ \ \ \ \  T = - {1 \over 2} \left( \partial\phi \right)^2\ ,
\end{equation}
where the minus sign follows standard conventions. In order to convince ourselves that this does indeed represent a conservation law, we can repeat the same procedure for the anti-holomorphic side and find, of course, 
\begin{align}\label{barT}
\partial \bar T = 0 \hspace{1cm},\hspace{1cm}  \bar T =- {1 \over 2} \left( \bar\partial\phi  \right)^2\ .
\end{align}
These two conservations laws can be now written in a more conventional form,
\begin{equation}
\bar \partial (\epsilon T) + \partial (\bar\epsilon \bar T) =
\partial_\mu J^{\mu}=0\ ,
\end{equation}
where the current has components $J^{\bar z} = \epsilon T$ and $J^{ z} = \bar\epsilon
\bar T$. However, since $\epsilon$ and $\bar\epsilon$ are arbitrary, this last equation
is in fact equivalent to  (\ref{T}) and (\ref{barT}).

The conservation equation (\ref{T}) gives rise to conserved charges in the standard sense
too (and analogously for \eqref{barT}). Recall that given a conserved current $J^\mu$, the integral $\int d^3x J^0$ is the
conserved charge, which in turn generates the symmetry through the Poisson bracket. We shall now describe the analogous of this procedure in our complex
notation.

 Since $T(z)$ is an holomorphic function (as a consequence of the conservation equation) we can expanded in Laurent modes in the form
\begin{equation}\label{TL}
T(z) = \sum_{n \in Z} {L_n \over z^{n+2}}\ ,
\end{equation}
where the $L_n$ are arbitrary coefficients. Soon we will see that these are the conformal
generators and satisfy the Virasoro algebra. Equation (\ref{TL}) can be inverted by
multiplying by $z^{m+1}$ and integrating on a circle containing the origin at both sides,
\begin{eqnarray}
  \oint {dz \over 2\pi i} z^{m+1} T(z) &=& \sum_{n \in Z} L_n \oint {dz \over 2\pi i} {1 \over z^{n-m+1}}\ .
\end{eqnarray}
The right hand side is non-zero only for $n=m$, so the sum reduces to $n=m$, and we can
solve for $L_m$
\begin{equation}\label{Ln}
L_m = \oint {dz \over 2\pi i} z^{m+1} T(z).
\end{equation}
The coefficients $L_n$ are the conserved charges associated to the conformal symmetry.
The reason is the following.  By Cauchy's theorem, the integral in (\ref{Ln}) does not
depend on the radius of the circle. Recall now that the time coordinate is equal to
$t=\log|z|$. Hence, the independence on $|z|$ is exactly telling us that the $L_n$ are
conserved in time.  This is un full analogy with the conservation of $\int_{\Sigma} d^3x J^0$
whose key property is that the value of the integral does not depend on which surface
$\Sigma$ is chosen.

One last comment regarding the existence of an infinite number of conserved charges.
The conformal symmetry contains a parameter $\epsilon(z)$ which is an arbitrary (analytic) function of
$z$. This is quite different from standard global symmetries, like space translations $x
\rightarrow x+ a  $, involving a constant parameter $a$. There is one
Noether charge for each parameter in the symmetry. Making a Laurent series of
$\epsilon$ as we did in (\ref{epsilonn}), we see that the conformal symmetry does in fact
contain an infinite number of parameters.  As we shall see in the next paragraph, the
charge $L_n$ can be seen as the Noether charge associated to the transformation in which only the coefficient of $\epsilon^n$ is different from zero; in other words, that $[\phi(z),L_n]=z^{n+1}\partial \phi$. 

As an exercise, the reader may show that the action  (\ref{CFT0}) in also invariant under
the {\it affine} symmetry, \begin{equation} \delta \phi = \epsilon(z)\ ,
\end{equation}
where $\epsilon$ only depends on $z$. Find the associated conserved charge. What is the
algebra of these charges?

\subsubsection{Light cone quantization: canonical realization of the conformal symmetry}

Our last step in the classical description of the conformal symmetry associated to the
action (\ref{CFT0}) is to extract the relevant Poisson brackets and check explicitly
that the currents $T(z)$ and their associated charges $L_n$ do generate these symmetries
in the canonical picture.

The Poisson bracket associated to the action 
\begin{equation}\label{Ic}
I = {1 \over 4\pi} \int d^2z\ \partial\phi \bar\partial\phi 
\end{equation}
can be extracted by standard methods, via the introduction of a canonical 
momentum $\pi(x,t)$. Here we shall follow an alternative route 
which gives the correct answer in a much quicker fashion, and is better adapted to the
corresponding quantum calculation via OPE's.

The idea is use the coordinate $\bar z$ as time, and derive an equal ``time" basic Poisson bracket between $\phi(z)$ and $\partial \phi(z)$. The general method needed is the following. 

Consider an action of the form $I[w] = \int dt (\ell_a(w)\dot w^a - H(w)) $ where $w^a$ denotes collectively all fields in the theory and $\ell_a(w)$ are some functions of $w^a$. The equations of  motion are
\begin{equation}
\sigma_{ab} \dot w^b = \partial _ a H \ \ \ \mbox{where} \ \ \   \sigma_{ab} =\partial_a l_b - \partial_b l_a.
\end{equation} 
If $\sigma_{ab}$, called the symplectic matrix, is invertible one can define a Poisson bracket structure as follows:
\begin{equation}
[w^a,w^b] = J^{ab}, \ \ \ \ \ \ \  J^{ab}\sigma_{bc} = \delta^{a}_{\ c} \ ,
\end{equation} 
and the equations of motion take the usual form $\dot w^a = [w^a,H]$. Furthermore, this Poisson bracket satisfies the Jacobi identity thanks to the fact that $\sigma_{ab}$ is an exact form. See \cite{wittenbos,Jackiw-Faddeev} for more details and examples of this procedure. 

Let us apply this procedure to the scalar field action (\ref{Ic}) where we interpret $\bar z$ as time. The function $\ell(\phi)$ in this problem is,
\begin{equation}
\ell(\phi(z)) = {1 \over 4\pi} \partial\phi(z) \ ,
\end{equation}
and the symplectic matrix $\sigma(z,z')$ is, 
\begin{eqnarray}
  \sigma(z,z') &=& {\delta \ell(\phi(z)) \over \delta \phi(z')} - {\delta \ell(\phi(z')) \over \delta \phi(z)} \nonumber \\
               &=& {1 \over 2\pi} \partial' \delta(z,z')\ .
\end{eqnarray}
The inverse of $\sigma(z,z')$ is
\begin{equation}
J(z,z') = 2\pi {1 \over \partial'} \delta(z,z') \ ,
\end{equation}
where $1/\partial$ is the inverse of the operator $\partial$. We then find the ``equal-time" Poisson bracket
\begin{equation}\label{PB}
[\phi(z,\bar z),\phi(z',\bar z)] = 2\pi  {1 \over \partial'} \delta(z,z')\ .
\end{equation}
The presence of $1/\partial$ in this expression may look intimidating.
Fortunately, we shall never need the Poisson bracket of $\phi$ with itself, but only
with its derivatives. Differentiating (\ref{PB}) we deduce,
\begin{equation}\label{PB1}
[  \phi(z,\bar z),\partial'\phi(z',\bar z)] = 2\pi  \delta(z,z') \ ,
\end{equation}
which does not have any inverse derivatives, and is the expression we shall need\footnote{As a footnote we observe that it is incorrect to declare, from (\ref{Ic}), that ${1 \over 4\pi} \partial \phi$ is conjugate to $\phi$. The correct formula is (\ref{PB1}).}.  

As a first check of our newly built Poisson structure, let us check that the Noether charges associated to conformal transformations have the right properties under it. We have proved that $T = -1/2 (\partial \phi)^2$ is the conserved charge. An arbitrary conformal transformation with parameter $\epsilon(z)$ is generated by 
\begin{equation}\label{Q_e}
Q_\epsilon = \oint {dz \over 2\pi i} \epsilon(z) T(z)\ .
\end{equation}
Now by Noether's inverse theorem \eqref{invNoe}, we expect $Q_\epsilon$ to generate the conformal transformation of the field through the commutator. To prove this, we use \eqref{PB1} to compute
\begin{eqnarray}
\delta \phi&=& i[\phi(z,\bar z),Q_\epsilon] \nonumber \\
 &=& -\oint {dz' \over 2\pi i} \epsilon(z')  [\phi(z,\bar z), \partial\phi(z',\bar z)] \partial\phi(z',\bar z)  \nonumber\\
&=& - \oint {dz' \over 2\pi i} \epsilon(z') \ 2\pi i \delta(z',z) \partial\phi(z',\bar z)\nonumber\\
&=& - \epsilon(z)\, \partial\phi  \ ,
\end{eqnarray}
which is, as expected, the correct conformal map of the field, \eqref{infCT}, for the holomorphic sector. As an immediate consequence, we see that in particular for the Virasoro charges $L_n$ \eqref{Ln} (which are just $Q_\epsilon$ for $\epsilon=z^{n+1}$) we have
\begin{eqnarray}\label{dphiLn}
i[\phi(z,\bar z),L_n] &=& - z^{n+1}\, \partial\phi\ .
\end{eqnarray}

The charge $Q_\epsilon$ is a linear
combination of all conserved charges $L_n$: by inserting the Laurent expansions
(\ref{epsilonn}) and  (\ref{TL}) for $T(z)$ and $\epsilon(z)$, the integral over $z$ is performed
and we find,
 \begin{equation}\label{Qepsilon}
Q_\epsilon = \sum_{n\in Z} \epsilon_n \, L_n.
\end{equation}

As an exercise, recall that the free action is also invariant under affine
transformations, $\delta\phi = \epsilon(z)$, whose associated charges are $i\partial\phi
= \sum {a_n \over z^{n+1}}$. Show that the inverse transformation is $a_n = \oint {dz
\over 2\pi i} z^{n} i\partial\phi$. Show that the Poisson bracket of $a_n$ gives
$[a_n,a_m] = n \, \delta_{n,m}$. Find  $L_n$ in terms of $a_n$. Compute the Poisson
bracket $[L_n,a_m]$.

We are now ready to compute the algebra of generators $L_n$, now as Poisson brackets of canonical fields. This is totally analogous to computing the algebra of angular momenta in quantum mechanics, $[L_i,L_j]=i\epsilon_{ijk} L_k$, that is, the algebra obeyed by the generators of the symmetry. We have 
\begin{eqnarray}
i[L_n,L_m]
&=& \oint {dz \over 2\pi i}\oint {dz' \over 2\pi i} z^{n+1} z'^{m+1}i \left[ {1 \over 2} (\partial\phi(z))^2,{1 \over 2} (\partial'\phi(z'))^2\right]\nonumber\\
&=& -\oint {dz \over 2\pi i}\oint {dz' \over 2\pi i} z^{n+1} z'^{m+1} \partial\phi(z)\partial'\phi(z')\ 2\pi i \, \partial'\delta(z,z')\nonumber\\
&=&  \oint {dz \over 2\pi i} \oint dz'\, z^{n+1} \partial'[z'^{m+1}\partial'\phi(z')] \partial\phi(z)\, \delta(z,z')\nonumber\\
&=& \oint {dz \over 2\pi i} z^{n+1} \left[ (m+1) z^m (\partial\phi)^2 + z^{m+1} {1 \over 2 } \partial(\partial\phi)^2\right]  \nonumber\\
&=& \oint {dz \over 2\pi i}  z^{n+m+1}  \left[ (m+1) - (n+1)  \right] {1 \over 2}(\partial\phi)^2  \nonumber\\
&=& (n-m) L_{n+m} \label{LnLm}\ .
\end{eqnarray}
Thus, indeed the conserved charges provide a canonical representation for the conformal algebra \eqref{classV}.

Note:   The charges $L_n$ do not depend on the contour (they are conserved).  The
commutator can be easily calculated when both contours (for $L_n$ and $L_m$) are
the same, but the result of the calculation does no depend on that.

\subsubsection{Liouville field }

Our next example of a \textit{classical} CFT will be Liouville theory. This is
a non-trivial example because, in contrast to the free scalar field, it is a fully interacting theory. For references on Liouville
theory we refer, for example, to \cite{Seiberg,Ketov}.  

Liouville theory is defined by the action of a field $\phi$ interacting through an exponential potential,
\begin{equation}\label{ILiouville}
I = {1 \over 4\pi} \int d^2z \left( \partial \phi \bar\partial\phi + \Lambda e^{
\phi/\gamma} \right) \ .
\end{equation}
Here $\Lambda$ and $\gamma$ are constants.  The constant $\Lambda$ is somehow trivial because
it can be redefined by shifting $\phi$. We thus expect this constant not to play any
important role.   The equation of motion following from (\ref{ILiouville}) is,
\begin{equation}\label{eomLio}
\partial\bar\partial \phi = {\Lambda \over 2\gamma} e^{\phi/\gamma}.
\end{equation}

At first sight, Liouville theory is not conformally invariant because the potential is
not invariant if the field transforms as a scalar. However, we note that the potential
is invariant under a different transformations for the field, namely,
\begin{equation}\label{Liou}
e^{\phi(z,\bar z)/\gamma} \rightarrow e^{\phi'(z',\bar z')/\gamma} = \left({\partial z'
\over \partial z}\right)^{-1} \left({\partial \bar z' \over \partial \bar z}\right)^{-1}
e^{ \phi(z,\bar z)/\gamma}.
\end{equation}
In the language of CFT, this is the transformation for a field of conformal weights $(h,\bar h) = (1,1)$, but is not \textit{primary field}\footnote{A field is called \textit{primary} with conformal weights $h,\bar h$ if, under any conformal map $z\rightarrow f,\bar z\rightarrow \bar f$ it transforms as 
\begin{align}\label{weight}
\phi'(z',\bar z')=\left( \frac{df}{dz} \right)^{-h} \left( \frac{d\bar f}{d\bar z} \right)^{-\bar h}  \phi(z,\bar z).
\end{align}
But in \eqref{Liou} it is $e^{\phi}$ which transforms as a primary field, not $\phi$. } properly. As we discuss below, it turns out that the kinetic term is also invariant under this transformation and hence
this is a symmetry of the whole action. This is of course again the conformal
symmetry, although it is being represented in a very different way.

Let us first check that the kinetic terms is also invariant. We shall do so by looking
at the infinitesimal transformations. This will also prove to be useful when computing the conserved
charge, and hence the generator of conformal transformations in Liouville theory.  We
know in advance that this generator cannot be $(\partial\phi)^2$ 
because now the transformations for the field has changed.

As we did for the non interacting field, for simplicity we shall deal only with half of the
transformations, namely $z \rightarrow z'=f(z)$. From \eqref{Liou}, the transformed field is
\begin{equation}
\phi'(z') = \phi(z) - \gamma \log (\partial f ) \ ,
\end{equation}
and its infinitesimal version $f(z)=z+\epsilon(z)$, in terms of $\delta\phi =
\phi'(z)-\phi(z)$, becomes,
\begin{equation}\label{dL}
\delta \phi = -\epsilon \partial \phi - \gamma \partial \epsilon\ .
\end{equation}
The first piece is clearly the same as that for the scalar field. The full
transformation is not that of a primary field. Our goal now, as usual, is to find the canonical
Noether charge that reproduces this transformation.

We start by checking the invariance of the action under the infinitesimal transformation
(\ref{dL}). We keep in this calculation all boundary terms.

We start by computing the variation of the potential,
\begin{eqnarray}
  \delta \left( \Lambda e^{\phi/\gamma} \right)  &=& \Lambda e^{\phi/\gamma}\, { -\epsilon \partial\phi - \gamma\partial\epsilon \over \gamma} \nonumber \\
   &=& -\partial \left( \Lambda\epsilon\, e^{\phi/\gamma} \right)   \label{dLP} 
\end{eqnarray}
obtaining a boundary term, as needed. This is of course not a surprise because we already knew
 that the potential was invariant under the finite transformation. 

Now we want to check that the kinetic part of the action is also invariant under
(\ref{dL}), which is not so evident. The transformation (\ref{dL}) has two pieces $\delta\phi=\delta_1\phi +
\delta_2\phi$. The first piece corresponds to the original transformation of the free action for the scalar field \eqref{dphi2}, and we already showed in \eqref{dsym} that acting on the kinetic term this transformation gives
\begin{equation}\label{dK1}
\delta_1 \left(\partial\phi\bar\partial\phi\right) = -  \partial \left( \epsilon
\partial\phi \bar\partial\phi \right) \ .
\end{equation}
On the other hand, the transformation $\delta_2\phi$ is new, and we need to check that
it leaves the kinetic term invariant too.  Using $\bar\partial\epsilon=0$ we find,
\begin{eqnarray}
  \delta_2 \left(\partial\phi\bar\partial\phi\right)  &=& -\partial(\gamma \partial\epsilon)\bar\partial\phi-\partial\phi \bar\partial(\gamma\partial\epsilon) \nonumber \\
   &=&  -\bar\partial (\gamma \phi \partial^2 \epsilon  ) \label{dK2} \ ,
\end{eqnarray}
which also is a boundary term, as required.

 We have proved that the Liouville action is conformally invariant, provided the field transforms as in (\ref{dL}). We would like now to extract the associated Noether charge. As we have done all along this review, to this end we need to compare the boundary terms that we obtained by varying the action with respect to the symmetry, with the boundary terms that arise in the on-shell variation of the action. The on-shell variation \eqref{fov} evaluated at the symmetry \eqref{dL} gives
\begin{align}\label{dLos}
\delta I & = -\frac{1}{4\pi}\int d^2z \left[ \partial \left(  \left( \epsilon \partial \phi+\gamma \partial \epsilon \right) \bar \partial \phi \right) + \bar \partial \left( \partial\phi \left( \epsilon \partial \phi +\gamma \partial \epsilon \right) \right) \right] \ ,
\end{align}
while the symmetry-variation is just the recollection of \eqref{dLP}-\eqref{dK2}:
\begin{align}\label{dLsym}
\delta  I&= -\frac{1}{4\pi} \int d^2z \left[ \partial \left( \epsilon \partial \phi \bar \partial \phi \right) + \bar \partial \left( \gamma  \phi \partial^2 \epsilon  \right) + \partial \left( \Lambda \epsilon e^{\phi/\gamma} \right) \right] \ .
\end{align}
Equating \eqref{dLos} and \eqref{dLsym} is the meaning of Noether's theorem. Simplifying terms and using the e.o.m. \eqref{eomLio}, we find
\begin{eqnarray}
\partial\left( (\epsilon\partial\phi + \gamma \partial\epsilon)\bar\partial\phi   \right) + \bar\partial \left(\partial\phi( \epsilon \partial\phi + \gamma  \partial\epsilon )\right)   &=&  \partial \left( \Lambda\epsilon\, e^{\phi/\gamma} \right) +  \partial \left( \epsilon \partial\phi \bar\partial\phi \right)  +\bar\partial (\gamma \phi  \partial^2 \epsilon ) \nonumber\\
\cancel{\partial\left( \epsilon\partial\phi \bar\partial\phi \right)} + \gamma \partial\left( \partial\epsilon \bar\partial\phi \right) + \bar\partial \left(\partial\phi( \epsilon \partial\phi + \gamma  \partial\epsilon )\right)  &=&  \partial\left( 2\gamma\epsilon \partial\bar\partial\phi  \right)   + \cancel{\partial \left( \epsilon \partial\phi \bar\partial\phi \right)} + \gamma\partial^2 \epsilon \bar\partial\phi \nonumber\\
\cancel{\gamma\partial^2\epsilon\bar\partial\phi} + \bcancel{\gamma \partial\epsilon \partial\bar\partial\phi} + \bar\partial\left(\epsilon (\partial\phi)^2 + \bcancel{\gamma \partial\phi\partial\epsilon} \right)   &=&  \bcancel{2\gamma\partial\epsilon \partial\bar\partial\phi} + 2\gamma \epsilon \bar\partial \partial^2\phi  + \cancel{\gamma\partial^2 \epsilon \bar\partial\phi}  \nonumber\\
\epsilon \bar\partial (\partial\phi)^2   &=& 2\gamma\epsilon \bar \partial \partial^2\phi \ ,
\end{eqnarray}
from where we read Noether's conservation law
\begin{equation}\label{impT}
 \bar\partial T=0, \ \ \ \ \ \ \ \ T =- {1 \over 2}(\partial\phi)^2  + \gamma \partial^2 \phi \ .
\end{equation}
In the literature \cite{Jackiw}, this expression for $T$ is called ``improved" energy momentum tensor. Additionally, it is direct to show by explicit calculation that $\bar \partial T=0$ by use of the e.o.m.

As always, we must now know check that $T$ indeed generates the right conformal transformation of $\phi$ via Noether's inverse theorem \eqref{invNoe}.
First of all we note that the Poisson bracket in Liouville theory is not different from the free case, because the potential does
not introduce any derivatives of the field, so it is still given by \eqref{PB1}. Furthermore, in total analogy to what was done for the free scalar in section \ref{scalarCFT}, the conserved $T(z)$ gives rise to an infinite number of conserved currents, with associated conserved charge $Q_\epsilon=\oint \frac{dz}{2\pi i} \epsilon(z) T(z) $ with the correspondent $T$ from \eqref{impT} for the Liouville action. Then, the symmetry transformation generated by this Noether charge is 
\begin{eqnarray}
  \delta \phi(z) &=&   [\phi(z),Q_\epsilon] \nonumber \\
&=& \oint {dz' \over 2\pi i} \epsilon(z')  \left[\phi(z) ,-{1 \over 2} (\partial'\phi(z'))^2 + \gamma \partial^{'2}\phi(z') \right ] \nonumber  \\
   &=& -\epsilon \partial\phi  + \gamma \oint dz' \epsilon(z') \partial' \delta(z,z') \nonumber \\
   &=& -\epsilon \partial\phi - \gamma\partial\epsilon \ ,
\end{eqnarray}
which is the correct transformation law for Liouville fields \eqref{dL}. 

The final step is to compute the resulting algebra of the Virasoro generators. Just as for the free scalar in \eqref{Ln}, the Virasoro modes are defined as the charges $Q_\epsilon$ associated specifically to the $\epsilon(z)=z^{n+1}$ transformations, namely $L_n=\oint {dz \over 2\pi i} \, z^{n+1}\, T(z)$ so we have
\begin{eqnarray}
L_n 
&=& -\oint {dz \over 2\pi i} \, z^{n+1}\, {1 \over 2} (\partial\phi)^2  + \gamma \oint {dz \over 2\pi i} \, z^{n+1}\, \partial^2\phi\nonumber\\
&=& L_n^{(0)} - \gamma (n+1) \oint {dz \over 2\pi i} \, z^{n}\,
\partial\phi
\end{eqnarray}
where we have called $L^{(0)}_{n}$ the free scalar Virasoro
charges, which we already know satisfy the Virasoro algebra with no central charge \eqref{LnLm}. We also
made an integral by parts in the other term, which is useful because
$\partial\phi$ is primary with conformal dimension 1 with respect to
$L_n^{(0)}$. By direct application of these definitions we first see by differentiating \eqref{dphiLn} that
\begin{equation}
i[L^{(0)}_{n},  \partial\phi ] = \partial( z^{n+1} \partial\phi) \ .
\end{equation}
So now we compute the full commutator:
\begin{eqnarray}
  i[L_n,L_m] &=& i[L^{(0)}_n,L^{(0)}_m] + \gamma(m+1) \oint {dz \over 2\pi i}z^m  i[L_n^{(0)},\partial\phi] + \gamma (n+1) \oint {dz \over 2\pi i}z^n  i[\partial\phi,L_m^{(0)}]  \nonumber\\
  & & \ \ \ \ \ \ \ \ \ + \ \gamma^2(n+1)(m+1) \oint {dz \over 2\pi i}  \oint {dz' \over 2\pi i} z^n z'^{m}i [\partial\phi,\partial'\phi']    \nonumber\\
   &=& (n-m) L^{(0)}_{n+m}\ + \  \left[(n+1)-(m+1)\right] \oint {dz \over 2\pi i} z^{n+m+1}\left(-\gamma  \partial^2\phi \right)\\
  && \ \ \ \ \ \  \  + \ \gamma^2(n+1)(m+1) \oint {dz \over 2\pi i}  \oint {dz' \over 2\pi i} z^n z'^{m} \partial\delta(z,z')  \nonumber\\
   &=& (n-m) L_{n+m} + \gamma^2 n(n^2-1) \delta_{n+m,0} \nonumber \ .
\end{eqnarray}
The central charge is then,
\begin{equation}
c =  12 \gamma^2,
\end{equation}
and depends only on $\gamma$, not on $\Lambda$, as we anticipated at the beginning. The presence of central charges is not an exclusive feature of quantum theories; as this example proves, they can also appear in totally classical theories!


 \subsubsection{Dirac fields in two dimensions}

In this subsection we consider another example of a 2d CFT, namely, the massless fermionic field in Euclidean spacetime described by the action,
\begin{equation}\label{dirac}
I = {1 \over 4\pi }\int d^2x\ \bar\Psi \gamma^\mu\, \partial_\mu \Psi \ .
\end{equation}
Here $\Psi$ is a two dimensional object whose components will be denoted as
\begin{equation}
\Psi = \left(\begin{array}{cc}
  \psi  \\
  \bar\psi  \\
\end{array}\right).
\end{equation}
The $\gamma^\mu$ are the Dirac matrices which satisfy the Euclidean Clifford algebra,
\begin{equation}
\{ \gamma_\mu,\gamma_\nu \} = 2 \delta_{\mu\nu}.
\end{equation}
There are many ways in which to represent the $\gamma'$s. A useful two dimensional representation is
\begin{equation}
\gamma^0 = \left( \begin{array}{cc}
  0 & 1 \\
  1 & 0 \\
\end{array}\right),
\ \ \ \ \ \ \gamma^1 = \left( \begin{array}{cc}
  0 & i \\
  -i & 0 \\
\end{array}\right).
\end{equation}
Finally, the Dirac conjugated that enters in the action \eqref{dirac} is defined as
\begin{equation}
\bar \Psi = \psi^\dagger \gamma^0.
\end{equation}

Just as we did for the scalar field, we shift to complex ``light-cone" coordinates $ z = {1 \over 2}( x+it),\bar z = {1 \over 2}( x-it)$ implying $\partial = \partial_x -i\partial_t,\bar\partial = \partial_x +i\partial_t$. We now observe that the Lagrangian contains
 \begin{eqnarray}
    \gamma^0 \gamma^0 \partial_0 + \gamma^0\gamma^1 \partial_1  &=&
           \partial_0 + \left( \begin{array}{cc}
            -i & 0 \\
            0 & i \\
          \end{array}\right) \partial_1 \nonumber  \\
    &=&  -i\left( \begin{array}{cc}
      \bar\partial & 0 \\
      0 & -\partial \\
    \end{array} \right)\ .
 \end{eqnarray}
 The action for this free fermion system becomes
 \begin{equation}\label{Iphi}
 I = {1 \over 4\pi i} \int d^2z\ \left( \psi \bar\partial \psi - \bar\psi \partial \bar\psi \right)\ ,
 \end{equation}
 and the problem splits once again in two parts, one holomorphic $\psi$ and one anti-holomorphic.


Before carrying on with the analysis we note that our action as it stands does not
make much sense because, naively, $\psi\partial\psi = {1 \over 2} \partial( \psi^2)$.
The action is a total derivative and there is no classical dynamics at all. So what's going on? The answer is that in order for Dirac's theory to make sense, even at the classical level we must demand that the fields are not ordinary $\mathbb C$ numbers but instead \textit{Grassman numbers}. These are by definition anticommuting quantities, so that $\psi^2=0$ and therefore $\psi\partial\psi \neq {1 \over 2} \partial( \psi^2)$. 

\bigskip

\textit{The conformal symmetry} 

\bigskip

Although the action \eqref{Iphi} is not evidently conformally invariant, in fact it is. The trick is not
to insist that field should transform as scalar under conformal mappings. As we did with
the scalar field, we shall treat each sector in the action separately. We then consider only the first part of the action \eqref{Iphi}
\begin{equation}
I = {1 \over 4\pi i}\int dz d\bar z\ \psi(z,\bar z) \bar\partial \psi(z,\bar z)\ ,
\end{equation}
which is invariant under the following holomorphic transformation:
\begin{align}\label{dpsi}
  z' = f(z) \hspace{.5cm},\hspace{.5cm}  \bar z'=\bar z \hspace{.5cm},\hspace{.5cm}
  \psi'(z',\bar z') = [\partial f(z)]^{-1/2} \psi(z,\bar z)  \ .
\end{align}
(note we have left $\bar z$ invariant). Indeed, 
\begin{eqnarray*}
\int dz' d\bar z'\ \psi'(z',\bar z') \bar\partial \psi'(z',\bar z')  &=& \int dzd\bar z [\partial f(z)] [\partial f(z)]^{-1} \psi(z,\bar z) \bar\partial \psi(z,\bar z)\\
   &=& \int dzd\bar z\ \psi(z,\bar z) \bar\partial \psi(z,\bar z) \ .
\end{eqnarray*}
The full action \eqref{Iphi} is invariant under the map \eqref{dpsi} provided $\bar \psi$ remains unchanged, i.e. transforms as a scalar. $\psi$  and $\bar\psi$ have the following conformal weights:
\begin{equation}
\psi \ \ : \ \ \left( {1 \over 2}, 0 \right)\hspace{1cm},\hspace{1cm} \bar \psi \ \ :\ \ \left( 0, \frac{1}{2} \right)\ .
\end{equation}

\bigskip

\textit{The Fermionic Virasoro operator and Canonical structure}

\bigskip

We now derive the generator of conformal transformations
acting on spinor fields. As usual, we first formulate the symmetry as an operation acting only
on the field, not on the coordinates. Expanding \eqref{dpsi} for the infinitesimal transformation $z' = z + \epsilon(z)$ to first order (we omit the $\bar z$ dependence which plays no role)
\begin{eqnarray*}
  \psi'(z + \epsilon(z)) &=& (1 + \partial \epsilon(z) )^{-1/2} \psi(z) \\
  \psi'(z) + \epsilon(z) \partial\psi(z) &=& \left( 1 - {1 \over 2}\partial \epsilon(z) \right) \psi(z)\ ,
\end{eqnarray*}
we derive the transformation rule:
\begin{equation}\label{dpsi z}
\delta_s\psi(z,\bar z) = -\epsilon(z) \partial\psi(z,\bar z) - {1 \over 2}\partial \epsilon(z)
\psi(z,\bar z) \ .
\end{equation}
The first term in this transformation is the same as of the scalar field, while the second
clearly comes from the extra factor $\left(\partial  f \right )^{-1/2} $ in the transformation rule.

Compute now the symmetry-variation of the action under the
infinitesimal symmetry keeping all boundary terms:
\begin{eqnarray}
\delta I[\psi,\delta_s\psi] &=& - {1 \over 4\pi i} \int  d^2z \left[ \left(\epsilon \partial \psi + {1 \over 2}  \partial \epsilon \,\psi \right) \bar \partial\psi + \psi \bar \partial \left( \epsilon \partial\psi + {1 \over 2} \partial \epsilon \ \psi \right) \right]\nonumber\\
&=& - {1 \over 4\pi i} \int d^2z\ \partial (\epsilon \psi \bar\partial\psi) \ ,
\end{eqnarray}
which is a boundary term, as expected. Note however that this boundary term is zero on shell and there it will not contribute to the charge. 

Compute now the on-shell variation, 
\begin{eqnarray}
\delta I[\bar\psi,\delta\psi] &=&  {1 \over 4\pi i} \int d^2z\ \bar\partial (\psi \delta \psi)\nonumber\\
&=&  - {1 \over 4\pi i} \int d^2z\  \bar\partial \left( \psi \left( \epsilon \partial\psi
+ {1 \over 2}\partial \epsilon\ \psi \right)\right)
\nonumber\\
&=& - {1 \over 4\pi i} \int d^2z\  \epsilon\ \bar\partial \left( \psi
\partial\psi \right) \ ,
\end{eqnarray}
where we have used that $\epsilon$ is independent of $\bar z$, and
$\psi^2=0$ (Grassman variable).  We have also dropped the bar of the field $\psi$.

The fermionic conserved current associated to conformal symmetry is then,
\begin{equation}\label{Tdirac}
T = -{1 \over 2} \psi \partial \psi \hspace{1cm}\mbox{with}\hspace{1cm} \bar \partial T=0 \ .
\end{equation}

Our next problem is to find the Poisson bracket associated to this action. We proceed as
we did for the scalar field and treat $\bar z$ as the ``time" direction. The function $\ell$ is then simply 
\begin{equation}
\ell(z) = {1 \over 4\pi i} \psi(z),
\end{equation}
 and the associated ``symplectic" form becomes
 \begin{equation}
 \sigma(z,z') = {1 \over 2\pi i} \delta(z,z') \ .
 \end{equation}
 The equal time (i.e. equal $\bar z$) anticommutator associated to these fermions is then
 \begin{equation}
 \{\psi(z,\bar z), \psi(z',\bar z)\} = 2\pi i \delta(z,z')
 \end{equation}
 And we observed that it is {\it symmetric} as it should for fermionic fields.

Finally, let us check that $T$ from \eqref{Tdirac} generates the conformal transformations \eqref{dpsi z}. The conserved charge is again given by $Q_\epsilon=\oint \frac{dz}{2\pi i} \epsilon(z) T(z)$ as in \eqref{Q_e}, and the inverse Noether theorem gives,
\begin{eqnarray}
\delta \psi &=& \left\{ \psi(z), Q_\epsilon  \right\} \nonumber\\
&=&\left\{ \psi(z), -\int {dz' \over 2\pi}  {1 \over 2}\epsilon(z') \psi(z') \partial' \psi(z') \right\} \nonumber\\
&=& -{1 \over 2} \int {dz' \over 2\pi} \epsilon(z') \left( \left\{ \psi(z),\psi(z')\right\} \partial' \psi(z') - \psi(z')\left\{ \psi(z),\partial' \psi(z') \right\} \right)\nonumber\\
&=&-  {1 \over 2} \epsilon(z) \partial\psi(z) - {1 \over 2}\partial( \epsilon \psi ) \nonumber\\
&=&   -\epsilon(z) \partial\psi(z) - {1 \over 2}\partial \epsilon \ \psi\ .
\end{eqnarray}
 as expected.

 For the Dirac field, with conformal dimension $1/2$ the correct Laurent expansion is
 \begin{equation}
 \psi(z) = \sum_{n=-\infty}^\infty {\psi_n \over z^{n+1/2}}\ ,
 \end{equation}
 and its inverse is
 \begin{equation}
 \psi_n = \oint {dz \over 2\pi i}\, z^{n-1/2}\, \psi(z)\ .
 \end{equation}
 Let us compute now the basic anticommutator of the charges $\psi_n$,
 \begin{eqnarray}
   \{\psi_n,\psi_m\} &=& \oint {dz \over 2\pi i}\, z^{n-1/2} \oint {dw \over 2\pi i}\, w^{m-1/2} \{\psi(z),\psi(w)\}  \nonumber\\
   &=&  \oint {dz \over 2\pi i}\, z^{n-1/2} \oint dw \, w^{m-1/2} \delta(z,w)  \nonumber\\
   &=&  \oint {dz \over 2\pi i}\, z^{n-1/2 + m - 1/2} \nonumber\\
 &=& \delta_{n+m,0}.
 \end{eqnarray}
 Note that this is symmetric, as expected. As an exercise, the reader can proceed forward and compute the Virasoro algebra for the fermionic fields.


\subsubsection{The bosonic and fermionic $bc$ systems}

Our last example of a CFT in two dimensions is the $bc$ \textit{system}. These fields appear in the context of bosonic string theory, in particular when the string is quantised via the path integral method. The $b$ and $c$ fields are the ghosts associated to the gauge fixing when we apply the Faddeev-Popov procedure. One of the peculiarities of this system is that the fields $b,c$ could be either bosonic or fermionic. As before, the full $bc$ action contains two terms conjugate to one another, so we will focus only on one pice:
\begin{equation}\label{Ibc}
I = {1 \over 2\pi i  }\int d^2z\ b\bar\partial c\ .
\end{equation}
It is direct to prove that this action is conformally invariant provided $b$ has conformal
weights\footnote{See \eqref{weight}} $(\lambda,0)$, and $c$ has $(1-\lambda,0)$, which give rise to the respective infinitesimal transformation for holomorphic $\epsilon(z)$,
\begin{align}\label{}
b'(z')&=\left( \frac{dz}{dz'} \right)^\lambda b(z) \hspace{1cm}\Rightarrow \hspace{1cm} \delta b(z)=-\epsilon(z) \partial b - \lambda (\partial \epsilon) b \label{db} \\
c'(z')&=\left( \frac{dz}{dz'} \right)^{1-\lambda} c(z) \hspace{.7cm}\Rightarrow \hspace{.7cm} \delta c(z)=-\epsilon(z) \partial c - (1-\lambda) (\partial \epsilon) c \label{dc}
\end{align}
What is the Noether current associated to this conformal symmetry? First, we must prove infinitesimally that \eqref{db}-\eqref{dc} are indeed a symmetry, and find the boundary term:
\begin{align}\label{}
\delta  I&=\frac{1}{2\pi i} \int d^2z \left( \delta b \bar \partial c + b\bar \partial \delta c \right) \nonumber   \\
&=-\frac{1}{2\pi i} \int d^2z \left( \left( \epsilon \partial b+\lambda \partial\epsilon\, b \right) \bar \partial c + b\bar \partial \left( \epsilon \partial c+(1-\lambda) \partial\epsilon\, c \right) \right) \nonumber   \\
&=-\frac{1}{2\pi i} \int d^2z \left( \left( \epsilon \partial b+b \partial\epsilon \right) \bar \partial c + b\epsilon \partial \bar \partial c \right)\nonumber \\
&=-\frac{1}{2\pi i} \int d^2z\, \partial \left( \epsilon b \bar \partial c \right) \ ,
\end{align}
which is indeed a boundary term, but moreover its contribution to the current is proportional to $\bar \partial c$ which vanishes on shell. Consider next the generic variation of \eqref{Ibc} for arbitrary $\delta b,\delta c$:
\begin{align}\label{}
\delta I&=\frac{1}{2\pi i} \int d^2z \left( \delta b \bar \partial c + b\bar \partial \delta c \right) \nonumber   \\
&=\frac{1}{2\pi i} \int d^2z \left( \delta b \bar \partial c\ - \bar \partial b\, \delta c + \bar \partial \left( b\delta c \right)  \right) \nonumber \\
&=\frac{1}{2\pi i} \int d^2z \left( \mbox{eom} + \bar \partial \left( b\delta c \right)  \right) \ ,
\end{align}
and evaluating along the e.o.m. and using the particular symmetry \eqref{db}-\eqref{dc} we get the on shell variation,
\begin{align}\label{}
\delta I &=-\frac{1}{2\pi i} \int d^2z\,  \bar \partial \left( b\epsilon \partial c+(1-\lambda) b \partial\epsilon\, c   \right) \nonumber \\
&=-\frac{1}{2\pi i} \int d^2z\,  \bar \partial \big( b\epsilon \partial c+(1-\lambda)\left(  \partial \left( b\epsilon c \right) - \partial b\, \epsilon c - b \epsilon \partial c \big)   \right) \nonumber \\
&=-\frac{1}{2\pi i} \int d^2z\, \big( \epsilon \bar \partial \left( \lambda \partial(bc)-\partial bc \right) + (1-\lambda) \partial \bar \partial(b\epsilon c) \big) \nonumber \ ,
\end{align}
where the last boundary term vanishes on-shell. The Noether current is therefore
\begin{equation}\label{Tbc}
T = \partial b\, c - \lambda \partial(bc).
\end{equation}

Finally, let's quickly check that \eqref{Tbc} generates the correct transformation of the fields. For simplicity, let's assume the fields are bosonic. The Poisson bracket for \eqref{Ibc} is $[c(z),b(z')]=2\pi i\delta(z-z')$, and the conserved charge is $Q_\epsilon=\oint \frac{dz}{2\pi i}\epsilon(z) T(z)$, so
\begin{align*}\label{}
\delta b & = \left[ b(z),Q_\epsilon \right]\\
&=\oint \frac{dz'}{2\pi i} \epsilon(z') \left[ b(z) , (1-\lambda) \partial' b(z')\ c(z')-\lambda b(z') \partial' c(z')  \right]\\
&=\oint {dz'} \epsilon(z')  \left( -(1-\lambda) \partial' b(z')  \delta(z-z') + \lambda b(z') \partial ' \delta(z-z')  \right) \\
&=-\partial b \epsilon -\lambda \partial \epsilon \ b \ ,
\end{align*}
and similarly for $c$. One can then show that $T$ does satisfy the Virasoro algebra.




\newpage
\chapter{Gauge theories in Hamiltonian form, through examples} \label{Hamgauge}

\section{Introduction}

Gauge theories play a central role in the development of fundamental theories of physical laws. All three fundamental interactions are described by Lagrangian possessing a gauge invariance: the electroweak theory, strong interactions and the theory of gravity. 

The most well-known  gauge theory is Maxwell's action
 \begin{align}\label{I[A]Maxwell}
I[A]= -\frac{1}{16\pi}\int d^4x\ F^{\mu\nu}F_{\mu\nu} \ ,
\end{align}
having the gauge symmetry
\begin{equation}\label{da0}
\delta A_\mu = \partial_\mu \Lambda(x) \ .
\end{equation} 

We shall study this theory in detail below. As often happens, though, there are easier ways to understand the basics of gauge theories resorting to simple problems in mechanics.  

Gauge theories have three important features. First of all the gauge symmetry itself, that is, a transformation containing \textit{arbitrary} functions of (space-)time. It follows immediately from the gauge symmetry that that there must be relations amongst the various equations of motion of the system, and thus not all degrees of freedom in the action will be determined by the e.o.m. Finally, in a Hamiltonian description, a gauge theory has \textit{constraints} which act as the generators of the corresponding gauge symmetries. 

\begin{figure}[ht]
\begin{center}
\includegraphics[scale=0.5]{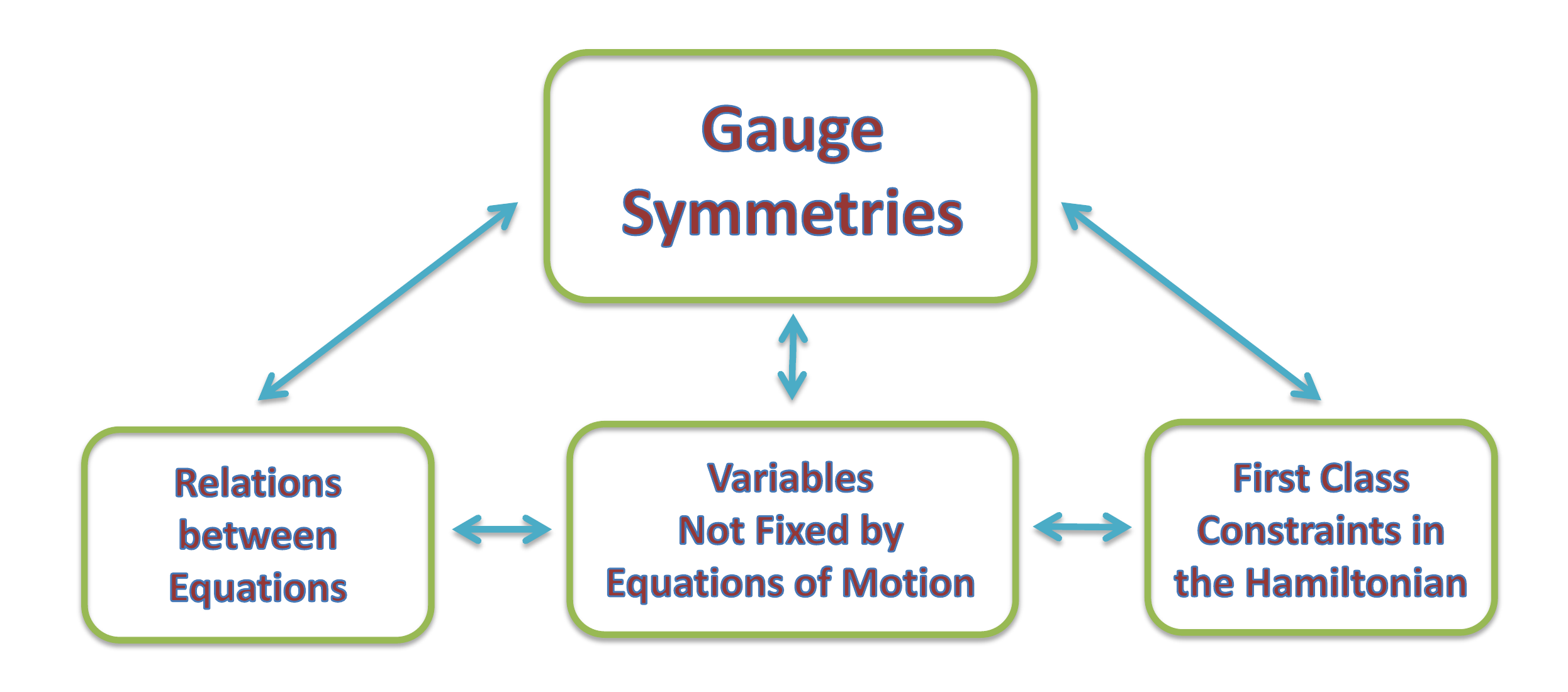} 
\caption{Properties of equations of  motion of gauge theories}
\label{triangle}
\end{center}
\end{figure} 
A theory whose equations of motion do not fully determine the evolution of its variables may seem pathological and unphysical at first sight. The great discovery of gauge symmetries uncovered the concept of equivalent classes of configurations, that is, fields that may differ in their mathematical presentation but represent the same physical reality: if $A_\mu$ is a given solution to Maxwell's equations, $A_\mu + \partial_\mu\Lambda$ represents the same physical reality, for any $\Lambda(x)$. 

\section{A quick tour to the classical aspects of gauge theories }

Before studying gauge theories in general we shall illustrate the salient points with a ``mechanical" model (motivated by electrodynamics). Consider the following action
\begin{align}\label{Laggauge}
\boxed{I[A_0(t),\psi(t)]= {1 \over 2}\int dt\,(\dot{\psi}-A_0)^2}
\end{align}
This simple mechanical model exhibits all classical features of a gauge theory: 
\begin{enumerate}
\item \textbf{Gauge symmetry:} the action \eqref{Laggauge} is invariant under 
\begin{align}\label{}
\psi\rightarrow \psi+\epsilon(t)\hspace{1cm}\mbox{and}\hspace{1cm} A_0\rightarrow A_0+\dot{\epsilon}(t) \ ,
\end{align}
where $\epsilon(t)$ is an arbitrary function of time. Because $\epsilon(t)$ is arbitrary, we call it a gauge symmetry, to distinguish it from global or Noether symmetries discussed in the previous chapter. 

The existence of this gauge symmetry implies the following other properties: 

\item \textbf{ The equations of motion are not independent.} For \eqref{Laggauge} we have
\begin{align*}\label{}
\frac{\delta I}{\delta A_0}=0\Rightarrow\ \ \ \ \ \ (\dot{\psi}-A_0)&=0\\
\frac{\delta I}{\delta \psi}=0\Rightarrow\ \ \ \frac{d}{dt}(\dot{\psi}-A_0)&=0,
\end{align*}
thus, the equation for the field $A_0$ already contains the equation for $\psi$. Since the equations are not all independent (there is only one equation, not two), the solution contains arbitrary functions.

\item \textbf{The general solution contains arbitrary functions}. The solution to the equations of motion is
\begin{equation}\label{gensol}
\psi(t) = f(t), \ \ \ \ \ \    A_0(t) = \dot f(t) \ ,
\end{equation} 
and contains an arbitrary function of time. No matter what the initial conditions are, one can always modify the evolution at later times. 

\item \textbf{The Hamiltonian possesses constraints.} To pass to the Hamiltonian one should define momenta for all variables. However, for this action this is not necessary since $A_0$ enters with no derivatives\footnote{A variable having no time derivatives is not enough, in general, to claim it will not have a relevant momentum. We shall be more specific about this point below based on examples. In the example described here it is the case. The general formalism to deal with Hamiltonian version of gauge theories is Dirac's constraint analysis. When possible we shall avoid the general analysis and derive the right results in a more economical way. It is left as an exercise to prove that defining momenta for both variables yields no new information, but makes the analysis longer.}. We define
\begin{align}\label{}
p_\psi=\frac{\partial L}{\partial\dot{\psi}}=\dot{\psi}-A_0 \ ,
\end{align}
and the Hamiltonian is
\begin{align}\label{}
H(p_\psi,\psi,A_0)&=p_\psi \dot{\psi}-L\\
&=
\frac{1}{2}p_\psi^2+A_0p_\psi \ .
\end{align}
The corresponding Hamiltonian action is
\begin{align}\label{Igenericgauge}
\boxed{I[p_\psi,\psi,A_0]=
\int \left( p_\psi\dot{\psi}-\frac{1}{2}p_\psi^2-A_0 p_\psi \right)dt}
\end{align}
The equations that follow from this action varying $A_0,\psi$ and $p_\phi$ are totally equivalent to those following from (\ref{Laggauge}). We see that $A_0$ appears as a \textit{Lagrange multiplier},
\begin{align}\label{di/da}
\frac{\delta I}{\delta A_0}=p_\psi=0 \ ,
\end{align}
and we find our first example of a \textit{constraint}, an equation involving no time derivatives. The two remaining e.o.m. are
\begin{eqnarray}\label{}
\frac{\delta I}{\delta p}&=&\dot{\psi}-p_\psi-A_0=0\\
\frac{\delta I}{\delta \psi}&=&-\dot{p}_\psi=0\label{di/dphi} \ ,
\end{eqnarray}
where again one equation \eqref{di/dphi} is already contained in another one \eqref{di/da}, so clearly from the Hamiltonian point of view we again have less equations than unknowns.

\end{enumerate}

We shall end the description of this model by briefly mentioning the issue of gauge fixing. A gauge theory has more undetermined functions than equations. To make sense of the evolution one could either look for gauge invariant functions, like electric and magnetic fields in electromagnetism, or fix the gauge by imposing one condition per gauge freedom. Fixing the gauge is not always the best procedure because often is breaks locality. The powerful BRTS methods has been developed to this end.   

Here we shall only make some remarks on how one would fix the gauge in this model, stressing a subtle point. In this example we have one gauge symmetry so we can impose one condition.  Is it the same to impose $A_0=0$ or $\psi=0$ as gauge condition?  It is not. The reason is that, as we saw before, the equation of motion for $\psi$ is contained in the equation for $A_0$, but not the other way around. This means that one can dispose of $\psi(t)$ and no information will be lost because its equation is already there. But one cannot dispose of $A_0$. To see this more explicitly, if we fix the gauge $\psi=0$ in the action \eqref{Laggauge} we have
\begin{align}\label{}
I_{\psi=0}[A(t)]&=\int dt\ A_0^2\ \ \overset{\mbox{\tiny eom}}{\to}\ A_0(t)=0\ ,
\end{align}
which uniquely fixes $A_0(t)$, and shows that the pure-gauge action \eqref{Laggauge} in fact has no degrees of freedom at all.
However, consider now what happens if instead we fix $A_0(t)=0$ (incorrect!) as gauge condition. The action becomes
\begin{align}\label{}
I_{A_0=0}[\psi(t)]&=\int dt\  \dot{\psi}^2\ \ \overset{\mbox{\tiny eom}}{\to}\ \ddot{\psi}=0 \ ,
\end{align}
which represents a free field $\psi$ that does carry one degree of freedom! Setting $A_0=0$ inside the action is incorrect because, as we discussed above, the equations of motion for $A_0$ are not contained in those of $\psi$.

~

The model we have just described captures all the characteristic features of a gauge theory. 
In the following section we shall present the general Hamiltonian structure of gauge theories, and then proceed to show how the most important cases (Maxwell's theory, Yang-Mills theory, and General Relativity) all fall into this classification. We shall also consider 2+1 Chern-Simons theories basically because they are extremely simple and of great pedagogical value. 

For other applications of Noether's second theorem, see \cite{lusanna1991second,lusanna1993shanmugadhasan}.

\section{General structure of gauge theories in Hamiltonian form}\label{genst}

There are many actions possessing a gauge symmetry; from electrodynamics to general relativity, from a relativistic particle to the string worldsheet. One of the nice features of the Hamiltonian formulation is that it provides a unified framework for all of them. In this way, many features can be studied in a general setting valid for all examples. 

In most cases the Hamiltonian is derived from the Lagrangian. In the presence of gauge invariance this process, devised by Dirac long ago, can be long and painful. The final result is however neat and clear. 

In this section we shall present gauge theories in the reverse order. We start with the Hamiltonian action from scratch describing its properties and the role of the different ingredients. Afterwards, we present the classic examples starting as usual from the Lagrangian and going through the Legendre transformation into the Hamiltonian.

All\footnote{Here we mean most well-known and useful examples: Maxwell's theory, Yang-Mills theories, General Relativity,  and many others.}  gauge theory actions, when written in Hamiltonian form, have the following generic form 
\begin{align}\label{IH}
\boxed{I[p_i,q^i,\lambda^a]=\int dt \left( p_i \dot{q}^i - H_0(p_i,q^i)  + \lambda^{a}\phi_a(p,q)  \right)}
\end{align}
This will be our ``paradigm" of a Hamiltonian gauge theory. Here $p_i,q^i$ and $\lambda^a$ are independent fields varied in the action. The total (or ``extended") Hamiltonian $H=H_0-\lambda^a \phi_a$ is decomposed in two terms: $H_0$ denotes that part of the Hamiltonian (if any) that is not a constraint, while $\lambda^a \phi_a$ includes the contributions from the constraints $\phi_a$. Note that this action contains no derivatives of the $\lambda$'s. Varying with respect to $p_i,q^j$ and $\lambda^a$, the equations of motion are 
\begin{eqnarray}
\dot q^{i} &=& {\partial H_0 \over \partial p_i} - \lambda^a {\partial \phi_a \over  \partial p_i},\label{h1}\\
\dot p_{i} &=& -{\partial H_0 \over\partial q^i} + \lambda^a {\partial \phi_a \over \partial q^i },\label{h2}\\
\phi_a(p,q) &=&0. \label{c3}
\end{eqnarray} 
This system of equations is very interesting.  All characteristic properties of a gauge theory are contained in the presence of the Lagrange multipliers $\lambda^a(t)$ and constraints the $\phi_a(p,q)$. 

The first two equations determine the evolution of $p,q$ given initial conditions $p_0,q_0$. One notices, however, that the initial conditions cannot be totally arbitrary because they must satisfy the constraints (\ref{c3}); also to actually integrate (\ref{h1}) and (\ref{h2}) one needs the functions $\lambda^{a}(t)$. Two questions immediately come to one's mind (i) How do we choose the functions $\lambda^{a}(t)$? (ii) Given initial conditions that do satisfy the constraints (\ref{c3}), is it guaranteed that the time evolution dictated by (\ref{h1}) and (\ref{h2}) will preserve the constraint (\ref{c3})?

Both questions can be answered at once by computing the time derivative of the constraints. Using (\ref{h1}) and (\ref{h2}) to express $\dot q^i$ and  $\dot p_i$ in terms of derivatives of $H_0$ and the constraints we find,  
\begin{eqnarray}
{d \over dt} \phi_a(p,q) &=& [\phi_a,H_0]  - [\phi_a,\phi_b]\lambda^b. 
\end{eqnarray} 
Now, the constraints (\ref{c3}) must hold for all times, this means that the evolution will be consistent provided $\dot \phi_a$ vanishes at all times, that is, 
\begin{equation}\label{con1}
[\phi_a,H_0]  - [\phi_a,\phi_b]\lambda^b  \approx 0. 
\end{equation} 
The symbol $\approx 0$ (``weakly zero") means that we do not require $\dot \phi_a$ to be strictly zero, it is enough if it vanishes when $\phi_a$ is zero. Equation (\ref{con1}) is a consistency condition for the time evolution. Calling $C_{ab}=[\phi_a,\phi_b]$, the following situations may occur:  

\begin{enumerate}
\item \textbf{Non-gauge theories:} If the matrix $C_{ab}$ is invertible, then (\ref{con1}) fixes completely the functions $\lambda^a(t)$, 
\begin{equation}\label{l0}
\lambda^a(t) = C^{ab}\,[\phi_b,H_0]  \ .
\end{equation} 
In this case, the role of the functions $\lambda^{a}(t)$ is to preserve (\ref{c3}). The system (\ref{h1})-(\ref{c3}) is then well understood: one gives initial conditions satisfying (\ref{c3}); the functions $\lambda^a$ are determined by (\ref{l0}), ensuring that (\ref{c3}) is satisfied at all subsequent times. In Dirac's language this is a theory with \textit{second class} constraints. The first applications of Lagrange multipliers fall in this class of systems well before Dirac's theory! 

But we are interested in gauge theories.  These arise in the opposite case.

\item \textbf{Gauge theories:} If the matrix $C_{ab}$ is zero, or more generally, if this matrix is weakly zero $C_{ab}\approx 0$, i.e. zero on the surface $\phi_a=0$, then (\ref{con1}) imposes no restrictions on the functions $\lambda^a(t)$ (again on the surface $\phi_a=0$) which therefore remain undetermined by the e.o.m. We see the first signal of a gauge theory: the dynamical evolution contains arbitrary functions of time. Now, 
if $[\phi_a,\phi_b]=0$, then  (\ref{con1}) imply also that $[\phi_a,H_0]=0$. A set of constraints are defined as \textit{first class} if they satisfy 
\begin{align}\label{1cc}
[ \phi_a,H_0 ]=\tensor{C}{_a^b} \phi_b\hspace{1cm},\hspace{1cm} [ \phi_a,\phi_b ] =\tensor{C}{_a_b^c} \phi_c \ ,
\end{align}
which are both weakly zero.

A remarkably general result is now the following: if the Hamiltonian and the constraints satisfy (\ref{1cc}), then the action (\ref{IH}) is invariant under the following transformations, 
\begin{align}
\delta q^i&=[ q^i, \phi_a ] \epsilon^a(t) \label{dqg} \\
\delta p_i&=[ p_i, \phi_a ]\epsilon^a(t) \label{dpg} \\
\delta \lambda^c&=\dot{\epsilon}^c+\epsilon^a(t) \tensor{C}{_a^c} -\lambda^a \epsilon^b(t) \tensor{C}{_a_b^c}  \ ,\label{dlg}
\end{align}
where $\epsilon^a(t)$ are arbitrary functions of time. These transformations are the gauge symmetry of the system. We have thus inverted the logic. All actions of the form (\ref{IH}) where the constraints and Hamiltonian satisfy (\ref{1cc}) have a gauge symmetry. (This is the analogue of Noether's inverse theorem discussed above; for an extension, see \cite{deriglazov2010}). We prove this result below. Before we mention the mixed case:

\item {\bf Mixed case.} Of course one may find mixed cases where some constraints are first class and some other are second class.  See \cite{BanadosGarayHenneaux1,BanadosGarayHenneaux2} for a non-trivial example. One could also find systems where conditions (\ref{con1}) lead to new constraints, called \textit{secondary} constraints. From now on we shall focus on gauge theories, that is, theories for which (\ref{1cc}) holds. 
 
\end{enumerate}

~

\noindent {\bf Proof of gauge invariance:} Assuming (\ref{1cc}) holds we prove invariance of the action (\ref{IH}) under (\ref{dpg})-(\ref{dlg}).   The variations of the canonical variables are
\begin{align}\label{}
\delta q_i=\epsilon^a \frac{\partial \phi_a}{\partial p_i}		\hspace{1cm},\hspace{1cm} 
\delta p_i=-\epsilon^a \frac{\partial \phi_a}{\partial q^i}  \ ,
\end{align}
then ($B$ stands for boundary terms that we drop)
\begin{align}\label{}
\delta I&= \delta \int dt \Big( p_i \dot{q}^i -H_0-\lambda^a\phi_a \Big) \nonumber\\
&= \int dt\Big[ -\epsilon^a \frac{\partial\phi_a}{\partial q^i}\dot{q}^i-\dot{p}^i \epsilon^a \frac{\partial\phi_a}{\partial p_i}+B - \frac{\partial H_0}{\partial q^i}\epsilon^a \frac{\partial \phi_a}{\partial p_i}+\frac{\partial H_0}{\partial p^i}\epsilon^a \frac{\partial \phi_a}{\partial q_i}+ \nonumber\\
&\ \ \ \ \ \ \ \ \ -\delta\lambda^a\phi_a-\lambda^a\left( \frac{\partial \phi_a}{\partial q^j}\epsilon^b \frac{\partial\phi_b}{\partial p_j} -\frac{\partial \phi^a}{\partial p_j}\epsilon^b \frac{\partial\phi_b}{\partial q^j} \right)\Big]\nonumber\\
&=\int dt \left[ -\epsilon^a \frac{d}{dt}\phi_a-\epsilon^a [ H_0,\phi_a ]-\delta\lambda^a\phi_a-\lambda^a \epsilon^b [ \phi_a,\phi_b ] + B \right] \ .
\end{align}
If $\phi_a$ are all first class constraints they satisfy \eqref{1cc}, the variation becomes
\begin{align}\label{}
\delta I&=\int dt \left[ \dot{\epsilon}^a\phi_a+\epsilon^a \tensor{C}{_a^b}\phi_b -\delta\lambda^a\phi_a-\lambda^a \epsilon^b \tensor{C}{_a_b^c}\phi_c \right] + B \nonumber \\
&=\int dt\  \left( \dot{\epsilon}^c+\epsilon^a \tensor{C}{_a^c} -\delta\lambda^c -\lambda^a \epsilon^b \tensor{C}{_a_b^c} \right) \phi_c + B \nonumber\\
&=  B \label{dlagm} \ ,
\end{align}
where we have used (\ref{dlg}) (and have redefined $B$ several times). 

Comments:
\begin{itemize}

\item \underline{No charge}

The gauge symmetry is a symmetry of the action and one could in principle compute its Noether charge. It is left as a exercise to prove that the result is $\phi_a$. That is the charge exists but has a zero value on all solutions. See also \cite{deriglazov2010}.  

\item \underline{Degrees of freedom}: one must note the crucial difference in the number of d.o.f. of a non-gauge theory vs that of a gauge theory:

\begin{enumerate}
\item In a non-gauge theory $I=\int dt\left( p_i \dot{q}^i -H_0 \right)$, with $i=1,\hdots, N$, the full solution of the dynamics
\begin{align}\label{}
\dot{q}^i&=\frac{\partial H_0}{\partial p_i}\hspace{1cm},\hspace{1cm}\dot{p}_i=-\frac{\partial H_0}{\partial q^i} \ ,
\end{align}

requires $2N$ integration constants, so there there are simply ${1 \over 2}\times 2N$ physical d.o.f.

\item In contrast, a gauge theory $I=\int dt\left( p_i \dot{q}^i -H_0-\lambda^a\phi_a \right)$ containing the same $2N$ canonical variables has the additional constraints $\phi_a$ with, say, $a=1,\hdots, g$. The equations are (\ref{h1}),(\ref{h2}),(\ref{c3}). In principle we need $2N$ intergartion constants. However, $\phi_a=0$ subtracts $g$ of them. There are also $g$ gauge symmetries which imply that the $p^i$ and $q_i$ are not by themselves \textit{physically meaningful}, but rather there exists some combinations of them that are. This reduces another $g$ constants as physically meaningful. The total number of d.o.f. is ${1 \over 2}(2N-2g)$.
 \end{enumerate}
\item In gauge theories, the time evolution (\ref{h1}),(\ref{h2}),(\ref{c3}) of the canonical variables is the usual one, $\dot{q}=\partial_p H_0 $ and  $\dot{p}=-\partial_q H_0$, plus an additional term, $\lambda^a \partial_p \phi_a$ and $\lambda^a \partial_q \phi_a$ which is a gauge transformation \eqref{dqg}-\eqref{dpg} with the gauge parameter being the Lagrange multipliers.

\end{itemize}

\section{Executive review of examples}\label{overview}

Many very important physical theories lie in this category. In this section we will consider the examples listed below, for which we exhibit their Hamiltonian action highlighting the constraints in {\color{blue} blue}:
\begin{enumerate}

\item The free spinless relativistic particle - in section \ref{relpar} we will show
\begin{align}\label{}
I[x,\lambda]&=\int d\tau \left[ p_\mu\dot{x}^\mu - \lambda {\color{blue}\left(  p^2 +m^2 \right)} \right]\ ,
\end{align}
which, when written in this gauge invariant form, has $H_0=0$. 

\item Free Electrodynamics -  in section \ref{subEM} we'll  write Maxwell's action as:
\begin{align}\label{Iemblue}
I[A_i,E_i,A_0]&=\int d^4x \left[ E_i \dot{A}^i - \left( \frac{1}{2} E_i^2+\frac{1}{4} F_{ij}^2 \right) +A_0 {\color{blue}\partial_i E^i} \right] \ .
\end{align}

\item Yang-Mills - although we won't review it here, it's the generalisation of \eqref{Iemblue}
\begin{align}\label{}
I[A_i,E_i,A_0]&=\int d^4x \left[ E_a^i \dot{A}^a_i - \left( \frac{1}{2} E^{a2}_i+\frac{1}{4} F^{a2}_{ij} \right) +A^a_0 {\color{blue}D_i E^a_i} \right]\ .
\end{align}

\item General Relativity: in section \ref{subHADM}, we study the Einstein-Hilbert action in its ADM form,
\begin{align}\label{}
I[g_{ij},\Pi^{ij},N,N_i]=\int d^4x\left[ \Pi^{ij}\dot{g}_{ij}-N{\color{blue}\mathcal{H}}-N_i {\color{blue}\mathcal{H}^i} \right]\ ,
\end{align}
which possesses four constraints. Here also, the Hamiltonian is zero (it's a constraint). 

\item Chern-Simons theory: in section \ref{3CS} we'll consider 3d gravity through
\begin{align}\label{}
I[A]=\frac{k}{2\pi} \int d^3x\  \epsilon^{ij} \eta_{ab} \left( - {A}{^a_i} {\dot{A}}{^b_j} + {A}{^a_0} {\color{blue} F^b_{ij}} \right)\ .
\end{align}

\end{enumerate}

For each one of these examples, we'll follow the same systematic procedure: 
\begin{itemize}
\item Introduce the Lagrangian form, 
\item Derive its Hamiltonian action, 
\item Show how the constraints generate the gauge symmetry. 
\end{itemize} 
We emphasise that all gauge theories can be understood within the general structure (\ref{IH}). This provides not only a greater conceptual clarity about the common features of all gauge theories, but also serves when dealing with more complicated systems.

\section{Examples in detail}

In this section we show in detail how to go from the Lagrangian actions into the general form 
 (\ref{IH}) for the examples just displayed. For some recent applications in connection to the Ward identities, see \cite{avery2015noether}.

\subsection{The relativistic point particle}\label{relpar}

The action for a spinless relativistic point particle parametrised as $x^\mu(\tau)$ is
\begin{align}\label{Irel}
I=-m\int ds=-m\int d\tau\sqrt{-\frac{dx^\mu}{d\tau} \frac{dx^\nu}{d\tau} \eta_{\mu\nu}} \ ,
\end{align}
where $\dot{x}^\mu=\frac{dx^\mu}{d\tau}$. Here $\tau$ stands for any parameter describing the curve (not necessarily proper time). As is well known and clear from \eqref{Irel}, the action is invariant under the unphysical reparametrisation: 
\begin{align}\label{repartransf}
\tau'= \tau'(\tau)\hspace{1cm},\hspace{1cm}x'^\mu(\tau')= x^\mu(\tau)\ .
\end{align}
As usual, one is also interested in expressing this symmetry as an infinitesimal transformation of $x^\mu$ (the ``field") rather than of the parameter $\tau$ (the ``coordinate"). Taking $\tau'=\tau+\epsilon(\tau)$, \eqref{repartransf} implies
\begin{align}\label{reparamgauge}
x'^\mu(\tau+\epsilon)&=x^\mu(\tau)\hspace{1cm}\Rightarrow\hspace{1cm} \delta x^\mu(\tau)=-\epsilon(\tau)\dot{x}^\mu \ ,
\end{align}
and one must now show explicitly that this variation is a symmetry of the action \eqref{Irel}:
\begin{align}\label{}
\delta I&=-m \int d\tau \frac{-\dot{x}^\mu \delta \dot{x}_\mu}{(-\dot{x}^2)^{1/2}}\nonumber \\
&=-m\int d\tau \frac{d}{d\tau} \left( \epsilon(\tau) \left( -\dot{x}^2 \right)^{1/2}  \right) \ ,
\end{align}
and thus the boundary term is $K=-m\epsilon(\tau) \left( -\dot{x}^2 \right)^{1/2}=\epsilon(\tau) L$. If one keeps going and tries to find the associated Noether's charge, this will be zero (``gauge" conserved quantities are always zero!). 

What about the Hamiltonian version? In principle, one would wish to define $H$ from this action and leave the Hamiltonian action in our favorite form $I=\int p\dot{x}-H_0+\lambda\phi$ as in \eqref{IH}. But after writing down the canonical momentum,
\begin{align}\label{p_nu}
p_\nu=\frac{m\dot{x}_\nu}{\sqrt{-\dot{x}^\mu \dot{x}_\mu}} \ ,
\end{align}
one is incapable of solving $\dot{x}^\mu$ in terms of $p_\mu$. This is because although \eqref{p_nu} \textit{seems} to represent 4 independent equations, they are only 3, since there is a constraint: contracting \eqref{p_nu} with itself,
\begin{align}\label{}
\phi\equiv  p_\nu p^\nu +m^2=0 \ ,
\end{align}
is satisfied without the e.o.m. This issue is taken as the starting point of Dirac's method of constraints, which we will not cover in these notes. Now, instead of defining the Hamiltonian version of the action \eqref{Irel}, we review the method of Polyakov, which does provide a straightforward Hamiltonian formulation.

\paragraph{The Polyakov action}

There is another way of writing the relativistic action, by introducing an auxiliary variable, the \textit{einbein} $e(\tau)$ which is treated as a dynamical field, and by defining the Polyakov action
\begin{align}\label{IPolyakov}
I_P[x^\mu(\tau),e(\tau)]=\frac{1}{2}\int \left( \frac{1}{e}\dot{x}^\mu\dot{x}_\mu-e\ m^2 \right)d\tau \ .
\end{align}
This action exhibits precisely the property highlighted in \eqref{IH}: the Lagrangian contains no derivatives of the einbein, so we can use its equation of motion
\begin{align}\label{}
\frac{\partial L}{\partial e}=0\hspace{1cm}\Rightarrow\hspace{1cm}e(x^\mu)=\frac{1}{m}\sqrt{-\dot{x}^2}\ ,
\end{align}
and replace it into the Polyakov action \eqref{IPolyakov}, which gives back the relativistic action \eqref{Irel}. Therefore the Polyakov action is classically completely equivalent to the original relativistic action \eqref{Irel}, but with a fundamental advantage: it is quadratic in the field $x$, something which is of high technical importance in order to quantize the theory. Since they are equivalent, the Polyakov action should also be invariant under reparametrization, but now we must determine the transformation law for the additional auxiliary field. In addition to \eqref{repartransf}, the transformation law
\begin{align}\label{}
e'(\tau')=e(\tau)\frac{d\tau}{d\tau'}\ ,
\end{align}
is easily seen to render the action invariant. The infinitesimal version of this as a transformation of the field instead of the coordinates is
\begin{align}\label{}
e'(\tau+\epsilon)=e (1-\dot{\epsilon})\hspace{1cm}\Rightarrow\hspace{1cm} \delta e(\tau)=-\frac{d}{d\tau}(\epsilon(\tau) e) \ ,
\end{align}
so the complete symmetry transformation of the fields under reparametrisation is
\begin{align}\label{}
\delta x^\mu(\tau)=-\epsilon(\tau)\dot{x}^\mu \hspace{.7cm},\hspace{.7cm} \delta e(\tau)=-\frac{d}{d\tau}(\epsilon(\tau) e)\ ,
\end{align}
and this is indeed a symmetry, for the variation of \eqref{IPolyakov} under this transformation is
\begin{align}\label{}
\delta I_P&=\frac{1}{2}\int d\tau \left( \frac{2e\dot{x}^\mu \delta \dot{x}_\mu -\dot{x}^2\delta e }{e^2}-m^2\delta e \right)\\
&=-\frac{1}{2}\int d\tau \frac{d}{d\tau} \left( \epsilon \left(  \frac{ \dot{x}^2}{e}  -m^2 e \right) \right)\ ,
\end{align}
that is, a boundary term $K=-\frac{1}{2}\epsilon(\tau) \left(  \frac{ \dot{x}^2}{e}  -m^2 e \right)=-\epsilon(\tau) L$.

The Hamiltonian for the Polyakov action is easily found to be
\begin{align}\label{}
H(p_\mu,x^\mu,e)=\frac{1}{2}e\left( p_\mu p^\mu +m^2 \right)\ ,
\end{align}
and therefore the Hamiltonian Polyakov action $\int p\dot q-H$ reads
\begin{align}\label{I_P}
\boxed{I_P[p,x,e]=\int d\tau \left[ p_\mu\dot{x}^\mu - \frac{1}{2}e\left( p_\mu p^\mu +m^2 \right) \right]}
\end{align}
This has precisely the form \eqref{IH} for a Hamiltonian action with one gauge symmetry, ``free" Hamiltonian $H_0=0$, the einbein $e$ playing the role of the Lagrange multiplier, and $\phi=\frac{1}{2}(p^2+m^2)$ is the constraint. The fact that $H_0=0$ here is due to the gauge invariance: the Hamiltonian is not conjugate to any physical \textit{time}, because we are using the arbitrary parameter $\tau$ to describe the evolution. 

Finally, in order to close the logical circle, let's check that indeed the constraint $\phi$ generates the gauge transformation \eqref{reparamgauge} through Poisson brackets. This must be so because $H_0=0$ and all the $C_{ab}=[\phi_a,\phi_b]=0$ are zero in this example because there's a single constraint so it's trivially first class; using \eqref{dqg}-\eqref{dlg},
\begin{align*}\label{}
\delta x^\mu &= [ x^\mu, \epsilon(\tau) \frac{1}{2}(p^2+m^2) ]=\epsilon(\tau) p^\mu\\
\delta p_\mu&=[ p_\mu,\epsilon(\tau) \frac{1}{2}(p^2+m^2) ]=0\\
\delta e&=\dot{\epsilon}(\tau)\ ,
\end{align*}
so the variation of the action is indeed a boundary term:
\begin{align}\label{}
\delta I&=\int d\tau \left[ p_\mu \delta \dot{x}^\mu - \frac{1}{2}\delta e \left( p^2+m^2 \right) \right]\nonumber\\
&=\int d\tau \frac{d}{d\tau} \left[ \frac{1}{2} \epsilon(\tau)\left( p^2-m^2 \right) \right] \ .
\end{align}

\subsection{Maxwell theory}\label{subEM}

To define a Hamiltonian action, we start by rewriting the electromagnetic Lagrangian by separating into its space+time components in order to define a canonical structure:
\begin{align}\label{}
I_{EM}&=-\frac{1}{4}\int d^4x\ F^{\mu\nu}F_{\mu\nu} \nonumber \\
&=\int d^4x \left[ -\frac{1}{2}F^{0i}F_{0i}-\frac{1}{4}F^{ij}F_{ij} \right] \nonumber \\
&=\int  d^4x \left[ \frac{1}{2}  (\dot{A}_i-\partial_i A_0)(\dot{A}^i-\partial^i A_0) -\frac{1}{4}F_{ij}F^{ij} \right] \nonumber \\
&=\int d^4x \left( \frac{1}{2} \dot{A}_i\dot{A}^i-\dot{A}_i\partial^i A_0+\frac{1}{2}\partial_iA_0 \partial^i A_0-\frac{1}{4}F_{ij}F^{ij} \right)\ .
\end{align}

As before, we define canonical momenta only for those variables that appear with temporal derivatives in the action; in this case, $A_i$,
\begin{align}\label{p_iMaxwell}
\pi_i\equiv \frac{\partial \mathcal{L}}{\partial \dot{A_i}}=\dot{A}_i
-\partial_i A_0\ \ \Rightarrow\ \ \dot{A}_i=\pi_i+\partial_i A_0\ .
\end{align}
Note from this that $\pi^i$ is the electrical field: $\vec{\pi}=\vec{E}$. The Hamiltonian is 
\begin{align}\label{}	
H(p,A)&=\pi_i\dot{A}^i -L \nonumber \\
&=\pi_i (\pi^i+\partial^i A_0)- \left[\frac{1}{2} (\pi_i+\partial_i A_0) (\pi^i+\partial^i A_0) - (\pi_i+\partial_i A_0) \partial^i A_0+\frac{1}{2}\partial_iA_0 \partial^i A_0-\frac{1}{4}F_{ij}F^{ij}  \right] \nonumber \\
&=\frac{1}{2}\pi_i\pi^i +\frac{1}{4}F_{ij}F^{ij}  -A_0 \partial_i \pi^i \ ,
\end{align}
where we have integrated by parts in the last line and dropped the boundary term (we postpone the important discussion of boundary terms for section \ref{boundary}!). 

Thus, the Hamiltonian action $\int p\dot q-H$ for electromagnetism is
\begin{align}\label{IHEM1}
\boxed{I_{EM}[A_i,\pi_i,A_0]=\int d^4x \left[ \pi_i \dot{A}^i - \left( \frac{1}{2} \pi_i\pi^i+\frac{1}{4} F_{ij}F^{ij} \right) +A_0 \partial_i \pi^i  \right]}
\end{align}
which again has precisely the general form \eqref{IH} $I=\int p\dot{q}-H_0+\lambda^a\phi_a$ , where $H_0$ is the ``dynamical"\ Hamiltonian (the energy) and $A_0$ the Lagrange multiplier
\begin{align}\label{}
H_0&=\frac{1}{2}\pi_i\pi^i +\frac{1}{4}F_{ij}F^{ij}=\frac{1}{2}\left( \vec{E}^2+\vec{B}^2 \right)\hspace{1cm},\hspace{1cm}\lambda=A_0 \ ,
\end{align} 
while the only constraint is Gauss' law:
\begin{align}\label{phiEM}
\phi=\partial_i \pi^i =\nabla \cdot \vec{E}=0 \ .
\end{align}
With regards to Hamilton's equations of motion
\begin{align}\label{}
\dot{\pi}_i&=-\frac{\partial H}{\partial A^i}\hspace{.5cm},\hspace{.5cm} \dot{A}_i=\frac{\partial H}{\partial \pi^i } \hspace{.5cm},\hspace{.5cm} \phi=0\ ,
\end{align}
one is confronted with the same issues discussed after \eqref{c3}, namely that we must guarantee that the e.o.m. will preserve the vanishing of the constraint, provided we choose some initial conditions that do. As we mentioned there, all we need to do is to demand that $\frac{d\phi}{dt}=[\phi,H]=0$. Since we are now in a field theory, we must actually compute $[ \phi(x), H(x') ]$ where $x$ and $x'$ are two distinct space-time points:
\begin{align}\label{}
[\phi(x),H(x')]&=\left[\partial_i \pi^i(x),\frac{1}{2}\left( \pi_i\pi^i(x')+F_{ij}F^{ij}(x')-\frac{1}{2}A_0 \partial_i \pi^i  (x') \right)  \right] \ .
\end{align}
Here we must use the field theory symplectic structure:
\begin{align}\label{}
[\pi_i(x),\pi_j(x')]=0\hspace{.5cm} , \hspace{.5cm}  [A_i(x),A_j(x')]=0 \hspace{.5cm} , \hspace{.5cm}  [A_i(x),\pi_j(x')]=\delta_{ij}\delta^3(x-x') \ ,
\end{align}
which yields
\begin{align}\label{}
\frac{d\phi}{dt}&=\left[\phi(x),H(x')\right] \nonumber \\
&=2\partial_k \partial'_i \left[  \pi_k (x), A_j(x') \right]F^{ij}(x')\nonumber\\
&=-2\left( \partial_k \partial'_i \delta^3(x-x') \right) F^{ik}(x')=0 \ ,
\end{align}
due to symmetry-antisymmetry in $(i,k)$. The constraint $\phi$ is therefore a conserved quantity along the classical evolution of the system, and must generate the gauge symmetry through the field-theoretic version of \eqref{dqg}-\eqref{dlg}, which is as follows. In the Poisson bracket, instead of considering simply the constraint function $\phi(x)=\partial_i \pi^i (x)$ we must take a \textit{functional} $\Phi[\Lambda(x)]$ depending on some arbitrary ``weight" test function $\Lambda(x)$ which will act as the gauge function: 
\begin{align}\label{PhiL}
\Phi[\Lambda(x)]=\int d^3x\ \Lambda(x)\partial_i \pi^i (x) \ ,
\end{align}
where we only integrate in space and not in time because we wish $\Phi$ to remain a function of time, so that we can compute an `equal time' commutation relation with something else. The gauge transformations, analogous to \eqref{dqg}-\eqref{dpg}, are defined as the bracket with the functional $\Phi$:
\begin{align}\label{}
\delta A_i(x)&=\left[A_i(x,t),\Phi[\Lambda]\right]\nonumber\\
&=\int d^3x' \Lambda(x',t) \partial'_j \left[A_i(x,t),\pi^j(x',t)\right]\nonumber\\
&=\int d^3x' \Lambda(x',t) \delta_{ji} \partial'_j \delta^{(3)}(x-x')\nonumber\\
&=-\partial_i \Lambda(x,t) \label{dA_ig} \ ,
\end{align}
which is precisely the form of the spatial gauge transformation of electrodynamics! On the other hand, the transformations of the momenta vanish:
\begin{align}\label{dp_ig}
\delta \pi_i(x,t)=\left[ \pi_i,\Phi[\Lambda] \right]=\int d^3x' \Lambda(x',t)\left[\pi_i,\partial_j \pi^j\right]=0 \ ,
\end{align}
which must be so - the momentum is the electric field, $\pi^i =E^i$, which must better be gauge invariant. Finally, the transformation law for the Lagrange multiplier $A_0$, just as in \eqref{dlagm}, is deduced from the requirement that, under the gauge transformations \eqref{dA_ig} and \eqref{dp_ig}, the variation of the action must be only in a boundary term (i.e. the transformations must be a symmetry)
\begin{align}\label{}
\delta I&=\int d^4x \left[ \pi_i  \delta \dot{A}^i +  \delta A_0\ \phi \right]\nonumber\\
&=\int d^4x \left[ -\pi_i \partial^i \partial_0 \Lambda  + \delta A_0\ \partial_i \pi^i \right]\nonumber\\
&=\int d^4x \left[  \partial_i \left( - \pi^i \partial_0 \Lambda \right)  + \partial_i \pi^i\ \partial_0 \Lambda  + \delta A_0\ \partial_i \pi^i \right] \nonumber \\
&=\int d^4x \left[  \partial_\mu K^\mu  + \partial_i \pi^i \left( \delta A_0 + \partial_0 \Lambda \right)  \right] \ ,\label{dA_0gauge}
\end{align}
with $K^\mu=(0,-\pi^i \partial_0 \Lambda)$, and we have used $\delta \pi_i=0,\delta \phi=0, \delta H_0=0$ since they are all gauge invariants. Thus we need $\delta A_0=-\partial_0 \Lambda$ in order for $\delta I$ to be a boundary term\footnote{Note that although the pre factor of $\delta A_0+\partial_0 \Lambda$ turned out to be $\phi=\partial_i \pi_i$, it's not zero because we cannot use the e.o.m.!}. Thus we conclude that the gauge transformation of the field $A_\mu$ generated by the constraint \eqref{phiEM} is indeed the expected one: 
\begin{align}\label{}
\phi=\partial_i \pi^i =\nabla \cdot \vec{E}=0\hspace{1cm}\mbox{generates} \hspace{1cm} \delta A_\mu=-\partial_\mu \Lambda(x) \ .
\end{align}


\subsection{General Relativity}\label{subHADM}

In this section we will focus on the ADM method. For other interesting applications of Noether's theorem to GR and supergravity see \cite{antoci2009,Solovev:1981,solov1982,avery2015}. 

In the previous sections we have seen that a theory with gauge symmetry actually possesses less degrees of freedom than one could have naively expected. In section \ref{subEM} we learnt that by performing a ``space+time" decomposition of the electromagnetic tensor and going to the Hamiltonian version of the action, we naturally arrived at the generic form for all gauge theories \eqref{IH}. We now follow an exactly analogous route for general relativity. In the case of gravitation, the dynamical field is the metric $g_{\mu\nu}$. Just as in electrodynamics we separated the field into $A_0,A_i$, here we shall work with $g_{00},g_{i0},g_{ij}$. Everything else is completely equivalent. This is called the ADM method \cite{ADM,Baez,Misner,Moore,Wald}. 

\begin{figure}[ht]
\begin{center}
\includegraphics[scale=0.4]{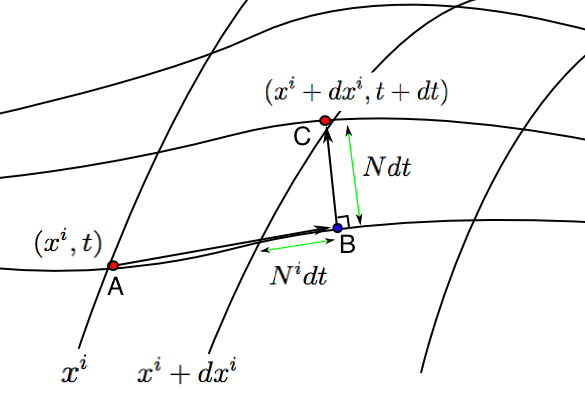} 
\label{fig:vecNADM}
\end{center}
\end{figure}

In general relativity, we work with a smooth but non trivial manifold $\mathcal{M}$, which we think as being composed by the set of $3-$dimensional surfaces $\Sigma^3_t$ one for each constant time $t$ (technically called a ``foliation"). For each time $t$, we call $g_{ij}(\vec{x},t)$ the \textit{intrinsic} metric on $\Sigma^3_t$. 
In order to relate foliations at infinitesimally close times, we define $N(\vec{x},t)$, the ``lapse" function, such that starting from $\vec{x}$ at $t$, if we advance a distance $N(\vec{x},t)dt$ in the (hyper)direction normal to $\Sigma^3_t$ at $\vec{x}$, we would reach exactly the surface $\Sigma^3_{t+dt}$. We must also define $N^i(\vec{x},t)$ such that $N^i(\vec{x},t)dt$ measures the ``shift"\ produced, at constant time, between $\vec{x}+d\vec{x}$ and the (blue) point that will eventually hit $(\vec{x}+d\vec{x},t+dt)$ by projecting with $Ndt$ (see Fig. \ref{fig:vecNADM}).  
The relation between the metric components and the lapse and shift functions $N$ and $N^i$ is obtained by simply writing the space-time interval between $A$ and $C$ in both forms, and a short calculation shows \cite{ADM}
\begin{align}\label{}
g_{00}=-N^2+N_i N^i\hspace{1cm},\hspace{1cm}g_{0i}=N_i \ ,
\end{align}
where spatial indices are raised/lowered using the spatial metric $g^{ij}$ and its inverse $g_{ij}$. 
By exchanging $g_{00},g_{0i}$ in favour of $N$ and $N^i$, ADM \cite{ADM} showed (this is the hard part of the calculation) that the Einstein-Hilbert action becomes, 
\begin{align}\label{IADM}
I_{EH}[N,N^i,g_{ij}]=\int d^4x\  N\sqrt{-^{(3)}g} \left( ^{(3)}R-K^2+K^{ij}K_{ij} \right)+B,
\end{align}
where $B$ is some boundary term that we forget for now, 
\begin{align}\label{}
K_{ij}&=\frac{1}{2N} \left[ -\dot{g}_{ij}+N_{i/j}+N_{j/i} \right]\hspace{1cm}\mbox{``extrinsic curvature"\ } \label{K_ijdot} \\
K&=g^{ij}K_{ij}\ ,
\end{align}
and $N_{i/j}$ is the covariant derivative of $N_i$ with respect to $x^j$ along the surface $\Sigma^3_t$. $^{(3)}g$ y $^{(3)}R$ indicate the determinant and Ricci scalar with respect to the $3-$dimensional metric $g_{ij}$. Thus $^{(3)}R$ contains no time derivatives; the only term with time derivatives is the $\dot{g}_{ij}$ contained in $K_{ij}$. Therefore, as we shall explain in brief, $g_{ij}$ is the true dynamical field, while $N$ and $N^i$ aren't; they will appear as Lagrange multipliers in the Hamiltonian approach. So, while in electrodynamics we saw that by making the space + time decomposition, $A_0$ was the Lagrange multiplier not fixed by the e.o.m., in the case of pure gravity $g_{00}$ and $g_{0i}$ are not fixed, and their combinations $N$ and $N^i$ will play the role of the Lagrange multipliers. 

To find the Hamiltonian form of the Einstein-Hilbert action, we start by defining momentum to the only field appearing with time derivatives in \eqref{IADM}:
\begin{align}\label{}
\Pi^{ij}= \frac{\partial \mathcal{L}}{\partial \dot{g}_{ij}} \ ,
\end{align}
which comes from the variation of the action due to variations in $\dot g_{ij}$:
\begin{align}\label{}
\delta I&=\int d^3x \sqrt{|g|} \left[ -2K\delta K+2K^{ij}\delta K_{ij} \right]\nonumber\\
&=\int d^3x \sqrt{|g|}\left[ Kg^{ij}-K^{ij} \right]\delta \dot{g}_{ij} \ ,
\end{align}
thus
\begin{align}\label{pi_ij}
\Pi_{ij}=\sqrt{|g|}\left[ Kg^{ij}-K^{ij} \right] \ .
\end{align}
In order to write down the Hamiltonian, we must invert this relation and solve for $K_{ij}$ (which contains $\dot{g}_{ij}$) in terms of $\Pi^{ij}$ and $g_{ij}$; taking the trace of \eqref{pi_ij}:
\begin{align}\label{}
K=\frac{\Pi}{2\sqrt{|g|}}\hspace{1cm}\mbox{where}\hspace{.5cm} \Pi\equiv \Pi^{ij}g_{ij} \ ,
\end{align}
so we can now clear $K_{ij}$ as:
\begin{align}\label{}
K^{ij}
&=\frac{1}{\sqrt{|g|}}\left( -\Pi^{ij}+\frac{\Pi}{2}g^{ij} \right)\label{K^ijHam} \ .
\end{align}
Finally, using \eqref{K_ijdot} and \eqref{K^ijHam} we solve $\dot{g}_{ij}$, and the Hamiltonian is:
\begin{align}\label{}
H&=\int d^3 x \left( \Pi^{ij}\dot{g}_{ij}-\mathcal{L} \right) \nonumber \\
&=\int d^3 x\left[ \Pi^{ij} \left( -\frac{2N}{\sqrt{|g|}} \left( \frac{\Pi}{2}g_{ij}-\Pi_{ij} \right) + N_{i/j}+N_{j/i} \right) -N\sqrt{|g|}\left( R-K^2+K^{ij}K_{ij} \right) \right] \nonumber\\
&=\int d^3x \left[ -\frac{2N}{\sqrt{|g|}}\left( \frac{\Pi^2}{2}-\Pi^{ij}\Pi_{ij} \right)+2\Pi^{ij}N_{i/j}-N\sqrt{|g|}R+N\frac{\Pi^2\sqrt{|g|}}{4|g|}+ \right. \\
&\left.-N\frac{\sqrt{|g|}}{|g|}\left( \frac{3\Pi^2}{4}-2\frac{\Pi}{2}\Pi+\Pi^{ij}\Pi_{ij} \right)  \right] \nonumber\\
&=\int d^3x \left[ N \left[ \frac{\Pi^{ij}\Pi_{ij}}{\sqrt{|g|}}-\frac{1}{2}\frac{\Pi^2}{\sqrt{|g|}}-\sqrt{|g|}R \right]-2\Pi^{ij}_{/j} N_i \right]+B\ ,
\end{align}
where again $\Pi^{ij}_{/j}$ stands for the covariant derivative along $x^j$ on $\Sigma^{3}_t$.
We thus now see the structure of the Hamiltonian and it's respective Hamiltonian densities:
\begin{align}\label{Hadm}
H= \int d^3x \left( N\mathcal{H}+N_i \mathcal{H}^i \right)\ ,
\end{align}
with
\begin{align}\label{HHi}
\mathcal{H}=\frac{1}{\sqrt{|g|}}\left( \Pi^{ij}\Pi_{ij}-\frac{\Pi^2}{2}   \right)-\sqrt{|g|}R\hspace{1cm},\hspace{1cm} \mathcal{H}^i=-2\Pi^{ij}_{/j}\ .
\end{align}
Finally, the Hamiltonian version of the Einstein-Hilbert action has precisely the expected form \eqref{IH} $I=\int p\dot{q}-H_0+\lambda^a \phi_a$
\begin{align}\label{IEH}
\boxed{I_{ADM}[g_{ij},\Pi^{ij},N,N_i]=\int d^4x \left[ \Pi^{ij}\dot{g}_{ij}-N\mathcal{H}-N_i \mathcal{H}^i \right]}
\end{align}
whose Hamilton's equations of motion are:
\begin{align}\label{}
\frac{\delta I}{\delta N}&=\mathcal{H}=0\hspace{1cm},\hspace{1cm}\dot{g}_{ij}=\frac{\delta H}{\delta \Pi^{ij}}\\
\frac{\delta I}{\delta N_i}&=\mathcal{H}^i=0\hspace{1cm},\hspace{1cm}\dot{\Pi}^{ij}=-\frac{\delta H}{\delta g_{ij}} \ ,
\end{align}
while the basic Poisson bracket for \eqref{IEH} is
\begin{align}\label{gpi}
\left[ g_{ij}(x),\Pi^{k\ell}(x') \right]=\frac{1}{2}\left( \delta^k_i\delta^\ell_j+\delta^\ell_i \delta^k_j \right)\delta^{3}(x-x') \ ,
\end{align}
due to the symmetry of both tensors. 

Let's review some of the main properties of this decomposition:
\begin{itemize}
\item There exist 4 constraints ($\mathcal{H}=0$ and $\mathcal{H}^i=0$) together with 4 Lagrange multipliers ($N$ and $N_i$).

\item The Hamiltonian \eqref{Hadm}, whose general form is $H=H_0+\lambda^a \phi_a$ is a pure constraint with $H_0=0$: the ``free"\ Hamiltonian is zero. This is generic for theories which are invariant under generalised coordinate transformations. This is of course what happened in the relativistic particle in \eqref{I_P}, because that action was invariant under one-dimensional diffeomorphisms (reparametrizations). 

\item The constraints are first class\footnote{They are indeed of first class since their Poisson brackets are proportional to themselves \eqref{1cc}, so they are a \textit{closed} subset under the bracket operation. However, their structure constants are not trivial. }. Indeed, they satisfy the Dirac algebra,
\begin{align*}\label{}
[\mathcal{H}(x),\mathcal{H}(y)]&=g^{ij}\mathcal{H}_j(x)\frac{\partial}{\partial x^i}\delta(x,y)-g^{ij}\mathcal{H}_j(y)\frac{\partial}{\partial y^i}\delta(x,y)\\
[\mathcal{H}_i(x),\mathcal{H}(y)]&=\mathcal{H}(y)\frac{\partial }{\partial x^i}\delta(x,y)\\
[\mathcal{H}_i(x),\mathcal{H}_j(y)]&=\mathcal{H}_j(x)\frac{\partial}{\partial x^i}\delta(x,y)-\mathcal{H}_i(y)\frac{\partial}{\partial y^j}\delta(x,y)\ .
\end{align*}

\item Degrees of freedom: in principle we would have $6+6$ integration constants coming from $\Pi^{ij}$ and $g_{ij}$ (they are symmetric $3\times 3$ matrices), but then we also have 4 constraints, and other 4 gauge symmetries implying that not all fields are independently physically measurable, which leaves us with $12-4-4=4$ d.o.f., the same that in the metric formulation. 

\item The gauge symmetries generated by the constraints via Poisson brackets correspond to \textbf{diffeomorphisms}, as it should be. For the $\mathcal{H}^i$ constraint we have (recall, as in \eqref{PhiL}, that in field theory we define the brackets via an arbitrary ``test function"\ $\xi_i$):
\begin{align}
\Phi_{\mathcal{H}^i}[\xi^i]&=\int d^3x\ \xi_i(x)\mathcal{H}^i \nonumber\\
&=-2\int d^3x\ \xi_i \Pi^{ij}_{/j}\nonumber\\
&=2\int d^3x\ \xi_{i/j}\Pi^{ij} \label{Phixi}\ ,
\end{align}
where we used \eqref{HHi}, and the integration by parts holds only if $\xi_i$ dies off fast enough in order for the boundary terms to vanish, which we assume. Hence the transformation generated by the Hamiltonian is precisely a diffeomorphism: using \eqref{Phixi} and \eqref{gpi},
\begin{align}\label{}
\delta g_{ij}(x)&= \left[ g_{ij}(x), \Phi_{\mathcal{H}^i}[\xi] \right] \nonumber  \\
&= 2\int d^3x'\ \xi_{k/\ell}(x') \left[  g_{ij}(x), \Pi^{k\ell}(x')     \right] \nonumber \\
&=\xi_{i/j}+\xi_{j/i}=\mathcal{L}_\xi g_{ij}\ ,
\end{align}
which corresponds exactly to the Lie derivative \eqref{Liegmunu}, evaluated along the $3-$dimensional surface $\Sigma^3_t$. Finally, one can proceed by computing the transformation $\delta \Pi^{ij},\delta N,\delta N^i$ in a similar way. 

\end{itemize}

\subsection{ \texorpdfstring{{$2+1$}}{Lg} Chern-Simons theory} \label{3CS}

As first noted by Ach\'ucarro and Townsend \cite{AT86} and subsequently developed by Witten \cite{Witten88}, General Relativity in $2+1$ dimensions with its usual metric representation can be reformulated as a Chern-Simons (CS) gauge theory, in very much the same way as an ordinary Yang-Mills theory, with a gauge group of $SL(2,\mathbb R)\times SL(2,\mathbb R)$ (when $\Lambda$ is negative). We will not review this connection here. Instead, we will go straight to the CS action to find its canonical structure, constraints and Hamiltonian representation of gauge symmetries.  

We start with the CS action, which is given by in component notation as
\begin{align}\label{Ics}
I[A]&=\frac{k}{4\pi} \int d^3x\ \epsilon^{\mu\nu\lambda}\ \mbox{Tr}\left( A_\mu \partial_\nu A_\lambda +\frac{2}{3} A_\mu A_\nu A_\lambda  \right)\ ,
\end{align}
where $\epsilon^{\mu\nu\rho}$ is the totally antisymmetric tensor. The connection $A_\mu=\tensor{A}{^a_\mu}(x)J_a$ takes values in some given algebra: $\tensor{A}{^a_\mu}(x)$ are numbers, while $J_a$ are some choice of generators in the algebra.

Performing a 2+1 splitting $A_\mu = (A_0,A_i)$ the Chern-Simons action acquires a  very simple form\footnote{Details:
\begin{align*}\label{}
CS[A]&= \epsilon^{\mu\nu\rho}\ \mbox{Tr}\left( A_\mu \partial_\nu A_\rho + \frac{2}{3} A_\mu A_\nu A_\rho \right)\\
&=\ \mbox{Tr}\left[  \epsilon^{tij}\left( A_0 \partial_i A_j + \frac{2}{3} A_0 A_i A_j \right) +  \epsilon^{itj}\left( A_i \partial_t A_j + \frac{2}{3} A_i A_0 A_j \right)  +  \epsilon^{ijt}\left( A_i \partial_j A_0 + \frac{2}{3} A_i A_j A_0 \right)  \right]\\
&=\ \epsilon^{ij}\mbox{Tr}\left[ - A_i \partial_t A_j +  A_0 \partial_i A_j + A_i \partial_j A_0+ \frac{2}{3} A_0 A_i A_j   - \frac{2}{3} A_i A_0 A_j     + \frac{2}{3} A_i A_j A_0   \right]\ ,
\end{align*}
where $\epsilon^{tij}=\epsilon^{ij}$. Next integrate one term by parts, use the cyclic property of the trace, and $\epsilon^{ij} A_j A_i=-\epsilon^{ij} A_iA_j$ to get 
\begin{align}\label{}
&=\ \epsilon^{ij}\mbox{Tr}\left[ - A_i \partial_t A_j +  A_0 \partial_i A_j - A_0 \partial_j A_i + \partial_j \left( A_i A_0 \right)+ \frac{2}{3} \left( 2A_0 A_i A_j   - A_0 A_j A_i     \right) \right]\\
&=\ \epsilon^{ij}\mbox{Tr}\left[ - A_i \partial_t A_j +  A_0 (\partial_i A_j - \partial_j A_i) + \partial_j \left( A_i A_0 \right)+  2A_0 A_i A_j   \right]\ ,
\end{align}
and finally, use $\epsilon^{ij}A_iA_j=\frac{1}{2}\epsilon^{ij}[A_i,A_j]$ so 
\begin{align}\label{}
&=\ \epsilon^{ij}\mbox{Tr}\left[ - A_i \partial_t A_j +  A_0 (\partial_i A_j - \partial_j A_i +  [A_i, A_j]) + \partial_j \left( A_i A_0 \right)  \right]\\
&=\frac{1}{2} \epsilon^{ij}\eta_{ab}\left[ - \tensor{A}{^a}_i \partial_t \tensor{A}{^b_j} +  \tensor{A}{^a_0} \tensor{F}{^b_{ij}} + \partial_j \left( \tensor{A}{^a_i} \tensor{A}{^b_0} \right)   \right]\ ,
\end{align}
where we have defined $\mbox{Tr}\left( J_aJ_b \right)=\frac{1}{2}\eta_{ab}$.},  
\begin{align}\label{IcsA}
I_{CS}[A]=\frac{k}{8\pi}\int d^3x\  \epsilon^{ij}\eta_{ab}\left[ - \tensor{A}{^a}_i \tensor{\dot A}{^b_j} +  \tensor{A}{^a_0} \tensor{F}{^b_{ij}} \right]\ ,
\end{align}
up to a boundary term. This action has the general structure (\ref{IH}): $\lambda^a=A^a_0$ are the Lagrange multipliers, $H_0=0$, and $\tensor{\phi}{^b_{ij}}=\tensor{F}{^b_{ij}}=0$ are the constraints.  The basic Poisson bracket is
\begin{align}\label{[A,A]}
\left[  \tensor{A}{^a_i}(x),\tensor{A}{^b_j}(x')  \right]&=\frac{4\pi}{k} \epsilon_{ij}\eta^{ab}\delta(x-x')
\end{align}
while the equations of motion read 
\begin{align}
\frac{\delta I}{\delta \tensor{A}{^a_0}}&=0\hspace{1cm}\Rightarrow \hspace{1cm}\tensor{F}{^a_{ij}}=0 \label{cseom1}\\
\frac{\delta I}{\delta \tensor{A}{^a_i} }&=0 \hspace{1cm}\Rightarrow \hspace{1cm} \tensor{\dot{A}}{^a_i}=D_i \tensor{A}{^a_0}\label{cseom2}\ ,
\end{align}
where $D_i \tensor{A}{^a_0}=\partial_i \tensor{A}{^a_0}+\tensor{\epsilon}{^a_b_c}\tensor{A}{^b_i}\tensor{A}{^c_0}$. 

It is left as a exercise to prove that the constraints $\int d^2x\, \epsilon^{ij} \xi_a(x) \tensor{F}{^a_i_j}$ generate the correct gauge symmetry transformations, and that they form a close algebra under the Poisson bracket (\ref{[A,A]}). We shall do this in detail in section \ref{bcCS} incorporating all boundary terms.

\chapter{Asymptotic boundary conditions and boundary terms}\label{boundary}

\section{Introduction and summary}

A great achievement of theoretical physics is the ``principle of least action" of classical mechanics.  A huge amount of phenomena are described by just one statement: the action must be stationary under arbitrary variations of the dynamical variables. Mathematically,
\begin{equation}
\delta I[q(t)] =0. 
\end{equation} 
This equation also tells us that initial and final condition must be held fixed. This leaves the road ready for a path integral formulation of quantum mechanics as the path integral of $e^{{i \over \hbar}S[q]} $ over all paths consistent with given initial and final conditions. 

Now, Nature is described by fields, and this elegant and powerful formulation of classical and quantum mechanics based on the action needs to be supplemented with a careful treatment of boundary conditions at infinity. The issue of boundary conditions is particularly important and interesting in the case of gauge theories where the assumption `all fields decay sufficiently rapidly at infinity' is not justified. The essence of this is captured by a quote by Fock, recently revided in \cite{vald}, when speaking about General Relativity: \textit{``The field equations and the boundary conditions are inextricably connected and the latter can in no way be considered less important than the former"}\cite{Fock}.

\begin{figure}[h]
\begin{center}
\includegraphics[scale=0.5]{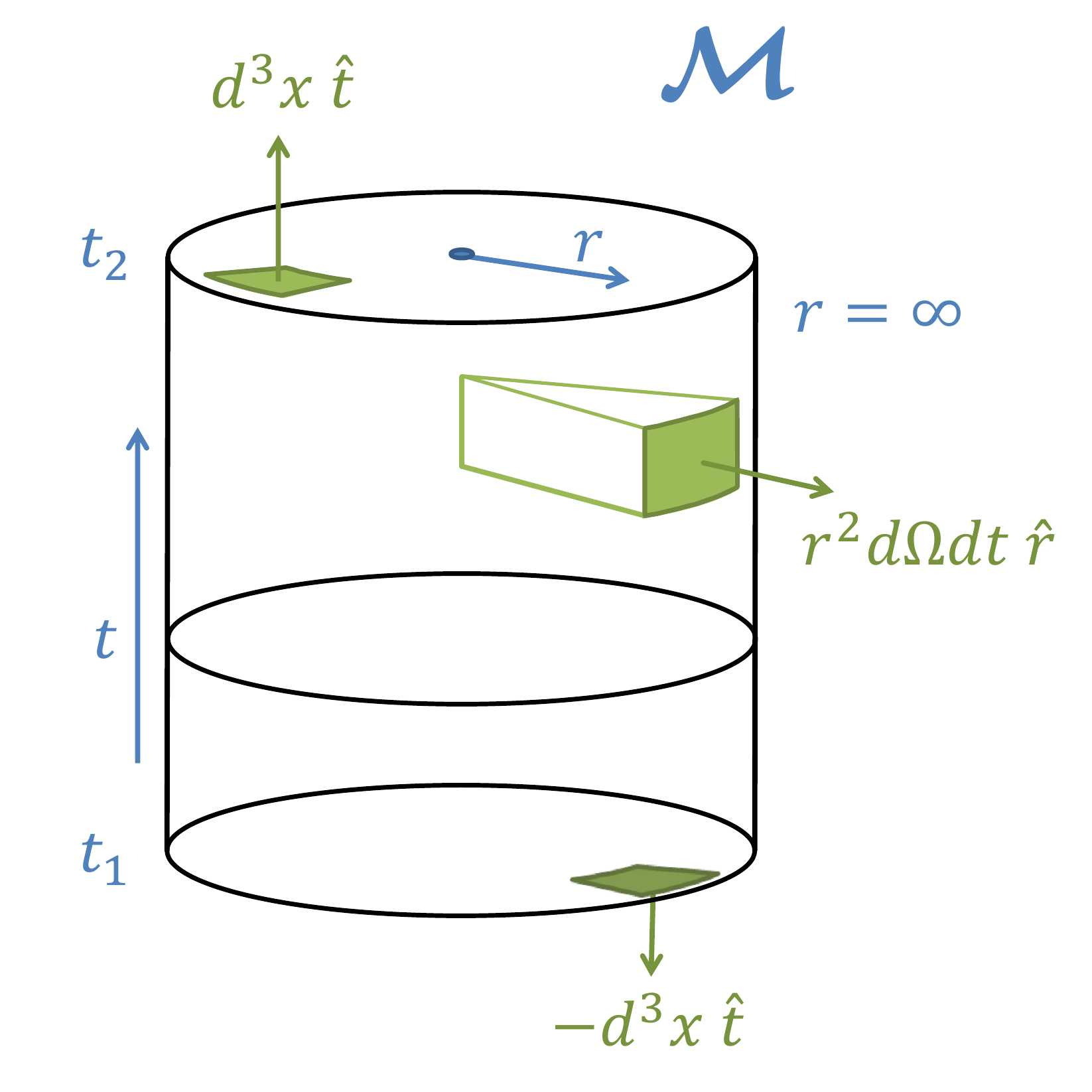} 
\caption{The manifold $\mathcal{M}$}
\label{Manifold}
\end{center}
\end{figure}

To appreciate the difference between gauge and non-gauge theories, let us first discuss a theory {\it without} gauge invariance. Take the simple case of a single real scalar field on some manifold $\mathcal{M}$ which we consider to be non-compact,
\begin{align}\label{}
I=\int_\mathcal{M} d^4x\ \mathcal{L}(\phi,\partial_\mu \phi) \ . 
\end{align}
The variation of the action is
\begin{align}\label{}
\delta I&=\int_\mathcal{M} d^4x\ \left( \frac{\partial \mathcal{L}}{\partial \phi}\delta \phi + \frac{\partial \mathcal{L}}{\partial \phi,_\mu } \delta \phi,_\mu \right) \nonumber \\
&=\int_\mathcal{M} d^4x\ \left( \frac{\partial \mathcal{L}}{\partial \phi}-\partial_\mu \left( \frac{\partial \mathcal{L}}{\partial \phi,_\mu } \right)   \right) \delta \phi + \int_\mathcal{M} d^4x\ \partial_\mu \left( \frac{\partial \mathcal{L}}{\partial \phi,_\mu} \delta\phi \right)\ .
\end{align}
For this system, the usual field equations of motion \eqref{fielde.o.m.} define an extremum provided the boundary term
\begin{align}\label{B}
B& =\int_\mathcal{M} d^4x\ \partial_\mu \left( \frac{\partial \mathcal{L}}{\partial \phi,_\mu} \delta\phi \right) = \int_{\partial \mathcal{M}} \frac{\partial \mathcal{L}}{\partial \phi,_\mu} \delta \phi\ d\Sigma_\mu
\end{align}
vanishes. As stated by Regge and Teitelboim, the action must possess well defined functional derivatives: this must be of the form $\delta I[\phi]=\int \left( \mbox{something} \right)\delta\phi$ with no extra boundary terms spoiling the derivative. The action must be \textit{differentiable} in order for the extremum principle to make sense \cite{Regge-Teitelboim}. In \eqref{B} we have rewritten it through the divergence theorem as an integral at spatial infinity. A typical situation is shown in Fig. \ref{Manifold} (with one dimension suppressed). 

The boundary $\partial M$ of $\mathcal{M}$ thus have three pieces. The two ``covers"\ at constant times $t_1$ and $t_2$, where $d\Sigma_\mu =d^3x$  pointing ``upwards"\ and ``downwards"\ in time respectively; and the cylinder at $r\rightarrow \infty$ where $d\Sigma_\mu = r^2d\Omega dt\, \hat r$ where $d\Omega$ stands for the solid angle. The boundary term is then
\begin{align}\label{BB}
B&=\int \frac{\partial \mathcal{L}}{\partial \phi,_0} \cancel{\delta \phi}\ d^3x\Big|^{t_2}_{t_1} + \int \frac{\partial \mathcal{L}}{\partial \phi,_r}\delta\phi\ r^2d\Omega dt \Big|_{r\rightarrow \infty}.
\end{align}
The first term, evaluated at $t_1$ and $t_2$, vanishes because $\delta\phi(t_1)=\delta\phi(t_2)=0$, i.e. the initial and final states are fixed. This is in full consistency with the equations of motion that require initial and final conditions for a unique solution. The last term is evaluated for large $r$ and one cannot assume that $\phi$ is also fixed there.  If one fixes the field for large $r$, the equations of motion may have no solution at all. 

In non-gauge fields theories one normally deals with fields with compact support where $\phi(r)\rightarrow 0$ fast enough for large $r$ (for instance, a massive field typically exhibits exponential decay).  This means that ${\partial {\cal L} \over \partial \phi,_r}$ falls fast enough at infinity (it suppresses the growth of $r^2$) making the boundary term \eqref{BB} indeed vanishing, so this does not become an issue and it is safe to omit the discussion.

For a gauge theory the situation is quite different. On the one hand there may be long range interactions (zero mass fields) and one cannot assume a compact support. On the other hand, even if the physical fields do vanish fast enough asymptotically, the presence of Lagrange multipliers, with no dynamical equations restricting them, makes this analysis delicate. 

Let's consider first Maxwell's Electrodynamics. To derive Maxwell's equations, one computes the variation of $I=-\frac{1}{4} \int F^2$ and then performs an integration by parts,
\begin{align}\label{}
\delta I&=- \int_M d^4x\ F^{\mu\nu}\partial_\mu \delta A_\nu\nonumber \\
&= \int_M d^4x\ \left( \partial_\mu F^{\mu\nu} \right) \delta A_\nu - \overbrace{\int_M d^4x\ \partial_\mu \left( F^{\mu\nu} \delta A_\nu \right)}^{B}\ .
\end{align}
So here the boundary term we pick up is, by the divergence theorem,
\begin{align}\label{}
B&=\int_{\partial M} F^{\mu\nu} \delta A_\mu d\Sigma_\nu \nonumber \\
&=  \int_{\partial M} d^3x\ F^{i0}\cancel{\delta A_{i} \Big|_{t_1}^{t_2}} + \int_{\partial M} dt d\Sigma_i\ F^{0i}\delta A_{0}   + \int_{\partial M} dt d\Sigma_j\  F^{ij}\delta A_i \ .
\end{align}
Just as before, the first term vanishes because the initial and final data are fixed, $\delta A_i(t_{1,2})=0$. The second and third terms need to be analyzed with care. The second one is more interesting because it involves $A_0$, a Lagrange multiplier. Since this field does not satisfy any equation of motion it can in principle take any value.  That means that the second term could be zero, finite, or even infinity.  One may be tempted to declare simply that $A_0$ must be such that this term vanish. As shown by Regge and Teitelboim \cite{Regge-Teitelboim}, this is not a wise choice. 

Our goal is to describe in detail the role and interpretation of boundary terms. We shall do so in the context of a very simple example, Chern-Simons theory. This is probably the simplest example where boundary conditions and boundary terms play a crucial role. And, despite being very simple, it has a huge structure. There is, however, one important aspect of boundary conditions that Chern-Simons theory does not cover, this is the analysis of radial fall-off conditions. Students who would like to master this important topic must read the classic paper by Regge and Teitelboim in order to properly understand how to deal with boundary terms in general. We present here a simplified version which, we believe, help to get the main ideas without the complicated math. But it must be kept in mind that our analysis does not cover all ingredients of the treatment of boundary conditions and boundary terms.   

The subject of boundary conditions and boundary terms has seen a huge activity since the discovery of the AdS/CFT correspondence \cite{Maldacena,Skenderis} (for a discussion along the ideas presented here see \cite{Andrade}). We shall not discuss the applications to AdS/CFT.

\section{Boundary terms in Chern-Simons theories} \label{bcCS}

Before proceeding, let us briefly review the general method to compute functional variations, which is the basic problem leading to the need of boundary terms. 

Let $f(x)$ a function and consider the functional $F[f] = \int dx\, f(x)^n$.  To compute the functional derivative of $F[f]$ one first computes the variation, 
\begin{equation}
\delta F[f] = \int dx\, n f(x)^{n-1} \delta f(x) \ .
\end{equation} 
With this formula for $\delta F$ at hand we ``divide by $\delta f(y)$" at both sides to obtain
\begin{eqnarray}
{\delta F[f] \over \delta f(y)} &=& \int dx\, n f(x)^{n-1} { \delta f(x)  \over \delta f(y)}\nonumber \\
                             &=&  \int dx\, n f(x)^{n-1} \delta(x-y) \nonumber\\
                              &=& nf(y)^{n-1}.                               
\end{eqnarray}  
A key step in this derivation was the appearance of $\delta(x-y) = { \delta f(x)  \over \delta f(y)} $ that reduces the integral.

This is a very simple example of a functional derivative. The trouble starts when the functional depends on derivatives of $f(x)$, for example $G[f]=\int dx\, f'(x)^n$. In this situation, one normally performs integration by parts until the the variation of the functional can be written in the form $\delta G[f] = \int dx \,$(something)$\delta f(x)$, from where the functional derivative is computed just as we did before.  Let us go through the details, 
\begin{eqnarray}
\delta G[f] &=& n\int dx\, f'(x)^{n-1} \delta f'(x) \nonumber\\
&=& n \int dx\,  {d \over dx} \left( f'(x)^{n-1}\delta f(x) \right) - n(n-1) \int dx\, f'(x)^{n-2} f''(x) \delta f(x)   
\end{eqnarray} 
The first term is a total derivative, and the hope is that it vanishes. If this is the case, then the functional derivative is simple to calculate, 
\begin{equation}
{\delta G[f] \over \delta f(y)} = - n(n-1) \, f'(y)^{n-2} f''(y).
\end{equation}  
The key question is then the value of the boundary terms that arise when we do these integrations by parts. The goal of this section is to explain how to deal and interpret these boundary terms.  We do this for the Chern-Simons action because it is the simplest and yet highly non trivial system where these phenomena can be investigated. 

As shown in detail above, the Chern-Simons action expanded in time + space is automatically Hamiltonian.  We only quote the results \eqref{IcsA}
\begin{align}\label{csh}
I_{CS}[A_0,A_i]=\frac{k}{8\pi} \int dt d^2x\ \epsilon^{ij} \eta_{ab} \left( \tensor{A}{^a_i} \tensor{\dot{A}}{^b_j} - \tensor{A}{^a_0}  \tensor{F}{^b_i_j} \right)\ .
\end{align}
Readers who have not gone through the derivation of \eqref{IcsA} can take (\ref{csh}) as a starting point action. This is an action with two dynamical fields $A^a_i$ and one-Lagrange multiplier $A_0^a$. This Lagrange multiplier imply the constraint, 
\begin{equation}
G_0^a = \epsilon^{ij} F_{ij}^a=0. 
\end{equation} 
This constraint is first class and generated the gauge transformations $\delta A^a _i = D_i \lambda^a$ of this theory. We shall see this in detail below. From (\ref{csh}) we derive the canonical Poisson bracket \eqref{[A,A]}
\begin{align}\label{pbcs}
[ \tensor{A}{^a_i}(x,t),\tensor{A}{^b_j}(x',t) ]=\frac{4\pi}{k} \epsilon_{ij}\eta^{ab}\delta(x-x'). 
\end{align}
The Poisson bracket of any two functions $L_1(A),L_2(A)$ of the canonical variable $A^a_i$ is then
\begin{equation}
[L_1(A), L_2(A) ] = {4\pi \over k}\int d^2 x {\delta L_1(A) \over \delta A^a_i } \epsilon_{ij}\eta^{ab} {\delta L_2(A) \over \delta A^b_j }.
\end{equation} 
We see that calculation of functional derivatives is very important. 

\section{Boundary terms for the generator of the gauge symmetry}

From the CS action \eqref{csh}, at first sight one could naively be led to conclude that the generator of gauge transformations with parameter $\xi^a$ is
\begin{align}\label{G0}
G_0[\xi]&=\frac{k}{8\pi} \int d^2x\ \epsilon_{ij}\xi_a \tensor{F}{^a_i_j}\ ,
\end{align}
that is, a smeared integral of the constraint $F^b_{ij}$ with a test function $\xi$, just as we did for Electrodynamics \eqref{PhiL} and GR \eqref{Phixi}. But this is not correct if one considers a manifold with boundaries. 

We know the correct gauge transformation with parameter $\xi^a$ acting on $A_i^a$,
\begin{equation}
\delta A_i^a = D_i \xi^a.
\end{equation} 
If the functional (\ref{G0}) was the true generator of gauge transformation with the Poisson bracket (\ref{pbcs}) then, the following identity should hold
\begin{equation}\label{qbt1}
 [A_i^a, G_0[\xi]] = \epsilon^{ij}\eta^{ab}{\delta G_0[\xi] \over \delta A^b_j}   \overset{?}{=} D_i\xi^a  \ .
\end{equation} 
The first equality is only the definition of Poisson brackets. The second equality is the tricky one. We now address the following three questions:

\begin{enumerate}
\item Is the functional derivative ${ \delta G_0[\xi]\over  \delta A^b_j}$ well-defined?
\item How one does compute it? 
\item Does it give the desired value (right hand side of (\ref{qbt1}))?. 
\end{enumerate}

To compute the functional derivative we proceed, as explained above, calculating the variation of (\ref{G0}). $G_0$ depends  on $A_i^a$ through $F_{ij}=[D_i,D_j]$. The generic variation of $G_0$ is, keeping all details, 
\begin{align}\label{}
\delta G_0[\xi]&=\frac{k}{8\pi} \int d^2x\ \epsilon^{ij}\xi_a \delta \tensor{F}{^a_i_j}\nonumber \\
&=\frac{k}{4\pi} \int d^2x\ \epsilon^{ij}\xi_a \ D_i \left( \tensor{\delta A}{^a_j} \right)  \\
&=\frac{k}{4\pi} \int d^2x\ \epsilon^{ij} D_i \left( \xi_a \tensor{\delta A}{^a_j}  \right) - \frac{k}{4\pi} \int d^2x\ \epsilon^{ij}\ D_i \xi_a\ \tensor{\delta A}{^a_j}  \label{5.74} \\
&=\delta \, B[\xi] - \frac{k}{4\pi} \int d^2x\ \epsilon^{ij}\ D_i \xi_a \ \tensor{\delta A}{^a_j}  \label{5.76} \ ,
\end{align}
where the boundary term is
\begin{align}\label{BCS}
B[\xi]&=\frac{k}{4\pi}\oint_{r\rightarrow\infty} d\varphi\ \xi_a\ \tensor{A}{^a_\varphi}\ .
\end{align}

Here, to obtain \eqref{5.76} we used that $D_i \left( \xi_a \delta A^a \right)=\partial_i \left( \xi_a \delta A^a \right)$ since $\xi_a \delta A^a $ a scalar under internal rotations, and then applied Stoke's theorem bringing in a boundary term. Note also that in the last step we have pulled out the variation since $\delta \xi^a=0$. This is not a naive step, as it may look, although correct in this case. We shall come back to this below. 

The variation $\delta G_0$, as shown in the last line (\ref{5.76}), has two pieces. One is a bulk contribution with the expected form $\int \mbox{(something)}^{i}_{\ a}\delta A^{a}_{\ j}$, ready to compute the functional derivative. But, there is also the unwanted boundary term $B$. 

At this point it becomes pertinent to ask whether the term $B[\xi]$ is different from zero or not. We note that it depends linearly on $\xi^a$, the gauge parameter, which in principle can take \textit{any} value. So, irrespective of the values of $A_\varphi$, that also enters in $B[\xi]$, one can always choose $\xi^{ a} $ such that this term is either zero or not zero. The arbitrariness of the values of these boundary terms in gauge theories prompted Regge and Teitelboim \cite{Regge-Teitelboim} to understand their meaning opening a key and seminal route to understand several issues related to gauge theories, including the definition of energy in general relativity. 

Before explaining the meaning of $B[\xi]$ we note the following. If $B[\xi]$ in \eqref{BCS} is not zero, one can always pass it to the other side and write equation (\ref{5.76}) in the form, 
\begin{equation}\label{dg}
\delta \bigg( G_0[\xi] - B[\xi]
 \bigg) = - \frac{k}{4\pi} \int d^2x\ \epsilon^{ij}  \ D_i \xi_a\ \tensor{\delta A}{^a_j}\ .
\end{equation} 
This equation suggest defining a new functional $G[\xi] \equiv G_0[\xi] - B[\xi] $, 
or explicitly,
\begin{align}\label{}
G[\xi] = \frac{k}{8\pi} \int d^2x\ \epsilon_{ij}\xi_a \tensor{F}{^a_i_j} -\frac{k}{4\pi}\oint_{r\rightarrow\infty} d\varphi\ \xi_a\ \tensor{A}{^a_\varphi}\ ,
\end{align}
whose variation is well-defined and, directly from (\ref{dg}),
\begin{align}\label{dG/dA}
 \frac{\delta G[\xi]}{\delta \tensor{A}{^a_j}}=- \frac{k}{4\pi}  \epsilon^{ij} D_i \xi_a \ .
\end{align}

The conclusion is now clear: in the presence of boundaries, the generator of gauge transformations is not $G_0[\xi]$ but $G[\xi]$.  Indeed, if we now go back to (\ref{qbt1}) but now we consider not $G_0[\xi]$ but $G[\xi]$, the functional derivative is well defined and we get
\begin{equation}\label{qbt}
 [A_i^a, G[\xi]] = \epsilon^{ij}\eta^{ab}{\delta G[\xi] \over \delta A^b_j}   \overset{\checkmark}{=} D_i\xi^a  \ ,
\end{equation} 
obtaining exactly the correct result for a gauge transformation. 

Of course, there are gauge transformations (choices of $\xi^a$) where the boundary term {\it is}  zero. Regge and Teitelboim noted that not all gauge transformations are on equal footing, and made the following classification:

\begin{itemize}
\item \textbf{Proper gauge transformations} are those choices of $\xi ^a$ such that $B[\xi]=0$. These form the class of `purely gauge' transformations that do not change the physics state of a system. Their generator is purely a constraint. Note that the value of $B[\xi]$ does not only depend on the choice of $\xi^a$ but also on the boundary values of the field $A_\varphi$. We discuss this is detail below.

\item \textbf{Improper gauge transformation} are those choices of $\xi ^a$ such that $B[\xi]\neq 0$.  This class of gauge transformations must not be considered ``pure gauge". They do change the state of the system. Let us put this on firm basis. Suppose we have a parameter $\xi^a$ and we act on $A^a_i$ generating the transformation $\delta A_a^a = D_i \xi^a$, then if the generator $G[\xi]$ is not zero ($B[\xi]$ is not zero) then this transformation is physical, it can be measured in an experiment.  
\end{itemize}

\section{Comments}

\begin{enumerate}

\item \textbf{Asymptotic Symmetries}. The most important aspect of the above calculation is the notion of ``asymptotic symmetries" \cite{Regge-Teitelboim}.   The question of whether $B[\xi]$ is zero or not, is formulated elegantly in terms of symmetries. The asymptotic symmetry group of a given field with specified boundary conditions is defined as the set of all the symmetry transformations of the field that preserve the asymptotic boundary conditions and possess a non-zero conserved charge. Normally, one does not choose asymptotic conditions trying to force $B$ to be zero or not. Instead, physical intuition for the problem at hand determines the behaviour of fields at infinity. 

As an example, consider the set of all connections $A_r,A_0,A_\varphi$ such that at the boundary satisfy, 
\begin{equation}\label{acond}
A_0=0, \ \ \ \ A_r=0 \hspace{2cm}\mbox{at}\ r\rightarrow \infty \ .
\end{equation} 
This choice of fields at infinity puts restrictions on the other component $A_\varphi$ through the equations of motion. Indeed, the Chern-Simons equations $F=0$ together with (\ref{acond}) imply that $A_\varphi(\varphi)$ is also only a function of $\varphi$. The next step is to look for the set of all gauge transformations $\delta A_\mu = D_\mu\xi$ that leave the boundary conditions (\ref{acond}) invariant.  That is, solve the equations $D_t\xi=D_r\xi=0$. These equations are very simple and the solution is given by parameters $\xi^a$ that depend only on $\varphi$, 
\begin{equation}\label{asimcs1}
\xi^a = \xi^a(\varphi).
\end{equation} 

The degrees of freedom remaining at the boundary have a very simple form. The only non-zero field is $A_\varphi(\varphi)$. The theory is also invariant under transformations whose parameters only depend on $\varphi$, 
$\delta A_\varphi = D_\varphi \xi$. The set of ``asymptotic symmetries" is generated by the improved generator and is thus non-trivial. 

A much more interesting example, widely used in CFT's and also in black hole physics\cite{Max99} as we shall see below, is the \textit{chiral condition}
 \begin{equation}\label{achiral}
A_0=A_\varphi, \ \ \ \ A_r=0 \hspace{2cm}r\rightarrow \infty\ .
\end{equation} 
The Chern-Simons equations $F=0$, together with (\ref{achiral}), imply that $A_\varphi(t+\varphi)$. Next we seek the set of all gauge transformations $\delta A_\mu = D_\mu\xi$ that leave (\ref{achiral}) invariant.  This time one finds that the solution is given by parameters, 
\begin{equation}\label{asimcs}
\xi^a = \xi^a(t+\varphi).
\end{equation} 
The degrees of freedom remaining at the boundary are the chiral fields $A_\varphi(t+\varphi)$ (and its holomorphic counterpart) and the theory is invariant under chiral transformations with parameters $\xi(t+\varphi)$. 

These examples exhibit the idea of a ``boundary symmetry" but overlooks the important problem of fall off conditions. In most cases, the asymptotic conditions (like (\ref{acond})) are expressed as fall offs in powers of $r$.  One then studies the set of all transformations leaving those conditions invariant.

\item {\bf The Asymptotic Algebra $ \left[ G[\xi],G[\rho] \right] $} .  An important consistency check  for the generator of gauge transformations is their algebra. Since $G[\xi]$ has well-defined variations we can compute directly its algebra. We provide all details\footnote{For readers familiar with differential forms this calculation can be made much shorter by noticing that $\delta G(\xi)/\delta A = D \xi$ is a 1-form. Thus  $[G(\xi),G(\rho)]=\int {\delta G(\xi)\over \delta A} \, {\delta G(\rho)\over \delta A} = \int D\xi D\rho = \int d(\xi D\rho) - \int \xi DD\rho =  G([\xi,\rho]) + \int \xi d\rho $. }, 
\begin{align}\label{}
\big[ G[\xi] , G[\rho] \big]&=\int d^2x\ \frac{4\pi}{k}\epsilon_{ij}\eta^{ab} \frac{\delta G[\xi]}{\delta \tensor{A}{^a_i}(x)} \frac{\delta G[\rho]}{\delta \tensor{A}{^b_j}(x)} \nonumber  \\
&=\frac{k}{4\pi}\int d^2x\ \eta^{ab} \underbrace{\epsilon_{ij} \epsilon^{in} \epsilon^{jm}}_{\epsilon^{nm}}\left( D_n \xi_a \right)  \left( D_m \rho_b \right) \nonumber \\
&=\frac{k}{4\pi}\int d^2x\ \epsilon^{nm} \partial_n \left(  \xi_a  D_m \rho^a \right) - \frac{k}{4\pi}\int d^2x\ \epsilon^{nm} \xi_a   D_n D_m \rho^a  \nonumber \\
&=\frac{k}{4\pi}\oint_{r\rightarrow \infty} d\varphi\ \xi_a D_\varphi \rho^a - \frac{k}{4\pi} \int d^2x\ \frac{1}{2} \xi_a \tensor{\epsilon}{^a_b_c} \tensor{F}{^b_i_j} \rho^c \epsilon^{ij} \nonumber \\
&=\frac{k}{4\pi}\oint_{r\rightarrow \infty} d\varphi\ \xi_a  \partial_\varphi \rho^a  -  \frac{k}{4\pi}\oint_{r\rightarrow \infty} d\varphi\,   \tensor{\epsilon}{_a_b_c}  \xi^a \rho^c\, \tensor{A}{^b_\varphi}  + \frac{k}{8\pi} \int d^2x\,  \tensor{\epsilon}{_a_b_c}  \xi^a \rho^c \epsilon^{ij} \tensor{F}{^b_i_j} \ ,
\end{align}
where we used that $\xi_a \rho^a$ is a scalar under the gauge group, $\epsilon^{nm}D_nD_m \rho^a= \frac{1}{2} \epsilon^{nm} F_{nm}\rho^a$ and also used Stokes theorem. The last two terms group to form $G$ again, and thus we finally find
\begin{align}\label{KMalgebra}
\boxed{ \big[  G[\xi],G[\rho] \big]  =G\big[ [\xi,\rho] \big] + \frac{k}{4\pi} \oint_{r\rightarrow\infty} d\varphi\ \xi_a \partial_\varphi \rho^a }
\end{align}
where we have used $\left( [\xi,\rho] \right)_b=\epsilon_{abc}\xi^a\rho^c$ for $\xi,\rho$ in the gauge algebra. The last term is called a ``central extension" of the algebra. By further imposing more restrictive boundary conditions, this affine algebra reduces to the Virasoro algebra, in accordance the famous result of Brown and Henneaux \cite{BH86}, but now in the Chern-Simons formulations instead of the metric one\cite{Max99}.

\item {\bf Boundary terms for the action: Hamiltonian and Energy.} 

Naively, when confronting the action   (\ref{csh}) one may conclude that the energy of this system is zero because its Hamiltonian is simply a combination of constraints.  This is however incorrect. The true Hamiltonian develops a boundary term, just like the generator of gauge transformations we have just described. Now, the Hamiltonian itself generates a gauge transformation with parameter $\xi=A_0$. In this sense everything we have said about the generator of gauge transformation could be applied to the Hamiltonian. There are however subtleties related to the boundary conditions, which we discuss below.   

The CS action (\ref{csh}) must be varied with respect to $A_0$ and $A_i$ independently, and for the equations of motion to constitute an extremum, all boundary terms in the variations leading to the e.o.m. must cancel. When we wrote down the CS e.o.m. \eqref{cseom1}-\eqref{cseom2}, we never actually checked that they define an extremum of that action. And in fact, they don't! So the problem we discussed for the generator of gauge transformations also appears in the action itself. We shall now analyse in detail these boundary terms, and find their physical interpretation.

Let's go through the same procedure as before to find the boundary term - we must compute the generic variation of the action \eqref{csh} and isolate the e.o.m.(for notation simplicity, we go back to the matrix form):
\begin{align}\label{}
\delta I_{CS}&=\frac{k}{8\pi}\delta \int d^3x\ \epsilon^{ij} \mbox{Tr} \left(A_i\dot{A}_j - A_0 F_{ij} \right)\nonumber\\
&=\frac{k}{8\pi}\int d^3x\ \epsilon^{ij} \mbox{Tr} \left( -\delta A_0 F_{ij} + \delta A_i \dot{A}_j - \dot{A}_i \delta A_j  - A_0 \delta F_{ij} + \cancel{\frac{d}{dt}\left( A_i \delta A_j \right)} \right) \nonumber \\
&=-\frac{k}{8\pi}\int d^3x\ \epsilon^{ij} \mbox{Tr} \left( \delta A_0 F_{ij} + 2\dot{A}_i \delta A_j  + A_0 \delta F_{ij}  \right) \ ,
\end{align}
where the time derivative vanishes since the initial/final configurations are held fixed, and we used the cyclic property of the trace. Moreover, it is direct to show that $\epsilon^{ij} \mbox{Tr}\left( A_0\delta F_{ij} \right)=2\epsilon^{ij} \mbox{Tr} \left[   \partial_i \left( A_0\delta A_j \right)- D_i A_0 \delta A_j \right]$, and thus we obtain
\begin{align}\label{}
\delta I_{CS}&= - \frac{k}{8\pi} \int d^3x\ \epsilon^{ij} \mbox{Tr} \left( \delta A_0 F_{ij} + 2 \left( \dot{A}_i -D_i A_0 \right) \delta A_j + 2 \partial_i \left( A_0 \delta A_j \right) \right)\nonumber \\
&= \frac{k}{8\pi} \int \mbox{(e.o.m.)}  -  \frac{k}{4\pi} \int dt \int d^2x\ \partial_i \mbox{Tr} \big(  \epsilon^{ij} \left( A_0 \delta A_j \right)  \big)\nonumber \\
&= \frac{k}{8\pi} \int \mbox{(e.o.m.)}  -  \frac{k}{4\pi} \int dt \int_{r\rightarrow \infty} d\varphi\  \mbox{Tr} \big( \left( A_0 \delta A_\varphi \right)  \big) \ , \label{dIcs}
\end{align}
where in the last step we used Stoke's theorem. Thus in order for the CS e.o.m. to be an extremum, we should pass the boundary term in \eqref{dIcs} to the left hand side and this should define a new action:
\begin{align}\label{}
\delta I_{CS} +  \frac{k}{4\pi} \int dt \int_{r\rightarrow \infty} d\varphi\  \mbox{Tr} \big( \left( A_0 \delta A_\varphi \right)  \big)  =\frac{k}{8\pi} \int \mbox{(e.o.m.)} \ .
\end{align}

But this is equivalent to a redefinition of the Hamiltonian in the action: instead of the `naive' Hamiltonian $H_0 = \frac{k}{8\pi}\int \epsilon^{ij} A_0 F_{ij}$, the true Hamiltonian must include an extra boundary term $E$,
\begin{equation}\label{hcs}
 H = H_0 + E \ ,
\end{equation} 
whose variation we know,
\begin{align}\label{dE}
\delta E&=\frac{k}{4\pi} \int dt \int_{r\rightarrow \infty} d\varphi\  \mbox{Tr} \big( \left( A_0 \delta A_\varphi \right) \big). 
\end{align}
This $E$ is by definition the boundary term that makes the functional $H$ or equivalently the action well-defined. We choose to call it $E$ referring to `energy', because it equals the value of $H$ on any solution of the equations of motion, since the bulk piece $H_0$ is a constraint. We therefore replace the Hamiltonian $H_0$ appearing in (\ref{csh}) by $H$ (\ref{hcs}). 

Equation \eqref{dE} defines the value of $E$ which depends on boundary conditions. Recall that in \eqref{5.76} we were able to simply pull the variation $\delta$ assuming that $\delta \xi=0$. In the present case this is not the correct procedure, because $A_0$ might depend on $A_\varphi$ at the boundary. So in order to find $E$, we must provide some more information about the system's behaviour at the boundary. 

Field theories are defined by their e.o.m. as well as their boundary conditions. Given one set of e.o.m., here the CS equations $F_{\mu\nu}=0$, one can define several different systems simply by choosing different boundary conditions, and $H$ then measures the energy for each one of those systems with particular boundary conditions. For example, one can choose the condition mentioned above $A_0=0$ (at the boundary), and under this condition, the boundary term is simply zero, and the energy is zero.  

A more useful choice for boundary conditions is the chiral condition referred above $A_0 = A_\varphi$, and then we can proceed with the calculation of $E$ from \eqref{dE},
\begin{align}
\delta E&=\frac{k}{4\pi} \int dt \int_{r\rightarrow \infty} d\varphi\  \mbox{Tr} \big( \left( A_\varphi \delta A_\varphi \right)  \big) = \frac{k}{4\pi} \delta  \int dt \int_{r\rightarrow \infty} d\varphi\  \mbox{Tr} \left( \frac{1}{2}A_\varphi^2 \right)  \ ,
\end{align}
from where we obtain 
\begin{equation}
E[A_\varphi] = \frac{k}{8\pi} \int dt\int_{r\rightarrow\infty} d\varphi\ \mbox{Tr} \left( A_\varphi^2 \right)  \ ,
\end{equation} 
which is now the energy of the system. 

\item {\bf 3d Gravity and BTZ black hole.}  In many situations, like the Chern-Simons formulation of 2+1 gravity \cite{Witten88}, the full action is actually the {\it difference} of two copies of the Chern-Simons action, 
\begin{equation}
I = I[A^+] - I[A^-] \ ,
\end{equation} 
for two independent fields $A^+$ and $A^-$. The corresponding choice of boundary conditions, leading to holomorphic and anti-holomorphic currents, is  
\begin{equation}
A^+_0= A_\varphi, \ \ \ \ \  A^-_0 = -A^-_\varphi \hspace{2cm}\mbox{for}\ \ r\rightarrow \infty \ .
\end{equation}   
The energy of the full system is then
\begin{align}\label{}
E[A_\varphi,\bar A_\varphi] =  \frac{k}{8\pi} \int dt\int_{r\rightarrow\infty} d\varphi\ \mbox{Tr} \left( \left( A^{+}_\varphi \right)^2-\left( A^-_\varphi \right)^2 \right)  \ .
\end{align}
A very illuminating example is to evaluate this energy for the BTZ black hole \cite{BTZ}. This is the analogue of the calculation of Regge and Teitelboim for the four dimensional Schwarzschild black hole \cite{Regge-Teitelboim}. The BTZ black hole with mass $M$ and angular momentum $J$, in the CS formulation, is described by a constant connection that obeys $\mbox{Tr} \left( A^\pm_\varphi \right) ^{2} =\pm \frac{2}{k}\left( M\pm J \right) $\cite{Max99}. Therefore in Euclidean time where $0 \leq t\leq 1$, the energy for the BTZ black hole is
\begin{align}\label{}
E_{BTZ}&=
  \frac{k}{8\pi} \int_{r\rightarrow\infty} d\varphi\ \left( \frac{2\pi}{k} \left( M+J \right)+\frac{2\pi}{k} \left( M-J \right) \right) = M \ ,
\end{align}
that is the black hole's mass, just as in the four dimensional case\cite{Regge-Teitelboim}.

\item {\bf The Hamiltonian generates time translations.} 

$H$ is now the correct generator of time translations, because since $H$ has well-defined functional derivatives, by the same steps leading to \eqref{qbt}, the e.o.m. now read
\begin{equation}
\tensor{\dot A}{^a_i} = [A_i^a, H] = \epsilon^{ij}\eta^{ab}{\delta H \over \delta A^b_j}  =  D_i \tensor{A}{^a_0}  \ .
\end{equation} 
This can be understood from an alternative perspective. As usual, we can regard the action of an infinitesimal change of coordinates ${x'}^{\mu}=x^\mu-\xi^\mu(x)$ over the fields as a variation of the fields themselves. It is not difficult to show that
\begin{align}\label{dAcs}
\delta \tensor{A}{^a_\mu}&=\mathcal{L}_\xi \tensor{A}{^a_\mu}
=\xi^\nu \tensor{F}{^a_\nu_\mu} + D_\mu \left( \xi^\nu \tensor{A}{^a_\nu} \right) \ ,
\end{align}
which, along a solution of the e.o.m. $F=0$ is simply a gauge transformation with parameter $\xi \cdot A^a$,
\begin{align}\label{}
\delta \tensor{A}{^a_\mu}&=D_\mu \left( \xi^\nu \tensor{A}{^a_\nu} \right) \ .
\end{align}
Thus we see that on shell there exist a correspondence between infinitesimal diffeomorphisms and gauge transformations of the field. In particular, a time translation will be associated with an infinitesimal gauge parameter $\xi^0 A_0$, which is perfectly consistent with \eqref{dAcs}.

\item {\bf Conserved Charges}. As we have mentioned, gauge symmetries do not carry Noether charges. In the presence of boundaries, however, the sub-group of gauge symmetries defining ``improper gauge transformations"  
is not generated by constraints, they do change the physical state, and have non-zero Noether charges.    

Nonetheless, improper gauge transformations are still symmetries of the action. The difference with a proper gauge symmetry is that the former are generated by non-zero quantities. But the transformations itself -irrespective of its generator-- is the same as always, and is a symmetry. 

As usual in Hamiltonian mechanics, the generator of a symmetry is conserved. The goal of this paragraph is to prove this statement explicitly. We shall prove that the improved generator $G[\xi]$ satisfies, 
\begin{equation}
{d \over dt}G[\xi] =0.
\end{equation}  
Since $G[\xi]$ is not zero, this equation does provide a non-trivial charge. 

The time derivative of any functional of the canonical variables $F[A_i]$ is computed by it's Poisson bracket with the Hamiltonian 
$\dot F = [F,H]$. The generator $G[\xi]$, however, depends not only on $A_i$ but also on the parameter $\xi$ which may also depend on time. The full time derivative of $G[\xi]$ is then  
 \begin{equation}\label{dotG}
 {d \over dt} G[\xi] = [G[\xi],H] + \int d^2x\ {\delta G[\xi] \over \delta \xi} \dot \xi  \ .
 \end{equation} 
Since the Hamiltonian is itself a gauge generator with a parameter $A_0$, i.e. $H=G[\xi=A_0]$, the first term in \eqref{dotG} is evaluated through \eqref{KMalgebra}. Restricting to a single sector of the CS theory, we get
\begin{eqnarray}
{d \over dt} G[\xi] &=& \int d^2x\, \epsilon^{ij} [\xi,A_0] F_{ij} + \int_{r\rightarrow \infty} d\varphi\ \xi \partial_\varphi \, A_0 + \int d\varphi\,  A_\varphi \dot \xi \nonumber\\
&=& 0 \ \  + \ \ \int d\varphi\, \xi\, ( \partial_\varphi - \partial_0) A_\varphi \nonumber\\
&=&  0  \ ,
\end{eqnarray} 
where we used the chiral boundary condition $A_0=A_\varphi$, and evaluated on shell $F_{ij}=0$.

This calculation is rarely performed explicitly because in most cases there are other ways to see that the charges are time independent. In this example it is easy to see that 
\begin{align}\label{Q_xi}
Q[\xi]=-\frac{k}{4\pi} \int_{r\rightarrow \infty} d\varphi\ \xi_a \tensor{A}{^a_\varphi}\hspace{1cm}
\end{align}
does not depend on time. (This is non trivial since $\xi^a$ and $A_\varphi$ do depend on time.) 

First we note that both $\xi(t+\varphi)$ and $A_\varphi(t+\varphi)$ are periodic in $\varphi$ and therefore also in $t+\varphi$. Thus we may expand them as Fourier series
\begin{align}\label{}
\xi_a(t+\varphi)&=\sum_n \xi_{na} e^{in(t+\varphi)} \hspace{.5cm},\hspace{.5cm} \tensor{A}{^a_\varphi}(t+\varphi)=\sum_m \tensor{A}{^a_\varphi_m} e^{im(t+\varphi)} \ , \nonumber
\end{align}
and the charge \eqref{Q_xi} gives
\begin{align}\label{}
Q[\xi]&=-\frac{k}{4\pi} \sum_{n,m} \xi_{na} \tensor{A}{^a_\varphi_m} \int_{0}^{2\pi} d\varphi\ e^{i(t+\varphi)(n+m)}\nonumber\\
&=-\frac{k}{2} \sum_m \xi_{a\ -m} \tensor{A}{^a_\varphi_m}\label{QFourier}
\end{align}
which is indeed time independent (we have used that  $\int_0^{2\pi}  d\phi e^{i\ell \phi} = 2\pi \delta_{\ell,0}.$). Of course there are infinite conserved charges since there are infinitely many $\xi_n$. Finally, one can proceed with Dirac's algorithm by defining the Dirac bracket and show that the components of the field $A$ satisfy, in the quantized version (where we replace the Dirac bracket by the commutator), the Kac-Moody algebra\cite{Max99}:
\begin{align}\label{KM}
[\tensor{A}{^a_n},\tensor{A}{^b_m}]&= i \tensor{\epsilon}{^a^b_c} \tensor{A}{^c_{n+m}} +\frac{nk}{2}\delta^{ab} \delta_{n+m,0}
\end{align}
For the sl(2,$\Re$) example, the final step is to impose further boundary conditions in order to ensure that the solution is asymptotically $AdS_3$, and one can show that \eqref{KM} reduces to the Virasoro algebra, with the central charge of Brown and Henneaux $c=-6k=\frac{3L}{2G}$\cite{BH86}.

\end{enumerate}

\chapter*{Acknowledgments } 
The authors would like to thank C. Armaza, G. Barnich, C. Bunster, A. Castro, A. Faraggi, P. Ferreira, A. Gomberoff, M. Henneaux, J.I. Jottar, B. Koch, M. Pino, A. Schwimmer and S. Theisen for many discussions on the topics covered in this review. Special thanks to Jegor Korovins for  many useful comments on the first version of this work.  The authors would also like to thank Ivan Muñoz for assistance and comments on the second version.  MB would like to thank specially C. Bunster and M. Henneaux, from whom he learned most of the material covered in this work. MB is partially supported by FONDECYT Chile, grant \# 1141 221. IR is partially supported by CONICYT (Chile) through CONICYT-PCHA/Doctorado Nacional/2014. 

\bibliography{refs}
\bibliographystyle{ieeetr}

\end{document}